\documentclass[fleqn,usenatbib]{mnras}
\usepackage{bm}

\usepackage{amssymb}
\usepackage{lmodern}
\usepackage[T1]{fontenc}

\usepackage{placeins}

\DeclareRobustCommand{\VAN}[3]{#2}
\let\VANthebibliography\thebibliography
\def\thebibliography{\DeclareRobustCommand{\VAN}[3]{##3}\VANthebibliography}

\usepackage{graphicx}	% Including figure files
\usepackage{amsmath}	% Advanced maths commands

\usepackage{array}
\newcolumntype{C}[1]{>{\centering\arraybackslash}p{#1}}

\defcitealias{standing26dmpp}{S26}
\defcitealias{pace13}{P13}
\defcitealias{murphy16}{M16}
\defcitealias{castro-gonzalez24}{CG24}
\defcitealias{muller24}{M24}

\newcommand{\degs}{$^{\circ}$}
\newcommand{\vsini}{$v\,\sin\,i$}{}
\newcommand{\ms}{m\,s$^{-1}$}{}
\newcommand{\kms}{km\,s$^{-1}$}{}
\newcommand{\logrhk}{log~$R^{\prime}_{\textrm{\textsc{HK}}}$}{}
\newcommand{\logrhkmax}{log~$R^{\prime}_{\textrm{\textsc{HK}}|\textrm{max}}$}{}

\newcommand{\mpsini}{$m_\textrm{p}\,\sin\,i$}{}

\newcommand{\hdsix}{\texorpdfstring{\hbox{HD\,67200}}{HD 67200}}
\newcommand{\hdtwo}{\texorpdfstring{\hbox{HD\,2134}}{HD 2134}}
\newcommand{\hdone}{\texorpdfstring{\hbox{DMPP-7}}{DMPP-7}}

{}

\title[DMPP exoplanet systems]{Three new exoplanet systems from the Dispersed Matter Planet Project}

\author[J. Barnes et al.]{
J.~R.~Barnes,$^{1}$\thanks{E-mail: john.barnes@open.ac.uk}
C.~A.~Haswell,$^{1}$
Z.~O.~B.~Ross,$^{1}$
E.~Rutherford,$^{1}$
M.~R.~Standing$^{2,1}$
\newauthor
~A.~T.~Stevenson$^{3,1}$
D.~Staab,$^{1}$
L.~Fossati,$^{4}$ %
J.~S.~Jenkins,$^{5,6}$ %
D.~Alves,$^{7}$ %
\\
$^{1}$School of Physical Sciences, The Open University, Walton Hall, Milton Keynes. MK7 6AA. UK \\
$^{2}$European Space Agency (ESA), European Space Astronomy Centre (ESAC), Camino Bajo del Castillo s/n, \\
~E-28692 Villanueva de la Ca\~{n}ada, Madrid, Spain \\
$^{3}$School of Physics \& Astronomy, University of Birmingham, Edgbaston, Birmingham B15 2TT, UK\\
$^{4}$Space Research Institute, Austrian Academy of Sciences, Schmiedlstra{\ss}e~6, 8042 Graz, Austria \\
$^{5}$Instituto de Estudios Astrof\'{i}sicos, Facultad de Ingenier\'{i}a y Ciencias, Universidad Diego Portales, Av. Ej\'{e}rcito 441, Santiago, Chile \\
$^{6}$Centro de Astrof\'isica y Tecnolog\'ias Afines (CATA), Casilla 36-D, Santiago, Chile\\
$^{7}$Departamento de Astronom\'ia, Universidad de Chile, Casilla 36-D, Santiago, Chile\\
}

\date{Accepted for publication in MNRAS. DOI: 10.1093/mnras/stag1486}

\pubyear{\the\year{}}

\begin{document}
\label{firstpage}
\pagerange{\pageref{firstpage}--\pageref{lastpage}}
\maketitle

\begin{abstract}
{We present a radial velocity analysis of three bright, low-activity stars identified by the Dispersed Matter Planet Project (DMPP). We use a Bayesian framework to compare purely Keplerian models with models incorporating stellar activity via a quasi-periodic Gaussian Process (GP). DMPP-7 (HD 118006) is a slightly evolved star that harbours a single $0.72$ Saturn-mass giant (\mpsini{}~$= 69$\,M$_\oplus$) with an orbital period of $P=4.93$\,d. A longer $21$\,d\,--\,$22$\,d period cannot be conclusively confirmed as a stellar rotation signature rather than a purely Keplerian signal. For \hdsix{}, which exhibits Ca}~\textsc{ii}~{H\&K variability, a model with only a GP is strongly favoured over a purely dynamical model. The GP model shows moderate evidence for a single Keplerian with \hbox{$P=2.67$\,d}. For \hdtwo{}, a $21$\,d\,--\,$32$\,d rotation period signal is associated with tentative FWHM variability. A model with a GP is not conclusively favoured, but all models considered show moderate evidence for an additional single Keplerian with $P=2.78$\,d. Despite our target selection favouring near edge-on orbital geometries, we find no evidence for transits in TESS photometry. \hdone{}\,b lies at the transition between the high-radius population and the Neptunian ridge and savannah regions. Further observations are required to establish whether the coherent short-period \hdsix{} and \hdtwo{} signals are stellar or dynamical in origin. If planetary, the signals correspond to minimum masses of \mpsini{}~$= 2.07$\,M$_\oplus$ and \mpsini{}~$= 2.86$\,M$_\oplus$.}
\end{abstract}

%We searched for stellar variability signatures using spectroscopic activity indicators to inform radial velocity (RV) modelling. 
%We examine the demographics of the candidate planets in these systems in the context of the broader exoplanet population and use mass–radius relationships to infer atmospheric properties.


%%%%%%%%%%%%%%%%%%%%%%%%%%%%%%%%%%%%%%%%%%%%%%%%%%%%%%%%%%%%%%%
\section{Introduction}
\protect\label{section:intro}
%%%%%%%%%%%%%%%%%%%%%%%%%%%%%%%%%%%%%%%%%%%%%%%%%%%%%%%%%%%%%%%

% Planet discovery numbers
% K2 (800-258) + 1 = 545
% Kepler (3427 - 822) + 1 = 2606
% K2 + Kepler = 3151 - i.e. 3151/4296 = 73.35%

%1818 with mass or mpsini (selected from pl_massj column of spreadsheet PS_2025.11.14_02.06.13_edit.odt)
% <= 6.2Mearth : 355 /  1818          = 0.19527 or 19.5%
% <= Mneptune  : (761 - 355) / 1818   = 0.22332 or 22.3%
% <= Mjup      : (1418 - 761) / 1818  = 0.36139 or 36.1%
% > Mjup       : (1818 - 1418) / 1818 = 0.22022 or 22.0% (use 22.1 so add up to 100%) 

% 4296 with measured radius (selected using pl_rade column from spreadsheet PS_2025.11.14_02.06.13_edit.odt)
% <= 1.8Rearth : 1383 / 4296          = 0.3219 or 32.2%
% <= Rneptune  : (3237 - 1383) / 4296 = 0.43156 or 43.2%
% <= Rjup      : (3672 - 3237) / 4296 = 0.101257 or 10.1%
% > Rjup       : (4296 - 3672) / 4296 = 0.145251 or 14.5%
Exoplanet research is increasingly focused on terrestrial planets and smaller gas giants. The statistics of these planets are dominated by transit-based surveys, particularly the Kepler and K2 missions, which comprise over 73\% of {discoveries \citep{christiansen2025}}\footnote{\label{note1}{Extracted from the NASA Exoplanet Archive on 01/06/26 \url{https://exoplanetarchive.ipac.caltech.edu}}}{.} Hence, the number of planets with a measured radius outnumber those with determined mass or minimum mass by a factor of $2.4$. This discrepancy arises largely because many of the transiting systems are too faint for efficient radial velocity (RV) follow-up.
The two detection methods have differing selection effects, so it is unsurprising that the subsets of known planets discovered by each of them have differing demographics.
\citealt{muller24} (\citetalias{muller24}) derived the mass-radius relationship using planets with reliable values for both mass and radius. {The relationships facilitate a comparison of planets in a given mass interval with those in the corresponding radius interval,
%This can be used to estimate a planet's approximate probable mass or radius when only the other parameter has been measured 
revealing} the RV method has been relatively unsuccessful at discovering and measuring masses for low-mass planets. The fraction of exoplanets with measured radii falling at or below $1.8$\,R$_\oplus$, which corresponds to the Radius Valley for FGK stars, is approximately 32\%, while the planets with measured corresponding masses (or minimum masses) of $\lesssim 6.2\,\textrm{M}_\oplus$ (\citetalias{muller24}) is only 20\%. 
{Similarly, for sub-Neptunian to Neptunian planets with probable $\textrm{H}_2$-dominated atmospheres, 43\% have measured radii in the $1.8-3.9,\textrm{R}\oplus$ range. In contrast, only 22\% have mass determinations in the corresponding $6.2-17.15,\textrm{M}_\oplus$ range.}
{Conversely, the} largest and most massive planets are relatively over-represented in the RV population: the fraction of planets with measured radius $> \textrm{R}_\textrm{Nept}$ is 24\%, while the corresponding fraction of planets with measured mass $> \textrm{M}_\textrm{Nept}$ is $\sim 48$\%. This last category of planet is the easiest to detect via either method, but the ratio of known large and small transiting planets is limited by the inherent rarity of large planets relative to small ones. This factor is less influential in RV surveys, where observers are encouraged by detections of RV variability resulting in a bias towards detections of more massive planets.

%Particularly, the fraction of planets represented within given measured radius ranges is not matched by the fraction of expected corresponding mass. The observed samples of planets discovered by the two methods are thus differently biased. The fraction of exoplanets with measured radii falling at or below $1.8$\,R$_\oplus$, which corresponds to the Radius Valley for FGK stars, is approximately 32\%, while the planets with measured corresponding masses (or minimum masses) of $\lesssim 6.2\,\textrm{M}_\oplus$ \citep{muller24} is only 20\%. Similarly for the  $1.8- 3.9\,\textrm{R}_\oplus$ or $6.2-17.15\,\textrm{M}_\oplus$ regime of sub-Neptune to Neptune-mass planets, most likely with $\textrm{H}_2$ dominated atmospheres, the relative fractions are 43\% and 22\%. The majority of planets in this regime have periods of $3-10$\,d and sit between the sparsely populated Neptune-Jupiter-mass regime \citep{mazeh16,castro-gonzalez24}. In contrast, for planets with $> \textrm{R}_\textrm{Nept}$ and $> \textrm{M}_\textrm{Nept}$ the respective fractions are reversed; the fraction of planets with measured radius $> \textrm{R}_\textrm{Nept}$ is 24\%, while the corresponding fraction of planets with measured mass $> \textrm{M}_\textrm{Nept}$ is $\sim 48$\%. Thus RV surveys are more biased than transit surveys as a result of relative insensitivity to less massive planets.

The sheer volume of measurements from high cadence space-based photometric surveys enable relatively low amplitude characteristic transits to be recovered from any background stellar signals that tend to vary on longer timescales. {Establishing detections with RV surveys is more difficult }since fewer observations must be able to discriminate between potentially multiple periodic  signals of dynamical origin and quasi-periodic stellar signals. The problem is most acute for low-mass planetary systems, where the induced dynamical amplitudes are close to the stellar activity amplitudes. The difficulty of disentangling signals is exacerbated because {the periods of} the vast majority of currently known exoplanets span the $1-100$\,d range, which coincides with stellar rotation signals. The most obvious route to identifying low-mass planets with RV surveys has been to {target} stars with relatively low activity.  {This does not necessarily guarantee that activity modelling can or should be circumvented.}

The Dispersed Matter Planet Project \citep{haswell20dmpp} (DMPP) selects stars with some of the lowest-known activity levels. The observed sub-basal chromospheric emission is attributed to the presence of close-orbiting, mass-losing exoplanets.
%The Dispersed Matter Planet Project \citep{haswell20dmpp} (DMPP) selects stars with some of the lowest-known activity levels. DMPP hypothesises that ongoing mass-loss from close-orbiting planetary systems leads to mass entrained in, or close to the orbital plane. For edge-on systems this circumstellar material has been shown to lead to excess absorption in sensitive resonance lines such as Mg \textsc{ii} h\&k and Ca \textsc{ii} H\&K \citep{haswell12,fossati13}. In the extreme case of the giant planet, WASP-12\,b, there is little to no flux in the cores of the Ca \textsc{ii} H\&K lines. It is generally observed that main-sequence stars exhibit a minimum chromospheric flux level \citep{schrijver87,rutten91}, which is generally known as the basal flux level. This level is consistent with an absence of significant magnetic activity, when only acoustic heating is present \citep{buchholz98}. Thus DMPP hypothesises that sub-basal activity may be telltale evidence for the presence of exoplanets. Stars at or below the basal limit of chromospheric emission were identified by \citet{haswell20dmpp} using the catalogue of \citet{pace13} (hereafter, \citetalias{pace13}). The DMPP survey has undertaken observations in several short observing runs, often with high cadence observations of a small subset of targets. 
An analysis of all observations to date, including survey completeness and planetary occurrence rates is detailed in \citet{standing26dmpp} (hereafter \citetalias{standing26dmpp}).  While \citetalias{standing26dmpp} includes all reported DMPP targets with sufficient single or multiple epoch observations, including those reported by \citet{haswell20dmpp,staab20dmpp1,barnes20dmpp3} and \citet{stevenson25hd28471}, {the study used purely Keplerian} signals to determine survey completeness.
%This is warranted, based on the lack of significant correlation of RV and activity and the small data sets available for most targets and enables both survey completeness and occurrence rates to be estimated.
This approach is warranted by the lack of significant correlation between RV and activity, as well as the limited data available for most targets, while still allowing estimation of survey completeness and occurrence rates.

{Here, we present a more detailed analysis of three of the previously unpublished systems with the largest and most extensive datasets that were reported in \citetalias{standing26dmpp}. We compare purely Keplerian models with models that also include a Gaussian Process (GP) to model stellar activity contributions. To avoid potential accumulative systematic biases arising from recursively adding signals, we use models that include the number of Keplerian signals as a free parameter. {Diffusive nested sampling} thus enables us to identify competing solutions within multimodal posterior space and compare models directly.} Section \S  \ref{section:observations_methods} details the observations and analysis methods and \S \ref{section:target_slection} gives details of DMPP target selection. In \S \ref{section:results} we present a detailed analysis of each target before discussing our findings further in \S \ref{section:discussion}.

%%%%%%%%%%%%%%%%%%%%%%%%%%%%%%%%%%%%%%%%%%%%%%%%%%%%%%%%%%%%%%%%%%%%%%%%%%%%%%%%%%%%
\section{Observations and Analysis Methods}
\protect\label{section:observations_methods}

%%%%%%%%%%%%%%%%%%%%%%%%%%%%%%%%%%%%%%%%%%%%%%%%%%%%%%%%%%%%%%%
\subsection{Spectroscopic observations and data extraction}
\protect\label{section:spectroscopy}

Regular RV observations were made with the 3.6m European Southern Observatory (ESO) telescope and the High Accuracy Radial Velocity Planet Searcher (HARPS) spectrograph. The observations were taken between 2015 and 2023 and were made in ESO Periods 095.C-0799(A), 097.C-0390(B), 098.C-0269(A), 098.C0499(A), 098.C-0269(B), 099.C-0798(A), 0100.C-0836(A) and 0110.248C.001. {We refer to these observing periods as P95--P110}. The observing runs spanned between 3 and 11 nights with up to 8 observations per target per night. {The P95--P100 ``post-2015'' observations (subsequent to the 2015 HARPS fibre upgrade) were made in HARPS High Accuracy Mode using a nightly master calibration and simultaneous object+ThAr observations to enable drift correction. The final P110 ``post-2020'' run (i.e. after the HARPS warm-up in early 2020) comprised 11 nights and} was made using
%the newer master LASER comb and 
simultaneous Fabry-Perot observations to correct drift. First epoch observations of \hdone{} were made in P110 and supplemented with a further 29 epochs spanning 71 nights in Period 114.27LM.001 (P114) in 2025 with the Echelle Spectrograph for Rocky Exoplanet and Stable Spectroscopic Observations (ESPRESSO) at ESOs 8.2m VLT facility \citep{pepe-espresso13}.

We re-extracted the standard HARPS and ESPRESSO Data Reduction Software (DRS) radial velocities (RVs) using the ``Semi-Bayesian Approach for RVs with Template-matching'' code (\textsc{s-bart}), which has demonstrated improved precision \citep{silva22sbart}. The same approach was used by \citet{stevenson25hd28471} and the same RVs that were derived in \citetalias{standing26dmpp} are used here.

The activity parameters for Bisector Inverse Span (BIS) and Full Width at Half Maximum (FWHM) were obtained from the HARPS DRS cross-correlation function (CCF) profiles and the {Ca} \textsc{ii} {H\&K} S-index measurements were obtained from \textsc{actin2} \citep{gomesdasilva18actin}. We also measured the line moments from the CCFs as outlined in \citet{barnes24moments} and include $M_3$, a measure of CCF symmetry, as a potential correlator in our analysis.

%%%%%%%%%%%%%%%%%%%%%%%%%%%%%%%%%%%%%%%%%%%%%%%%%%%%%%%%%%%%%%%%%%%%%%%%%%%%%%%%%%%%
%%%%%%%%%%%%%%%%%%%%% Monte Carlo Plots %%%%%%%%%%%%%%%%%%%%%%%
%%%%%%%%%%%%%%%%%%%%%%%%%%%%%%%%%%%%%%%%%%%%%%%%%%%%%%%%%%%%%%%
\begin{figure*}
    \begin{center}
        %\begin{tabular}{ccc}
            \includegraphics[trim=5mm 15mm 5mm 0mm, height=0.58\columnwidth]{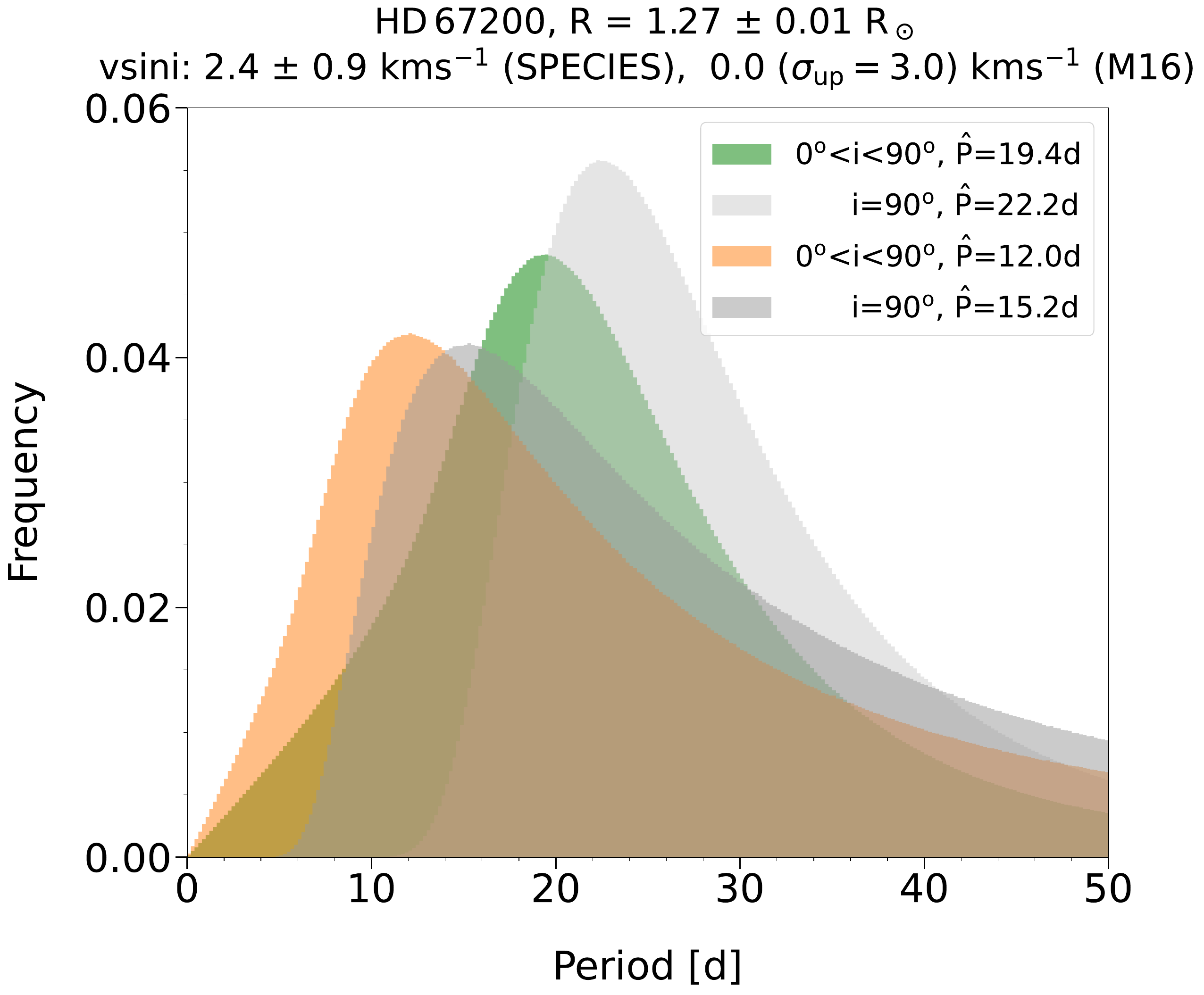}
            \includegraphics[trim=6mm 15mm 0mm 0mm, height=0.58\columnwidth]{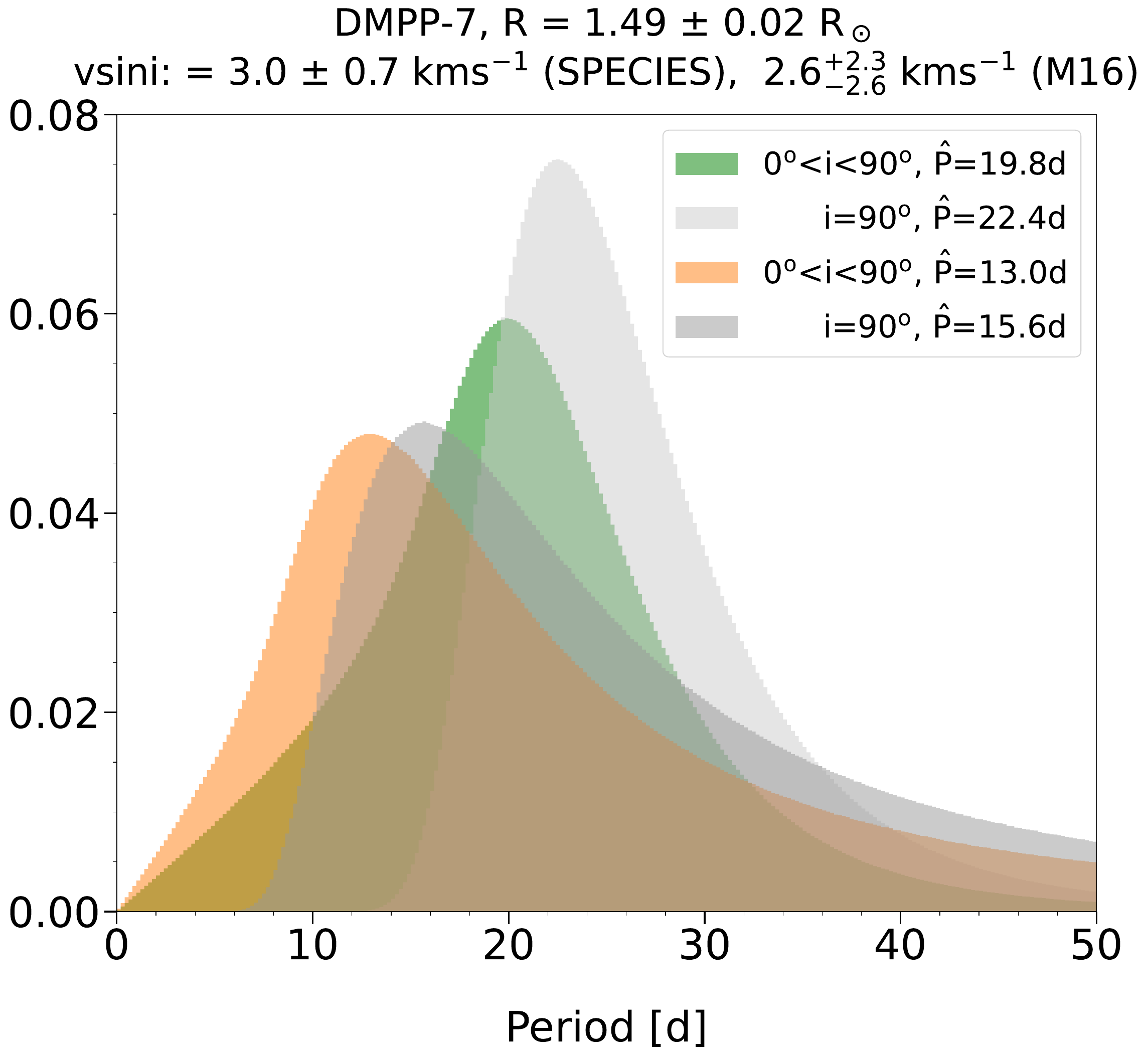}
            \includegraphics[trim=-4mm 15mm 5mm 0mm, height=0.58\columnwidth]{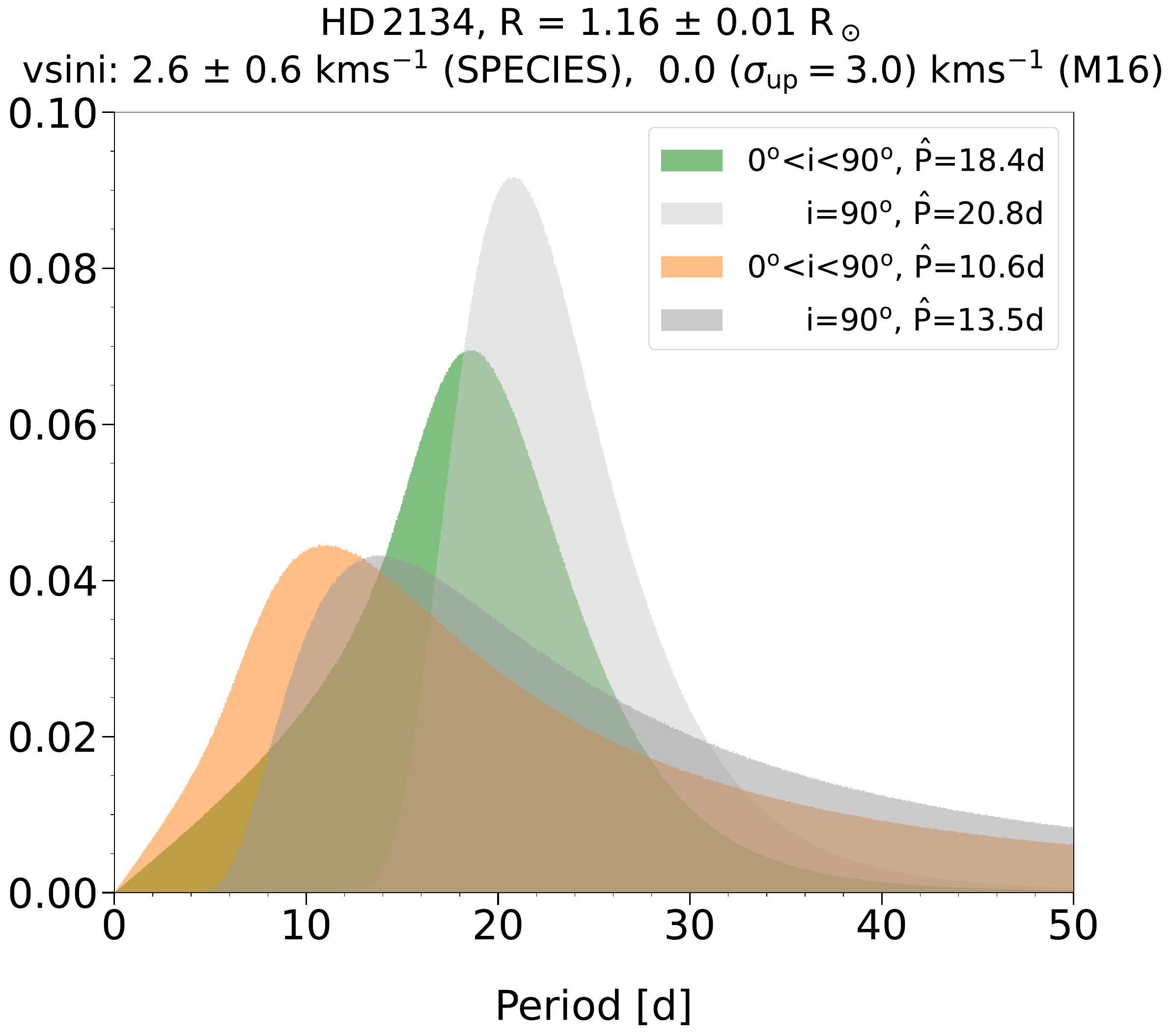} 
        %\end{tabular}
    \end{center}
    \caption{Monte Carlo period simulations for each target. The \vsini{} values derived from \textsc{species} are used for randomly oriented (sinusoidally distributed) axial inclinations (green) and for a fixed $i=90^{\textrm{o}}$ (light grey). The respective curves using the \citetalias{murphy16} \vsini{} estimates are also shown (orange and dark grey). The modal periods, $\hat{P}$, are shown for each distribution.}
    \label{fig:all_monte}
\end{figure*}

\subsection{Photometric observations with TESS}
\protect\label{section:tess_photometry}

Each of the three targets were observed with the Transiting Exoplanet Survey Satellite (TESS) in multiple sectors and observing cycles. The TESS Simple Aperture Photometry (SAP) and Pre-Conditioned SAP (PDCSAP) lightcurves \citep{Twicken2020TESS} based on the shorter cadence observations are optimised for transit searches. As discussed in \citetalias{standing26dmpp}, the co-trending basis vectors used to obtain the PDCSAP data can lead to removal of stellar variability, while the potentially more suitable SAP data contain large systematic effects that make them unsuitable for analysis of low-activity stars\footnote{see \url{https://archive.stsci.edu/missions-and-data/tess} and the TESS archive manual at \url{https://outerspace.stsci.edu/display/TESS/TESS+Archive+Manual}}. We instead performed photometric extraction of the TESS full frame images (FFIs) using the \texttt{unpopular} package \citep{hattori2022unpopular}. 
Reliable rotation periods have been reported from K2 photometry by \citet{reinhold20} at the $\gtrsim 1000$\,ppm level, while \citet{boyle25} found significant drop in reliability for periods beyond the {TESS} orbital period due to the $27$\,d sector length limitation. When used in conjunction with periodogram searches that account for zero point offsets for each sector, the photometric output from \texttt{unpopular} has the potential to retain signals at periods longer than the $\sim 27$\,d TESS sector length unlike the TESS PDCSAP data, which are optimised for transit searches. 
%\citet{boyle25} nevertheless found the period uncertainties could be improved by combining sectors, but that this did not improve reliability due to persistent systematics.

%%%%%%%%%%%%%%%%%%%%%%%%%%%%%%%%%%%%%%%%%%%%%%%%%%%%%%%%%%%%%%%%%%%%%%%%%%%%%%%%%%%%
%%%%%%%%%%%%%%%% STELLAR PARAMETERS TABLE %%%%%%%%%%%%%%%%%%%%%
%%%%%%%%%%%%%%%%%%%%%%%%%%%%%%%%%%%%%%%%%%%%%%%%%%%%%%%%%%%%%%%
\begin{table*}
    \renewcommand{\arraystretch}{1.2}
    \centering
    \begin{tabular}{lC{3.3cm}C{3.3cm}C{3.3cm}C{2.5cm}}
    \hline
                                                           & {HD\,67200}            & {HD\,118006} / DMPP-7            & \hdtwo{}                & Ref \\
    \hline                                                                  
        $B$                                                & $8.29$                 & $9.39$                   &  $9.50$                 &  1 \\
        $V$                                                & $7.70$                 & $8.80$                   &  $8.82$                 &  1 \\
        $G$                                                & $7.57$                 & $8.66$                   &  $8.66$                 &  2 \\
        Age [Gyr]                                          & $1.23_{-0.52}^{+0.39}$ & $3.87^{+0.28}_{-0.31}$   &  $6.63_{-0.86}^{+0.94}$ &  3 \\
        $\pi$ [mas]                                        & $18.37 \pm 0.01$       & $9.78 \pm 0.02$          &  $14.00 \pm 0.01$       &  2 \\
        $1/\pi$ [pc]                                       & $54.44 \pm 0.04$       & $102.3 \pm 0.2$          &  $71.44 \pm 0.06$       &  2 \\
        $M_*$ [M$_\odot$]                                  & $1.27 \pm 0.02$        & $1.23 \pm 0.02$          &  $1.05 \pm 0.02$        &  3 \\
        $R_*$ [R$_\odot$]                                  & $1.27 \pm 0.01$        & $1.49 \pm 0.02$          &  $1.16 \pm 0.01$        &  3 \\
        $T_\textrm{eff}$ [K]                               & $6140 \pm 50$          & $6096 \pm 30$            &  $5675 \pm 50$          &  3 \\
        log\,$g$                                           & $4.35 \pm 0.08$        & $4.17 \pm 0.09$          &  $4.22 \pm 0.10$        &  3 \\
        Fe/H                                               & $0.36 \pm 0.04$        & $0.13 \pm 0.05$          &  $0.24 \pm 0.04$        &  3 \\

        \multicolumn{5}{c}{\vspace{-2mm}} \\
        
        %$v\,\sin\,i$ $[\textrm{km\,s}^{-1}]$ (\textsc{species})   &  $4.0 \pm 0.3$          & $5.2 \pm 0.3$          & $3.0 \pm 0.7$            &  3 \\

        $v\,\sin\,i$ $[\textrm{km\,s}^{-1}]$ (\textsc{species})    & $2.4 \pm 0.9$          & $3.0 \pm 0.7$           &  $2.6 \pm 0.6$           &  3 \\

        $v_\textrm{mac}$ $[\textrm{km\,s}^{-1}]$ (\textsc{species}) & $4.7 \pm 0.3$          & $1.5 \pm 0.1$           &  $3.1 \pm 0.3$           &  3 \\
        $v\,\sin\,i$ $[\textrm{km\,s}^{-1}]$ (M16)        & $<3$                   & $2.6^{+2.3}_{-2.6}$     &  $<3$                    &  3 \\
        $v_\textrm{mac}$ $[\textrm{km\,s}^{-1}]$ (M16)     & $5.0 \pm 0.5$          & $4.8 \pm 0.3$           &  $3.7 \pm 0.5$           &  3 \\

        \multicolumn{5}{c}{\vspace{-2mm}} \\
        
        \logrhk{}                      & $-5.03 \pm 0.01$       & $-5.05 \pm 0.03$        & $-5.05 \pm 0.05$         &  3 \\
        \logrhk{} min, max                & $-5.08, -4.99$         & $-5.11, -5.00$          & $-5.28, -4.86$           &  3 \\
    \hline
    \end{tabular}
    \caption{Stellar parameters - references: \citet{hog2000tycho}$^1$, \citet{gaia23dr3}$^2$, this work using \textsc{species}$^3$. }%\citet{gomesdasilva2021activities}$^4$.}
    \label{tab:stellar_params}
\end{table*}

%%%%%%%%%%%%%%%%%%%%%%%%%%%%%%%%%%%%%%%%%%%%%%%%%%%%%%%%%%%%%%%
\subsection{Radial Velocity Data Analysis}
\protect\label{section:data_analysis}

We performed data analysis of the spectroscopic data using the likelihood periodogram analysis routine in the Astrometry and Radial velocity Software package, \textsc{ars} \citep{anglada13,anglada16proxima}. The likelihood model in \textsc{ars} can search for purely sinusoidal variability (e.g. circular orbits) or variability due to Keplerian orbits. It includes both offsets, {$\gamma$, and additional white noise (jitter) terms, $\sigma_\textrm{jit}$,} for individual datasets ({i.e. HARPS post-2015, HARPS post-2020, and ESPRESSO for the datasets we consider here}). More detailed RV and photometric data analysis employed \texttt{kima} \citep{faria18kima,faria23kimanote}. \texttt{Kima} uses \texttt{DNest4} \citep{brewer16dnest4} to perform {diffusive nested sampling}, enabling global model evidence to be calculated and assessed (e.g. \citealt{standing22bebpo2}). \texttt{Kima} employs a model that includes the number of Keplerians as a model parameter. The optimal number of Keplerians that are needed to adequately describe the data, $N_\textrm{p}$, can thus be obtained, avoiding potential biases that can arise when adding Keplerians recursively through more standard modelling. As with \textsc{ars}, \texttt{kima} allows individual data sets to be modelled simultaneously by including offset parameters {and white noise terms} in the likelihood model.

\subsubsection{Gaussian Process models in \texttt{Kima}}
\protect\label{section:gpkima}
\texttt{Kima} enables a GP to be included in the model, either solely on the primary data (i.e.\ the RVs) or jointly with
activity indicators. {In addition to spectroscopic data modelling, photometric data can be modelled with \texttt{kima} by using a purely GP model. The standard quasi-periodic kernel in \texttt{kima} is implemented as the squared-exponential periodic (SEP) kernel. \texttt{kima} also includes implementation of the} \textsc{s+leaf} {Exponential--Sine Periodic (ESP) kernel \citep{delisle20spleaf} for modelling RV timeseries. We adapted \texttt{kima} to implement the} \textsc{s+leaf} {ESP kernel for simultaneous RV and activity timeseries modelling. In this implementation, as in the standard SEP kernel, the RVs and the activity indicator share the hyperparameters $\eta_2$ (decay timescale), $\eta_3$ (characteristic period), and
$\eta_4$ (proxy for harmonic complexity), while allowing separate GP amplitudes for the RVs ($\eta_1$ in all models) and the
activity ($\eta_1^{\mathrm{act}}$). The model is configured so the ESP kernel is essentially a substitute for the standard SEP kernel, with the advantage that ESP kernel computes time scales as $\sim \mathcal{O}(N_\textrm{data})$ rather than $\mathcal{O}(N_\textrm{data}^3)$ \citep{ambikasaran15,foremanmackey17celerite,delisle20spleaf}. Further detail is given in Appendix \ref{section:appendixGP}, including an assessment of how choice of the ``harmonic complexity'', $\eta_4$, controls the structure within each quasi-periodic cycle.
}

\subsubsection{{Further details on modelling with} \texttt{Kima}}
\protect\label{section:kimamodel}
%Further, it is also possible to specify known Keplerians with priors, ensuring that any Keplerian signals detected with confidence can be retained, while exploring models with additional variability. 
%While this is more akin to recursively adding Keplerians, it is 
%This approach is useful when modelling known Keplerians with a GP, since an unconstrained GP (e.g. lack of prior knowledge to enable restriction of GP amplitude or characteristic periodicities) tends to mop up much of the signal (even with carefully controlled hyperparameters) by virtue of its greater flexibility and lower Bayesian penalty. 

%{With} \texttt{kima}, {models can} be selected based on the global evidence.

{With \texttt{kima}, models can be selected based on the global evidence $\mathcal{Z}$ (or, equivalently, the log-evidence $\log \mathcal{Z}$). The Bayes factor (BF) between two competing models $M_1$ and $M_2$ is defined as the ratio of their evidences}

\begin{equation}
\mathrm{BF}({M_1/M_2}) = \frac{\mathcal{Z}_1}{\mathcal{Z}_2}
                 = \exp\!\left(\log \mathcal{Z}_1 - \log \mathcal{Z}_2\right),
\end{equation}

\noindent
{where values $>1$ favour model $M_1$.
%Within a given model, a converged solution with \texttt{DNest4} also enables us to use the ratio of the number of available posterior samples as a proxy for the BF. For instance, 
In addition, a converged solution with \texttt{kima} enables the evidence for $N_\textrm{p}$ Keplerians in a model to be compared with the evidence for $N_\textrm{p}-1$ Keplerians by simply taking the ratio of the corresponding number of posterior samples to obtain $\textrm{BF}(N_\textrm{p}/(N_\textrm{p}-1))$.} Following \citet{trotta08bayes} and \citet{standing22bebpo2}, we assume $\textrm{BF} > 150$ implies strong evidence in favour of a model, while $\textrm{BF} \leq 3$ implies inconclusive evidence for a model. Intermediate values of $3 < \textrm{BF} \leq 12$ are interpreted as weak evidence and $12 < \textrm{BF} \leq 150$ as moderate evidence. Further details, including corresponding probabilities and sigma values, are given in Table 2 of \citet{standing22bebpo2}.

\texttt{Kima} enables angular momentum deficit (AMD) checks \citep{Laskar1997,laskar2000,petit17amd} to ensure that posterior samples describing only stable orbits are obtained (see section 4.2 in \citealt{stevenson25hd28471}). This can be especially important for compact multiplanet systems where high eccentricity or resonant orbits might quickly lead to physically unstable solutions. \texttt{kima} automatically selects wide priors for the model parameters based on the input data to ensure unbiased sampling. These can be adjusted if necessary. In particular, we follow the same procedure outlined in 
%\citet{stevenson25hd28471} and adopted in 
\citetalias{standing26dmpp} and use the \citet{kipping13} Beta distribution prior for eccentricities. \texttt{kima} uses the more flexible Kumaraswamy distribution for which the parameters $\alpha=0.881,\,\beta=2.878$ closely match the Beta distribution \citep{stevenson25eccentricities}. The AMD criterion and eccentricity distribution prior together tend to favour more stable and less eccentric orbits, but still enable eccentricity and closely packed orbits where there is evidence in the data.

%%%%%%%%%%%%%%%%%%%%%%%%%%%%%%%%%%%%%%%%%%%%%%%%%%%%%%%%%%%%%%%%%%%%%%%%%%%%%%%%%%%%
\section{Target Selection}
\protect\label{section:target_slection}

%The {DMPP targets were initially identified} by virtue of their low chromospheric emission from the 7864 stars were collated by \citetalias{pace13}. {S$_\textrm{MW}$} measurements for 7864 stars were collated by \citetalias{pace13} from a number of sources and derived \logrhk{} for a subset (2085) of the sample. We identified systematic inconsistencies in the \citetalias{pace13} \logrhk{} values because the \hbox{$B\,-\,V$} colours were obtained via $T_\textrm{eff}$ rather than the original sources used to compile the catalogue. We instead obtained the observed Johnson \hbox{$B\,-\,V$} colours by cross matching the \citetalias{pace13} sources with the extended Hipparcos compilation of \citet{anderson12xhip} in order to calculate \logrhk{} values \citep{noyes84} for the \citetalias{pace13} targets. Targets with $0.4 < B-V < 1.2$ and $\Delta M_\textrm{v} < 0.45$ from the main sequence were selected.

The DMPP survey candidates were identified by selecting targets {with
%\logrhk{}~$<-5.1$, or 
\logrhkmax{}~$<-5.1$ from} the catalogue compiled by  \citetalias{pace13}.  Further details of the selection are given in \citet{haswell20dmpp} (see methods section) and \citetalias{standing26dmpp}. Using the Phase 3 products for our targets, extracted with DRS Version 3.8 (i.e. with no sky subtraction), we obtain the {instrument specific} S-index converted to the Mount Wilson S-index (hereafter, {S$_\textrm{MW}$}) using \textsc{actin2}\footnote{https://github.com/gomesdasilva/ACTIN2} \citep{gomesdasilva18actin,gomesdasilva2021activities,gomesdasilva2022activities} to monitor activity variability {and calculate \logrhk{} directly for each measurement.}  Flux level measurements from \'{e}chelle spectra without sky subtraction are prone to systematic errors. The sky background contribution is greatest during bright time when the source is relatively close to the bright moon, particularly if thin cirrus cloud was also present during an observation. The problem becomes more of an issue when the Ca \textsc{ii} H\&K chromospheric core fluxes are very small or the signal-to-noise ratio (SNR) of the data are low (e.g. for fainter targets). Activities derived from HARPS DRS spectra may overestimate \logrhk{} since sky subtraction is not performed during extraction when a simultaneous reference spectrum is used on Fibre B\footnote{www.eso.org/sci/facilities/lasilla/instruments/harps/ doc/manual/HARPS-UserManual2.4.pdf}.

As Table \ref{tab:stellar_params} shows, the mean \logrhk{} values{ we obtain from our HARPS observations} are not strictly sub-basal, contrary to their survey selection criterion (i.e. they have \logrhk{}~$> -5.1$). However, with the exception of HD\,67200, they show activities that are at times sub-basal. For two of our targets, we find that observations taken at the smallest moon-object distance, $\theta_\textrm{sep}$, and highest fractional lunar illumination (FLI) have systematically elevated {S$_\textrm{MW}$} levels relative to observations at higher $\theta_\textrm{sep}$ and $\textrm{FLI}$.
%Assuming {S$_\textrm{MW}$} scales with $\theta_\textrm{sep}^2 \times \textrm{FLI}$, and taking observations with SNR > 10 at 3900\,\AA~of, we find a fractional change in {S$_\textrm{MW}$} at the extrema of the spanned $\theta_\textrm{sep}^2\textrm{FLI}(\textrm{max} - \textrm{min})$ to be $-1.0 \pm 0.3$\% ($\textrm{FLI} = 0-87$\%, $\theta_\textrm{sep} = 83-103$\degs) for HD\,67200, $0.76 \pm 1.7$\% ($\textrm{FLI} = 0-69$\%, $\theta_\textrm{sep} = 131-170$\degs)for \hdone{} and $\sim 2.6 \pm 1.2$\% for \hdtwo{} ($\textrm{FLI} = 0-89$\%, $\theta_\textrm{sep} = 64-112$\degs). Within the uncertainties, no effect is found for \hdone{} for the single HARPS epoch observations, likely as a result of the greater $\theta_\textrm{sep}$ at all epochs. The implied {S$_\textrm{MW}$} systematics of $\sim 1-3$\% for HD\,67200 and \hdtwo{} under brighter and closer moon conditions suggest that the tabulated log $R^\prime_\textrm{\textsc HK}$ values may be overestimated by $0.02$\,dex and $0.03$\,dex respectively, i.e. still above the basal flux limit.
Inspection of {S$_\textrm{MW}$} as a function of $\theta_\textrm{sep} \times \textrm{FLI}$ (as a first order approximation) suggests $\sim 1-3$\% higher {S$_\textrm{MW}$} for HD\,67200 and \hdtwo{} under the brightest and closest moon conditions compared with $\textrm{FLI} \sim 0$\% conditions. The corresponding tabulated \logrhk{} values may thus be overestimated by only $0.02$\,dex and $0.03$\,dex respectively, i.e. remaining above the basal flux limit after correction. No significant trend can be discerned for HD\,118006 from the single epoch HARPS observations, likely owing to a greater $\theta_\textrm{sep} = 131-170$\degs{} and smaller range of $\textrm{FLI} = 0-69$\% (for comparison, the HD\,67200 and \hdtwo{} observations were taken with $\theta_\textrm{sep} = 83-103$\degs{} and $64-112$\degs{} for $\textrm{FLI} \sim 0-90$\%).

{HD\,67200, HD\,118006 and \hdtwo{} show} \hbox{\logrhk{} $<-5.1$} for respectively 0\%, 16\% and 10\% of the HARPS observations. Accounting for the systematic overestimates of \logrhk{} does not yield any HD\,67200 observations with \logrhk{} $<5.1$, while 21\% of \hdtwo{} observations would fall below \logrhk{} $<5.1$. Irrespective of whether or not these stars show sub-basal chromospheric activity at all observation epochs, they exhibit very low-activity levels and are thus amenable to searches for low-amplitude dynamically induced signals.
%In the analysis and discussion that follows, {we use ``S-index'' when referring to all {S$_\textrm{MW}$} timeseries.}

%%%%%%%%%%%%%%%%%%%%%%%% HD67200 activity periodograms and correlations
\begin{figure}
 %\begin{tabular}{c}
    %\includegraphics[trim=3mm 4mm 2mm 0mm, width=1.01\columnwidth]{DMPPmulti/HD67200_periodograms_activities.pdf} &
    %\includegraphics[trim=0mm 4mm 0mm 10mm,width=0.97\columnwidth]{DMPPmulti/HD67200_activities_v3.pdf} \\
    \includegraphics[trim=0mm 6mm 0mm 0mm, width=1.0\columnwidth]{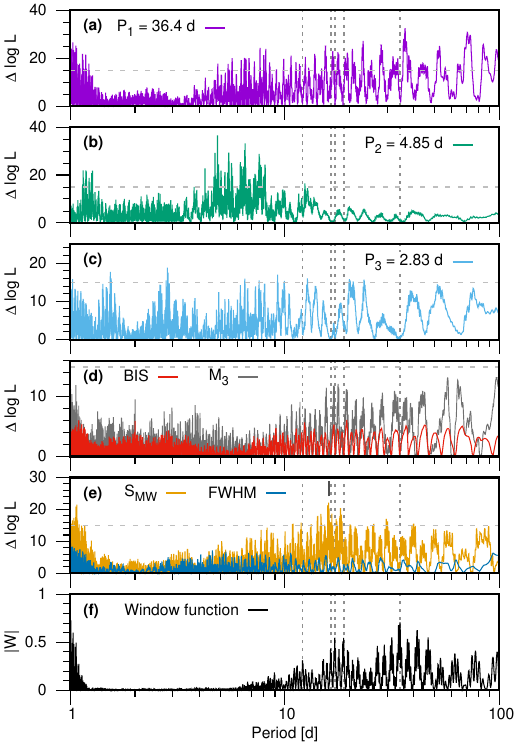}
    %\includegraphics[trim=4mm 5mm 0mm 0mm,  width=1.04\columnwidth]{DMPPmulti/HD67200_activities_v4.pdf}
 %\end{tabular}
    \caption{{\hdsix{} log-likelihood periodograms. Panels (a), (b) and (c) show the recursive RV periodograms. The corresponding activity periodograms are shown for (d) BIS and $M_3$ and (e) {S$_\textrm{MW}$} and FWHM. Significant power is only seen in {S$_\textrm{MW}$} centred around $15.99$\,d (grey tick). The window function in (f) shows significant peaks at 12.1, 16.3, 17.1, 18.8 and 34.5\,d (highlighted by the vertical dashed lines in all panels).}}
    \label{fig:hd67200_periodograms}
\end{figure}

%%%%%%%%%%%%%%%%%%%%%%%% HD67200 S-index 
\begin{figure}
	\includegraphics[trim=0mm 10mm 0mm 0mm, width=1.0\columnwidth]{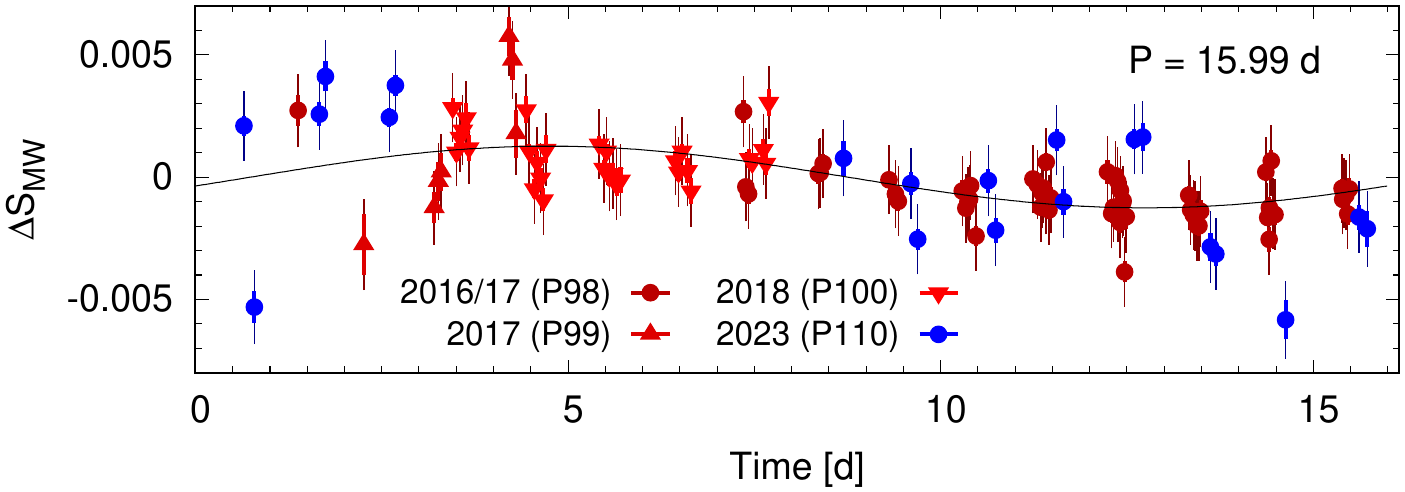}
    \caption{{S$_\textrm{MW}$} phased on the $P = 15.99$\,d period in Fig. \ref{fig:hd67200_periodograms}. The additive noise term from the maximum likelihood fit is shown by the darker and thinner error bars. The data from ESO Periods P98, P99, P100 {(post-2015)} and P110 {(post-2020)} are plotted with different colours and symbols.}
    \label{fig:hd67200_sindex_phased}
\end{figure}
%%%%%%%%%%%%%%%%%%%%%%%% HD67200 Activity correlations
\begin{figure}
 %\begin{tabular}{c}
    %\includegraphics[trim=3mm 4mm 2mm 0mm, width=1.01\columnwidth]{DMPPmulti/HD67200_periodograms_activities.pdf} &
    %\includegraphics[trim=0mm 4mm 0mm 10mm,width=0.97\columnwidth]{DMPPmulti/HD67200_activities_v3.pdf} \\
    %\includegraphics[trim=3mm -6mm 2mm 0mm, width=0.95\columnwidth]{DMPPmulti/HD67200_periodograms_activities_v2.pdf} &
    \includegraphics[trim=0mm 6mm 0mm 0mm,  width=1.0\columnwidth]{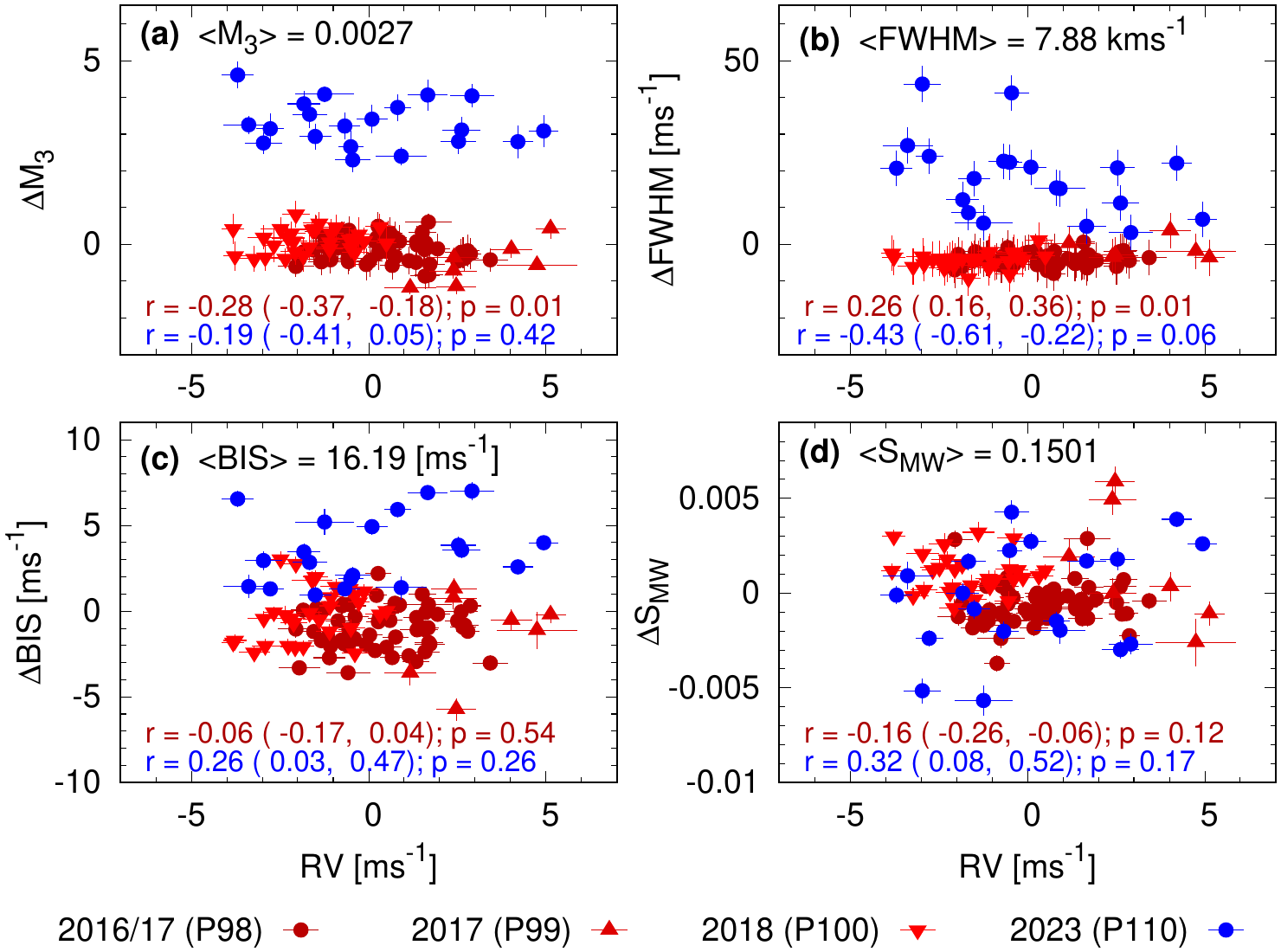}
 %\end{tabular}
    \caption{\hdsix{} RV correlations with activity indicator metrics. The data are colour coded as in Fig. \ref{fig:hd67200_sindex_phased}. {The Pearson's~$r$ linear correlation value, with uncertainties and the $p$-statistic, are indicated for the combined P98--P100, post-2015 epochs (red) and the P110, post-2020 epoch (blue)}.}
    \label{fig:hd67200_correlations}
\end{figure}
%%%%%%%%%%%%%%%%%%%%%%%%% HD67200 BIS 
%\begin{figure}
%    \begin{center}
%	\includegraphics[width=0.6\columnwidth]{DMPPmulti/HD67200_bisectors.pdf}
%    \end{center}
%    \caption{Mean bisector subtracted bisectors for all observations.}
%    \label{fig:hd67200_bis_phased}
%\end{figure}

%%%%%%%%%%%%%%%%%%%%%%%%%%%%%%%%%%%%%%%%%%%%%%%%%%%%%%%%%%%%%%%%%%%%%%%%%%%%%%%%%%%%
\section{Results}
\protect\label{section:results}
%%%%%%%%%%%%%%%%%%%%%%%%%%%%%%%%%%%%%%%%%%%%%%%%%%%%%%%%%%%%%%%%%%%%%%%%%%%%%%%%%%%%

The available photometric data and spectroscopic activity indicators were used to help inform our modelling of each target. We performed likelihood periodogram analysis and posterior sampling using both Keplerian-only and Keplerian+GP models.

%%%%%%%%%%%%%%%%%%%%%%%%%%%%%%%%%%%%%%%%%%%%%%%%%%%%%%%%%%%%%%%%%%%%%%%
%%%%%%%%%%%%%%%%%%%%%%%%% HD67200 %%%%%%%%%%%%%%%%%%%%%%%%%%%%%%%%%%%%%

%%%%%%%%%%%%%%%%%%%%%%%%% HD67200 RV periodograms %%%%%%%%%%%%%%%%%%%%%

\subsection{HD 67200}
\protect\label{section:hd67200}
%%%%%%%%%%%%%%%%%%%%%%%%%%%%%%%%%%%%%%%%%%%%%%
%%%%%%%%%%%%%%%%%%%%%%%%%%%%%%%%%%%%%%%%%%%%%%

\subsubsection{\hdsix{} stellar parameters and evolution status}
\protect\label{section:hd67200params}

HD\,67200 is a bright nearby F star with a visual magnitude of $V=7.7$. We used the Spectroscopic Parameters and atmosphEric ChemIstriEs of Stars code, \textsc{species} \citep{soto18species,soto21species} to derive stellar parameters for \hdsix{} (Table \ref{tab:stellar_params}). \citetalias{pace13} identified \hdsix{} as a low activity star from two observations of \logrhk{} = $-5.34$ and $-5.10$. The original observations were made by \citet{arriagada11activities} and \citet{jenkins11activities} who both used inter-order scattered-light subtraction during the extraction process. The need for sky background subtraction was not deemed necessary by \citet{arriagada11activities}. From their \hbox{Ca \textsc{ii} H\&K S-index} measurements, we found \logrhk{}~=~$-5.23$ and $-5.05$ after correcting the \logrhk{} calculation issue (see \S \ref{section:target_slection}) in the \citetalias{pace13} catalogue. Thus, although \hdsix{} does not meet {the \logrhkmax{}~$< -5.1$ requirement for DMPP selection}, we retained it in our sample owing to the low \logrhk{} = $-5.23$ value. We also note that the \logrhk{}~=~$-5.05$ from \citet{jenkins11activities} may be closer to the basal level if sky background was actually significant during the observations. With \textsc{actin2} and the HARPS {S$_\textrm{MW}$} conversion calibrated with 43 stars, we find \logrhk{}~$= -5.03 \pm 0.01$. {The full range of $-5.08 <$ \logrhk{} $< -4.99$ is not consistent with the sub-basal \citet{arriagada11activities} observation, indicating that \hdsix{} was in a persistently higher activity state during the period of our DMPP observations with HARPS.}
%For comparison, the older \citet{lovis2011activity} (L11) S-index calibration used only 7 stars and yields log $R^\prime_\textrm{HK} = -5.05 \pm 0.01$ with values in the range $-5.10 <$~log $R^\prime_\textrm{HK} < -5.01$. 
The main sequence age of $1.23 \pm 0.52$ Gyr is consistent with the estimate made by \citet{gomesdasilva2022activities}. 

The equatorial rotation velocity of \vsini{} = $5.2 \pm 0.3$\,\kms{} is consistent with the \textsc{species}-derived estimate of \citet{perdelwitz24}. However, \textsc{species} does not take into account the macroturbulent velocity contribution when obtaining rotational broadening estimates. The macroturbulent velocity contribution, $v_\textrm{mac} = 4.7 \pm 0.3$\,\kms{}, is obtained by \textsc{species} from the relationship in \citet{dossantos16}. A more reliable rotational velocity estimate is thus obtained by simply subtracting in quadrature  \vsini{} = 
\hbox{$
\sqrt{
v\,\textrm{sin}\,i_\textrm{\textsc{species}}^2 - 
v_\textrm{mac,\textrm{\textsc{species}}}^2
}
$}. 
The adjusted estimate of \vsini{}~=~$2.4 \pm 0.9$\,\kms{} is given in Table~\ref{tab:stellar_params}. An alternative approach, implemented by \citet{murphy16} (\citetalias{murphy16}), derives both \vsini{} and $v_\textrm{mac}$ directly from the spectra, but requires high spectral resolution for successful recovery of both parameters, particularly when \vsini{} is close to the spectral resolution. For \hdsix{}, an \citetalias{murphy16} two-parameter fit is optimised when $\textrm{\vsini{} = 0}$\,\kms{}; hence, while we obtain a finite positive $v_\textrm{mac}$ estimate, we can only determine a \vsini{} upper 68.3\% confidence limit of 3 \kms{} (Table~\ref{tab:stellar_params}). We thus prefer the \textsc{species} estimate.

Monte Carlo simulations with $R = 1.27 \pm 0.01$\,R$_\odot$, randomly oriented (i.e. sinusoidally distributed) stellar axial inclinations and the \textsc{species} \vsini{}~=~$2.4 \pm 0.9$\,\kms{} estimate (Fig. \ref{fig:all_monte}, green) suggest a modal or most probable stellar rotation period of 
$\hat{P}_\textrm{rot} = 19.4$~d, 
with a median and upper and lower $68.3$\% values of 
%\hbox{$13.5 < P_\textrm{rot,med} = 22.0 < 35.5$\,d}
\hbox{$\tilde{P}_\textrm{rot} = 21.9~(13.4; 35.5)$\,d}.
From our M16 estimate, we find  $\hat{P}_\textrm{rot}$~$= 11.8$~d with a distribution (Fig. \ref{fig:all_monte}, orange) that is skewed to longer rotation rates due to the half-Gaussian distributed values of $0 < \textrm{\vsini{}} \leq 3$\,\kms{}; here, we find
%$10.8 < P_\textrm{rot,med} = 24.5 < 83.9$\,d. 
$\tilde{P}_\textrm{rot} = 24.5 (10.8; 83.9)$\,d. The distributions and modal periods are also shown and indicated in Fig. \ref{fig:all_monte} (grey) for a fixed $i = 90^\textrm{o}$. 
The SPECIES age estimate for \hdsix{} suggests a rotation period of $10.6^{+2.2}_{-3.1}$\,d based on the gyrochronology relation of \citet{mamajek08age}. An age estimate from \textsc{actin2} of $3.7 \pm 1.7$\,Gyr similarly leads to a rotation period estimate of $19.9 \pm 5.6$\,d, but may be slightly biased to younger ages in the absence of sky subtraction and to older ages given the likelihood that chromospheric emission is depressed under the DMPP hypothesis.

%----------------- HD67200 TABLE RVs -------------------

%Re-run on 21/05/2026 - to ensure correct mstar for masses used
%cluster_hdbscan.py --nkep=2 --csize=300 --nparms=3  --sigma=5.0 --t0=58863.719874 --binsz=0.1 --gamma=0 --mstar="1.27 0.02" --scluster="1"
%cluster_hdbscan.py --nkep=3 --csize=2500 --nparms=3  --sigma=5.0 --t0=58863.719874 --binsz=0.1 --gamma=0 --mstar="1.27 0.02"

\begin{table*}
    
\renewcommand{\arraystretch}{1.25}
\centering
\setlength\tabcolsep{0.0\columnwidth}

\begin{tabular}{lccc}

\hline
\vspace{-4mm} \\ 

% ----------------- Model 1a -----------------

\multicolumn{4}{c}{{{Model~1a} (${N_\textrm{p}=2}$)}} \\
%\hline
\vspace{-4mm} \\
{Parameter}                         & {\hdsix{}\,b$^\prime$}                   & {\hdsix{}\,c$^\prime$}                    & \\
%\hline
$P$ [d]                       & ${7.60197}^{+0.02557}_{-0.00126} [7.60182]$            & ${36.44961}^{+0.49779}_{-0.06229} [36.51028]$ & \\
$K$ [ms$^{-1}$]               & ${1.90}^{+0.27}_{-0.25} [2.25]$                        & ${2.62}^{+0.28}_{-0.35} [2.63]$               & \\
$M_0$ [rad]                   & ${1.102}^{+3.095}_{-0.315} [0.995]$                    & ${2.943}^{+3.097}_{-2.704} [0.110]$           & \\
e                             & ${0.353}^{+0.094}_{-0.087} [0.476]$                    & ${0.193}^{+0.097}_{-0.126} [0.296]$           & \\
$\omega$ [rad]                & ${3.048}^{+0.333}_{-0.257} [3.069]$                    & ${3.090}^{+0.586}_{-0.493} [2.788]$           & \\ 

%\multicolumn{4}{c}{\vspace{-4mm}} \\ 

$\sigma_\textrm{p15,\,p20}$ [ms$^{-1}$]        & 
\multicolumn{2}{c}{${0.89}^{+0.09}_{-0.08} [0.82]$,~~${1.39}^{+0.36}_{-0.26} [1.25]$} & \\

%Offset of 15032.8398 m/s added to gamma values
$\gamma_\textrm{p15,\,p20} - 15032.84$ [ms$^{-1}$]    & 
\multicolumn{2}{c}{${0.22}^{+1.25}_{-1.07} [0.16]$,~~${14.01}^{+1.10}_{-1.15} [15.52]$} & \\

\multicolumn{4}{c}{\vspace{-3mm}} \\

$a$ [AU]                      & ${0.0820}^{+0.0005}_{-0.0005} [0.0822]$                & ${0.2332}^{+0.0019}_{-0.0015} [0.2356]$       & \\
$m_\textrm{p}\,\sin\,i$ [M$_\oplus$]   & ${6.389}^{+0.731}_{-0.817} [7.180]$                    & ${15.570}^{+1.688}_{-2.032} [15.611]$         & \\
$t_\textrm{c}$ [JD]           & ${58861.421}^{+0.308}_{-3.747} [58861.780]$            & ${58850.875}^{+5.986}_{-30.479} [58858.968]$  & \\

\multicolumn{4}{c}{\vspace{-3mm}} \\

BF($N_\textrm{p}=2/N_\textrm{p}=1$)                      & \multicolumn{2}{c} {$>2213$} & \\
$\ln{\mathcal{L}}$ [MAP]          & \multicolumn{2}{c} {$-161.20$} & \\

% ------------------ Model 1b ----------------
\vspace{-3mm} \\ 
\multicolumn{4}{c}{{{Model~1b} (${N_\textrm{p}=3}$)}} \\
\vspace{-4mm} \\
{Parameter} & {\hdsix{}\,b$^{\prime\prime}$} & {\hdsix{}\,c$^{\prime\prime}$} & {\hdsix{}\,d$^{\prime\prime}$}                                   \\
$P$ [d]                       & ${3.20794}^{+0.00475}_{-0.27300} [3.20803]$            & ${7.62757}^{+0.00085}_{-0.00130} [7.62780]$            & ${36.97461}^{+0.05330}_{-0.56404} [37.01817]$          \\
$K$ [ms$^{-1}$]               & ${1.01}^{+0.21}_{-0.19} [1.11]$                        & ${2.01}^{+0.22}_{-0.28} [2.30]$                        & ${3.08}^{+0.34}_{-0.44} [3.30]$                        \\
$M_0$ [rad]                   & ${3.494}^{+1.663}_{-2.625} [3.459]$                    & ${4.357}^{+0.368}_{-0.641} [4.478]$                    & ${3.826}^{+1.449}_{-2.696} [5.135]$                    \\
e                             & ${0.244}^{+0.147}_{-0.174} [0.335]$                    & ${0.221}^{+0.067}_{-0.064} [0.229]$                    & ${0.031}^{+0.039}_{-0.024} [0.039]$                    \\
$\omega$ [rad]                & ${3.666}^{+0.960}_{-1.033} [3.809]$                    & ${2.872}^{+0.388}_{-0.372} [2.869]$                    & ${2.532}^{+2.132}_{-1.308} [1.194]$                    \\

$\sigma_\textrm{p15,\,p20}$ [ms$^{-1}$]      & 
\multicolumn{3}{c}{${0.75}^{+0.09}_{-0.08} [0.63]$,~~${0.96}^{+0.35}_{-0.29} [0.71]$} \\

%Offset of 15032.8398 m/s will be added to gamma values
$\gamma_\textrm{p15,\,p20} - 15032.84$ [ms$^{-1}$]  & 
\multicolumn{3}{c}{${0.15}^{+1.22}_{-1.11} [0.03]$,~~${13.77}^{+1.06}_{-1.18} [14.48]$} \\

\multicolumn{4}{c}{\vspace{-3mm}} \\
$a$ [AU]                      & ${0.0459}^{+0.0004}_{-0.0025} [0.0460]$                & ${0.0821}^{+0.0005}_{-0.0005} [0.0826]$                & ${0.2347}^{+0.0016}_{-0.0020} [0.2356]$                \\
$m_\textrm{p}\,\sin\,i$ [M$_\oplus$]   & ${2.601}^{+0.506}_{-0.541} [2.814]$                    & ${7.085}^{+0.815}_{-1.037} [8.156]$                    & ${18.779}^{+2.102}_{-2.696} [20.216]$                  \\
$t_\textrm{c}$ [JD]           & ${58861.900}^{+0.719}_{-0.930} [58861.124]$            & ${58857.233}^{+0.500}_{-0.235} [58857.215]$            & ${58835.929}^{+18.859}_{-2.190} [58835.526]$           \\

\multicolumn{4}{c}{\vspace{-3mm}} \\

BF($N_\textrm{p}=3/N_\textrm{p}=2$) &  \multicolumn{3}{c} {$22.9$} \\

$\ln{\mathcal{L}}$ [MAP] &  \multicolumn{3}{c} {$-138.85$} \\

\multicolumn{4}{c}{\vspace{-4mm}} \\

Global evidence, $\ln \mathcal{Z}$  &  \multicolumn{3}{c} {$-213.25$} \\

\vspace{-2mm} \\
\hline

% -------------------- Model~2a and 2b -----------------

% !! Obtained Np=1 via  cluster_hdbscan.py --nkep=1 --csize=800 --nparms=101  --sigma=5.0 --eta3="15.0 1000" --t0=58863.7198743101 --scluster="1 3" !!

        & {{Model~2a (GP + ${N_\textrm{p}\,=\,0}$})} & {{Model~2b (GP + ${N_\textrm{p}\,=\,1}$})} \\
\vspace{-4mm} \\
{Parameter}                                 &                                             & {\hdsix{}\,b$^{\prime\prime\prime}$}          & \\
%\hline

$P$ [d]                                         &                                             & ${2.79853}^{+0.40939}_{-0.30363} [3.20815]$ & \\
$K$ [ms$^{-1}$]                                 &                                             & ${0.89}^{+0.36}_{-0.20} [1.43]$             & \\
$M_0$ [rad]                                     &                                             & ${3.273}^{+1.482}_{-1.788} [3.648]$         & \\
e                                               &                                             & ${0.212}^{+0.351}_{-0.170} [0.650]$         & \\
$\omega$ [rad]                                  &                                             & ${3.375}^{+1.389}_{-1.886} [3.393]$         & \\

$\eta_1$ [ms$^{-1}$]                            & {${  2.86}_{ -0.58}^{ +0.81} [  2.09]$}     &  {${2.90}^{+1.31}_{-0.69} [2.89]$}       & \\
$\eta_2$ [d]                                    & {${  9.130}_{ -3.402}^{+69.080} [  5.098]$} &  {${69.977}^{+164.569}_{-45.913} [568.287]$} & \\
$\eta_3$ [d]                                    & {${ 10.829}_{ -0.734}^{ +4.304} [ 10.046]$} &  {${14.741}^{+3.439}_{-0.592} [19.598]$}     & \\
$\eta_4$                                        & {${  0.671}_{ -0.120}^{ +0.216} [  0.503]$} &  {${0.677}^{+0.208}_{-0.127} [0.548]$}       & \\
$\sigma_\textrm{p15,\,p20}$ [ms$^{-1}$]              & {${  0.71}_{ -0.10}^{ +0.12} [  0.65]$, ${  1.15}_{ -0.35}^{ +0.44} [  0.81]$}     &  
${0.67}^{+0.08}_{-0.07} [0.59]$, ${0.82}^{+0.35}_{-0.28} [0.78]$ & \\

%# Add 15.0328398 km/s to above and switch around gamma_0 and gamma_1 as gamma_0 is the reference (vsys)
$\gamma_\textrm{p15,\,p20} - 15032.84$ [ms$^{-1}$]           & ${     0.02}_{    -1.30}^{    +0.93} [     1.04]$,~~${    13.50}_{    -1.70}^{    +1.77} [    13.30]$ &
${0.67}^{+0.08}_{-0.07} [0.59]$,~~${13.77}^{+2.00}_{-1.82} [13.04]$ \\

\multicolumn{4}{c}{\vspace{-3mm}} \\

$a$ [AU]                                        &                                             & ${0.0421}^{+0.0038}_{-0.0029} [0.0466]$       & \\
$m_\textrm{p}\,\sin\,i$ [M$_\oplus$]            &                                             & ${2.175}^{+0.605}_{-0.492} [2.997]$           & \\
$t_\textrm{c}$ [JD]                             &                                             & ${58862.203}^{+1.213}_{-1.119} [58861.602]$   & \\

\multicolumn{4}{c}{\vspace{-3mm}} \\

$\ln{\mathcal{L}}$ [MAP]                        & $-182.75$                                   &   $-164.62$                                             & \\
BF(GP/no-GP) | BF($N_\textrm{p}=1 / N_\textrm{p}=0$)    & $>1271$                                     &   $10.73$                                         & \\
%Bayes Factor                                    & $>1271$                                     &   $10.73(4.32)$                                         & \\
Global evidence, $\ln \mathcal{Z}$   & \multicolumn{2}{c}{$-207.67$}                                                              & \\

\end{tabular}
\caption{{Top: \hdsix{} parameters for the purely Keplerian models with $N_\textrm{p}=2$ (Model~1a) and $N_\textrm{p}=3$ (Model~1b). Bottom: Parameters for models containing a pure GP with $N_\textrm{p}=0$ (Model~2a) and a GP with $N_\textrm{p}=1$ (Model~2b). White noise ($\sigma_\textrm{p15,\,p20}$) and velocity offsets ($\gamma_\textrm{p15,\,p20}$) for post-2015 and post-2020 datasets are given. Maximum posterior (MAP) solutions are indicated in square parentheses. Derived $a$, $m_\textrm{p}\,\sin\,i$ and time of inferior conjunction, $t_\textrm{c}$, are given.}}
\label{tab:hd67200_solution}
%\end{adjustbox}
\end{table*}

%------------------------ HD67200 TABLE RV+Sindex  ------------------

% NB Model 3B obtained via: cluster_hdbscan.py --nkep=1 --csize=1000 --nparms=101  --sigma=5.0 --eta3="15.0 1000" --t0=58863.7198743101

\begin{table*}
    
\renewcommand{\arraystretch}{1.25}
\centering
\setlength\tabcolsep{0.02\columnwidth}

\begin{tabular}{lccc}
\hline
\vspace{-4mm} \\

% ------------------------ Model 3a and 3b ---------------------
        & {{Model 3a (GP + ${N_\textrm{p}=0}$})} & {{Model 3b (GP + ${N_\textrm{p}=1}$})} \\
{Parameter}                                 &                                             & {\hdsix{}\,b}    & \\
%\hline

$P$ [d]                                         &                                             & ${2.67380}^{+0.30804}_{-0.19540} [2.26190]$ & \\
$K$ [ms$^{-1}$]                                 &                                             & ${0.84}^{+0.29}_{-0.19} [1.16]$             & \\
$M_0$ [rad]                                     &                                             & ${3.350}^{+1.637}_{-1.828} [2.398]$         & \\
e                                               &                                             & ${0.178}^{+0.347}_{-0.137} [0.551]$         & \\
$\omega$ [rad]                                  &                                             & ${3.250}^{+1.829}_{-1.921} [2.796]$         & \\

$\eta_1$ [ms$^{-1}$]                            & ${     2.91 }_{    -0.62 }^{    +0.99 } [     2.83 ]$ &  {${2.82 }^{+0.84 }_{-0.61 } [1.89 ]$}     & \\
$\eta_1^\textrm{act}$({S$_\textrm{MW}$})                               & ${     1.95 }_{    -0.43 }^{    +0.59 } [     1.94 ]$ &  {${1.97 }^{+0.61 }_{-0.42 } [1.68 ]$}     & \\
$\eta_2$ [d]                                    & ${    57.723}_{   -34.227}^{   +40.720} [    83.942]$ &  {${66.151}^{+49.913}_{-32.654} [56.459]$} & \\
$\eta_3$ [d]                                    & ${    15.103}_{    -0.520}^{    +0.750} [    15.120]$ &  {${15.035}^{+0.553}_{-0.455} [15.006]$}   & \\
$\eta_4$                                        & ${     0.625}_{    -0.089}^{    +0.158} [     0.547]$ &  {${0.634}^{+0.164}_{-0.096} [0.511]$}     & \\

$\sigma_\textrm{p15,\,p20}$ (RV) [ms$^{-1}$]               &${     0.78 }_{    -0.09 }^{    +0.11 } [     0.81 ]$,~~${     1.24 }_{    -0.30 }^{    +0.36 } [     1.06 ]$     &
{${0.68}^{+0.09}_{-0.08} [0.59]$},~~{${0.82}^{+0.35}_{-0.27} [1.24]$} \\

$\sigma_\textrm{p15,\,p20}$ ({S$_\textrm{MW}$})                      &${     0.68 }_{    -0.08 }^{    +0.08 } [     0.73 ]$,~~${     1.99 }_{    -0.36 }^{    +0.49 } [     1.75 ]$     &
{${0.68}^{+0.08}_{-0.07} [0.69]$},~~{${1.97}^{+0.47}_{-0.36} [1.69]$} \\
%# Add 15.0328398 km/s to above and switch around gamma_0 and gamma_1 as gamma_0 is the reference (vsys)

$\gamma_\textrm{p15,\,p20}$ (RV) $- 15032.84$ [ms$^{-1}$]  & ${    -0.28 }_{    -1.35 }^{    +1.04 } [    -0.66 ]$,~~${    13.45 }_{    -1.65 }^{    +1.83 } [    13.31 ]$ &
{${0.05}^{+11.50}_{-2.68} [0.11]$},~~{${12.96}^{+2.19}_{-12.30} [-21.80]$} \\

$\gamma_\textrm{p15,\,p20}$ ({S$_\textrm{MW}$})                      & ${   147.97 }_{    -0.77 }^{    +0.83 } [   147.94 ]$,~~${   147.41 }_{    -1.13 }^{    +1.19 } [   146.75 ]$ &
{${150.41}^{+0.84}_{-0.61} [149.54]$},~~{${147.59}^{+1.23}_{-1.19} [147.65]$} \\

\multicolumn{4}{c}{\vspace{-3mm}} \\

$a$ [AU]                                        &                                             & ${0.0409}^{+0.0030}_{-0.0021} [0.0367]$                & \\
$m_p$\,sin\, $i$ [M$_\oplus$]                   &                                             & ${2.066}^{+0.518}_{-0.468} [2.353]$                    & \\
$T_\textrm{eq} (A_\textrm{B} = 0, 0.36)$ [K]  &                                               & $1650, 1475$                                           & \\
$R_\textrm{p}$ (predicted from \mpsini{}) [R$_\oplus$]       &                                & $1.24^{+0.10}_{-0.09}$                                 & \\
$R_\textrm{p}$ (predicted, $i=57.3$\degs) [R$_\oplus$]       &                                & $1.30^{+0.11}_{-0.10}$                                 & \\
$t_\textrm{c}$ [JD]                             &                                             & ${58862.236}^{+1.227}_{-1.130} [58862.727]$            & \\

\multicolumn{4}{c}{\vspace{-3mm}} \\

$\ln{\mathcal{L}}$ [MAP]                        & $-367.57$                                   &   $-346.43$                                            & \\
BF(GP/no-GP) | BF($N_\textrm{p}=1 / N_\textrm{p}=0$) & $>1013$                                     &   $10.32$                                         & \\
%Bayes Factor                                    & $>1013$                                     &   $10.32(5.3)$                                         & \\
Global evidence, $\ln \mathcal{Z}$   & \multicolumn{2}{c}{$-400.04$}                                                                                   & \\

\end{tabular}
\caption{{Top: \hdsix{} parameters for the GP + Keplerian model with RV and simultaneous S$_\textrm{MW}$. The GP + N$_\textrm{p}=0$ (Model~3a) and GP + N$_\textrm{p}=1$ (Model~3b) are tabulated. Parameters are given in as in Table \ref{tab:hd67200_solution} with the addition of $\eta_1^\textrm{act}$, $\sigma$ and $\gamma$ parameters for the S$_\textrm{MW}$ timeseries. The derived planet equilibrium temperature, $T_\textrm{eq}$, and predicted planet radii, $R_\textrm{p}$, (\citetalias{muller24}) are given for Model~3b.}}
\label{tab:hd67200_solution_rv_activity}
%\end{adjustbox}
\end{table*}

\subsubsection{{\hdsix{} search for $P_\textrm{rot}$ in TESS photometry}}
\protect\label{section:hd67200photometry}

{With a high negative declination of -70:01:25.96 (J2000), \hdsix{} has been observed by TESS during Years 1, 3, 5 and 7 in a total of 42 sectors. Recovery of periods beyond $\sim$10\,days or half the TESS sector length of 13.7\,d with standard period analyses is difficult or unreliable (e.g. see \citealt{colman24tessrotation,claytor22TESSdeeplearning,boyle25}). \cite{hattori25tess} had success recovering longer periods with periodogram analyses that employed separate zero-point offsets for each sector. Appendix \ref{section:appendixTESS} presents the} \texttt{unpopular} {lightcurves in Fig. \ref{fig:hd67200_lightcurve}. The corresponding sinusoidal periods, $P$, and the semi-amplitudes, $A$, that we discussed in \citetalias{standing26dmpp} are tabulated in Table \ref{tab:TESS_GP_params}. In addition, we give the $R_\textrm{var}$ variability, which measures the 5\%\,--\,95\% span of the sorted data after resampling into 3\,hr bins to reduce short-timescale noise in the TESS photometry \citep{basri11,reinhold20}. Using the same approach as \cite{hattori25tess}, with standard GLS periodogram analysis, \citep{zechmeister09gls} yields} \textit{inconsistent} {periods for \hdsix{}, from year to year and also when analysing Years 1 and 3 as separate odd and even sectors.
 We find periods of $3.65$\,d~--~$13.7$\,d, but with relatively stable sinusoidal amplitudes, $A$, of $30$--$49$\,ppm. The amplitudes are significantly below the span of the photometry, indicated by the $R_\textrm{var}$ statistic. We suspect that TESS instrumental systematics may be affecting the \hdsix{} photometry (see Appendix \ref{section:appendixTESS} for further details).
} 

{
To investigate whether the high $R_\textrm{var}$ to $A$ ratio arises from modulating activity levels, we attempted to recover stellar rotation periodicities by using the} \textsc{s+leaf} {ESP kernel \citep{delisle20spleaf}. The expected harmonic content of spot modulated light curves can be quantitatively justified using the Kepler analysis of \citet{basri18doubledip}, who show that “double-dip’’ lightcurve segments have photometric variability amplitudes approximately half those of “single-dip’’ segments. With respect to the fundamental characteristic rotation period, $\eta_3$, this implies a relative $\eta_3/2$ harmonic amplitude of $\sim$30–50\%. Table \ref{tab:harmonics} shows this corresponds to $(\eta_3/2)/\eta_3$ harmonic weighting ratios in the ESP kernel of $w_2/w_1 \approx 0.3$--$0.5$, which is achieved for $\eta_4 \approx 0.7$--$0.9$. We initially tried using a uniform prior of $\eta_4 : \mathcal{U}[0.8;10]$ on the harmonic complexity to ensure that double-dip morphology could be adequately captured in the lightcurves ($\eta_4 \sim 10$ essentially only allows the fundamental $\eta_3$ harmonic to appear). A wide characteristic period prior of $\eta_3 : \mathcal{U}[10;100]$ was adopted, with lower bound informed by our Monte-Carlo simulation. We assumed that photometric coherence should persist for at least as long as the time an active region is visible, with a prior of $\eta_2: \mathcal{U}[> 0.5\,\eta_3;t_\mathrm{span}]$, though more properly to recover periodicity on the characteristic rotation timescale, prior ranges of at least $\eta_2: \mathcal{U}[> \eta_3;t_\mathrm{span}]$ or $\eta_2: \mathcal{U}[> 2\,\eta_3;t_\mathrm{span}]$ are more appropriate as is typically seen in lightcurve data \citep{basri18doubledip}.
}

{
We re-binned the photometric observations obtained with \texttt{unpopular} into $0.5$\,d bins to ensure that effective posterior sampling was computationally tractable. As we suspected, runs with a prior of $\eta_2: \mathcal{U}[>0.5\eta_3;t_\mathrm{span}]$ tend to favour pile-up of the hyperparameter posteriors at the lower $\eta_3 = 10$\,d boundary with the correlated photometric variability described by the $\eta_2$ timescale at $\sim 0.5\eta_3$. i.e. the GP prefers to use the decay timescale correlation to fit the very short timescale variability that is probably due to data systematics. Requiring $\mathcal{U}[\geq\eta_3;t_\mathrm{span}]$, also tends to result in $\eta_3$ posterior sample pile up at the lower $10$\,d boundary or a preference for $\eta_3 \sim 13-14$\,d; which probably corresponds to the 13.5\,d TESS Earth-orbit signature. While the GP model with appropriate harmonic complexity has the potential to capture the observed variability better than a strict sinusoid, the presence of spacecraft systematics precludes successful recovery of stellar rotation signals for \hdsix{}. Restricting harmonic complexity to include only the fundamental characteristic periodicity does not offer any significant benefits over straightforward periodogram analysis in this instance.
}

%%%%%%%%%%% HD67200 RV PLOT (Keplerian and Keplerian + GP) %%%%%%%%%%%%
\begin{figure*}
    \includegraphics[trim=4mm 0mm 5mm 0mm, width=2\columnwidth]{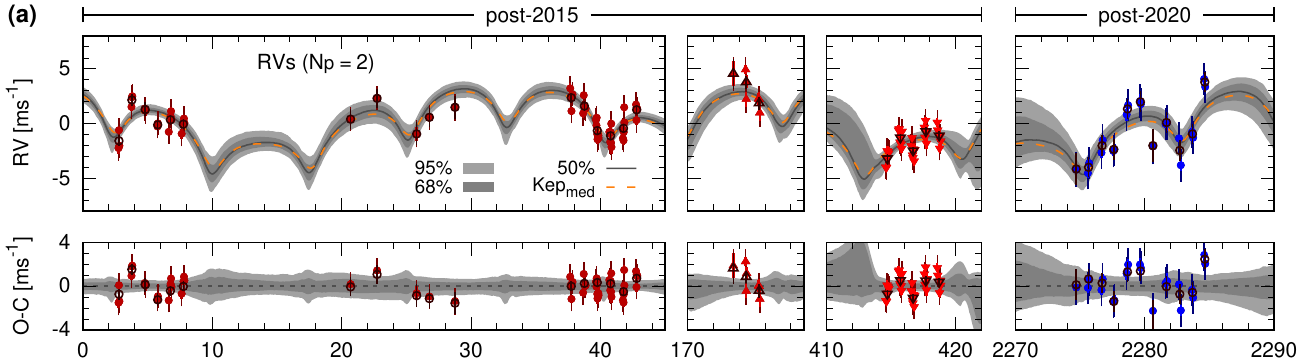}
        \vspace{0mm} \\
        \includegraphics[trim=4mm 0mm 5mm 0mm, width=2\columnwidth]{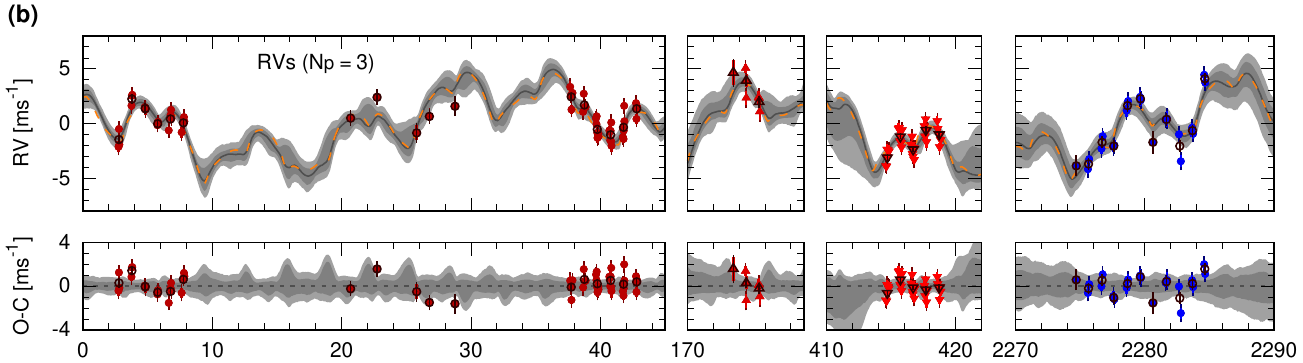}
        \vspace{0mm} \\
        \includegraphics[trim=4mm 0mm 5mm 0mm, width=2\columnwidth]{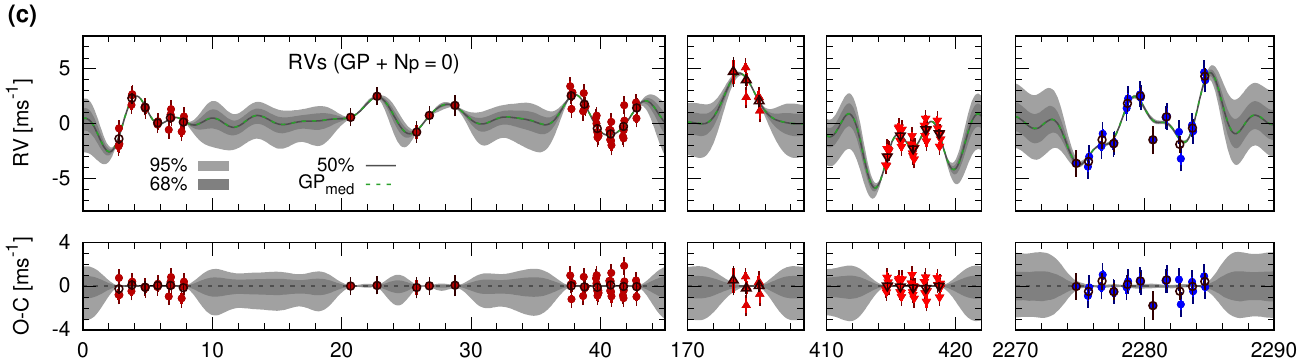}
        \vspace{0mm} \\
        \includegraphics[trim=4mm 0mm 5mm 0mm, width=2\columnwidth]{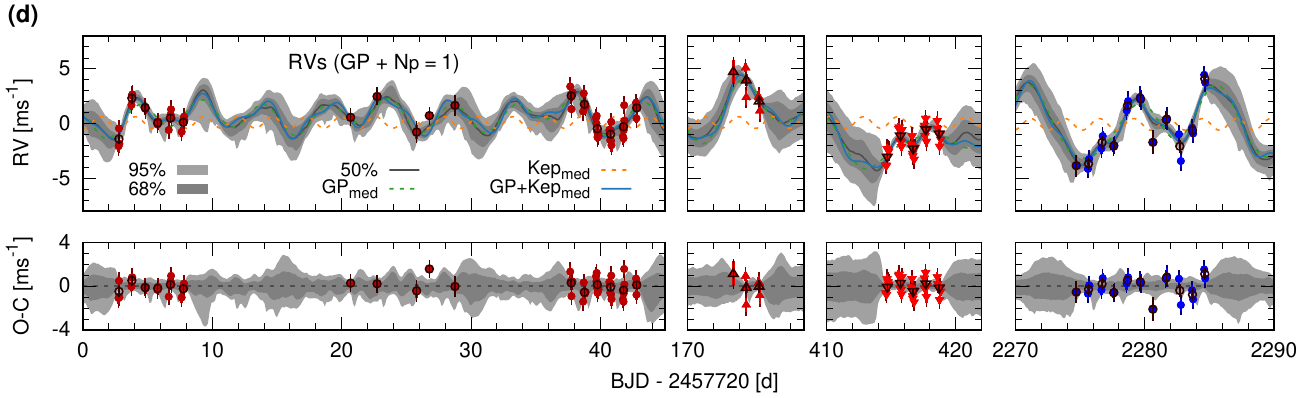}
    \caption{{Solution curves for the \hdsix{} RVs. The purely Keplerian solutions are shown in (a) for $N_\textrm{p} = 2$ (Model~1a) and in (b) for the $N_\textrm{p} = 3$ (Model~1b). The GP + Keplerian solutions are shown in (c) for $N_\textrm{p} = 0$ (Model~2a) and in (d) for $N_\textrm{p} = 1$ (Model~2b). See Table \ref{tab:hd67200_solution} for model solution details. The median 50\% curve (dark grey) from the posteriors and single most representative median curves (see legends) are shown along with the 68\% confidence (mid grey) and 95\% confidence (light grey) envelopes (see main text for details). Residuals and binned residual points for the representative median sample are shown in the lower sub-panels. The data uncertainties are plotted and also show the representative median model white noise terms added in quadrature (darker shades, thinner and longer error bars). Nightly binned data points are also plotted as darker open symbols.}}
    %(b) The model with GP included, which prefers $N_\textrm{p} = 0$.}
    \label{fig:hd67200_rvs}
\end{figure*}

%%%%%%%%%%%%%%%%%%%%%%%% HD67200 RV PLOT (Keplerian and Keplerian + GP)
\begin{figure*}
    \includegraphics[trim=4mm 0mm 5mm 0mm, width=2\columnwidth]{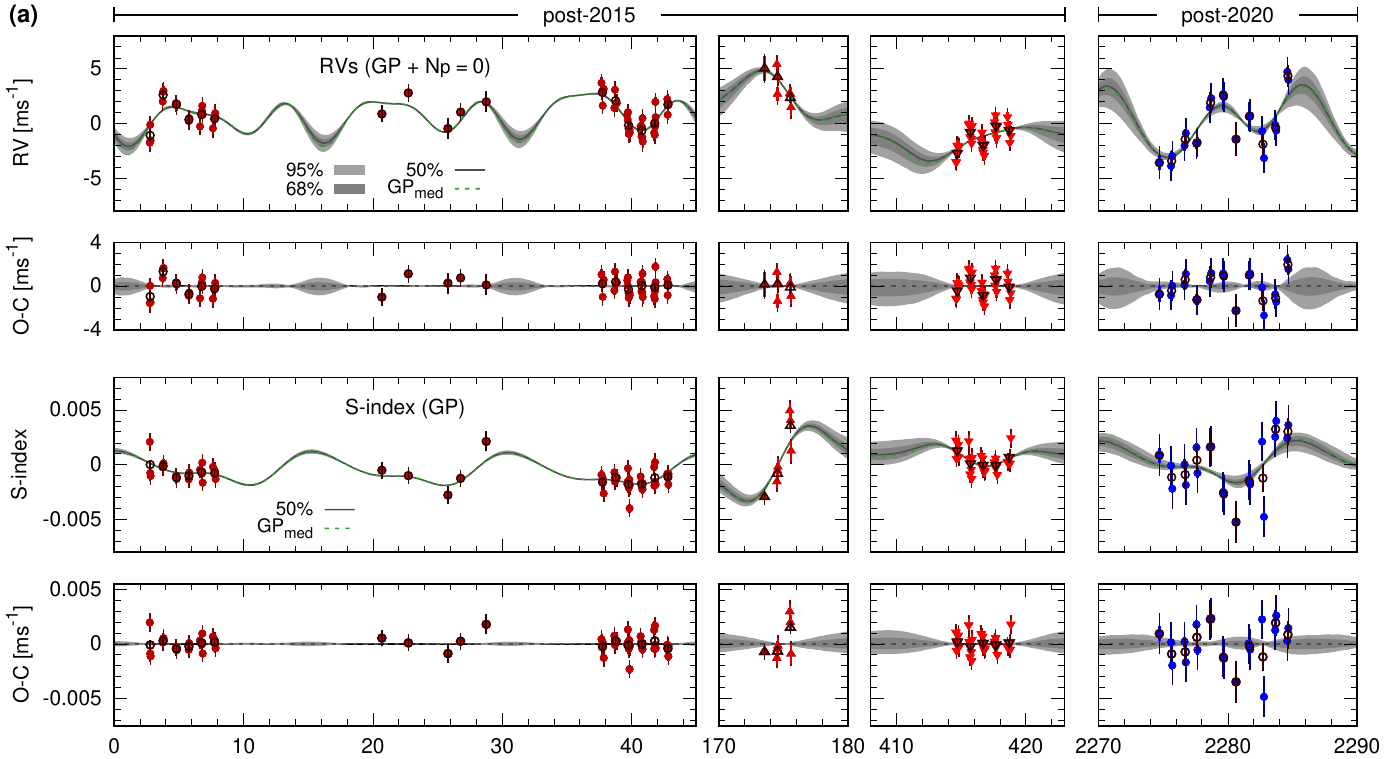}
        \vspace{0mm} \\
        \includegraphics[trim=4mm 0mm 5mm 0mm, width=2\columnwidth]{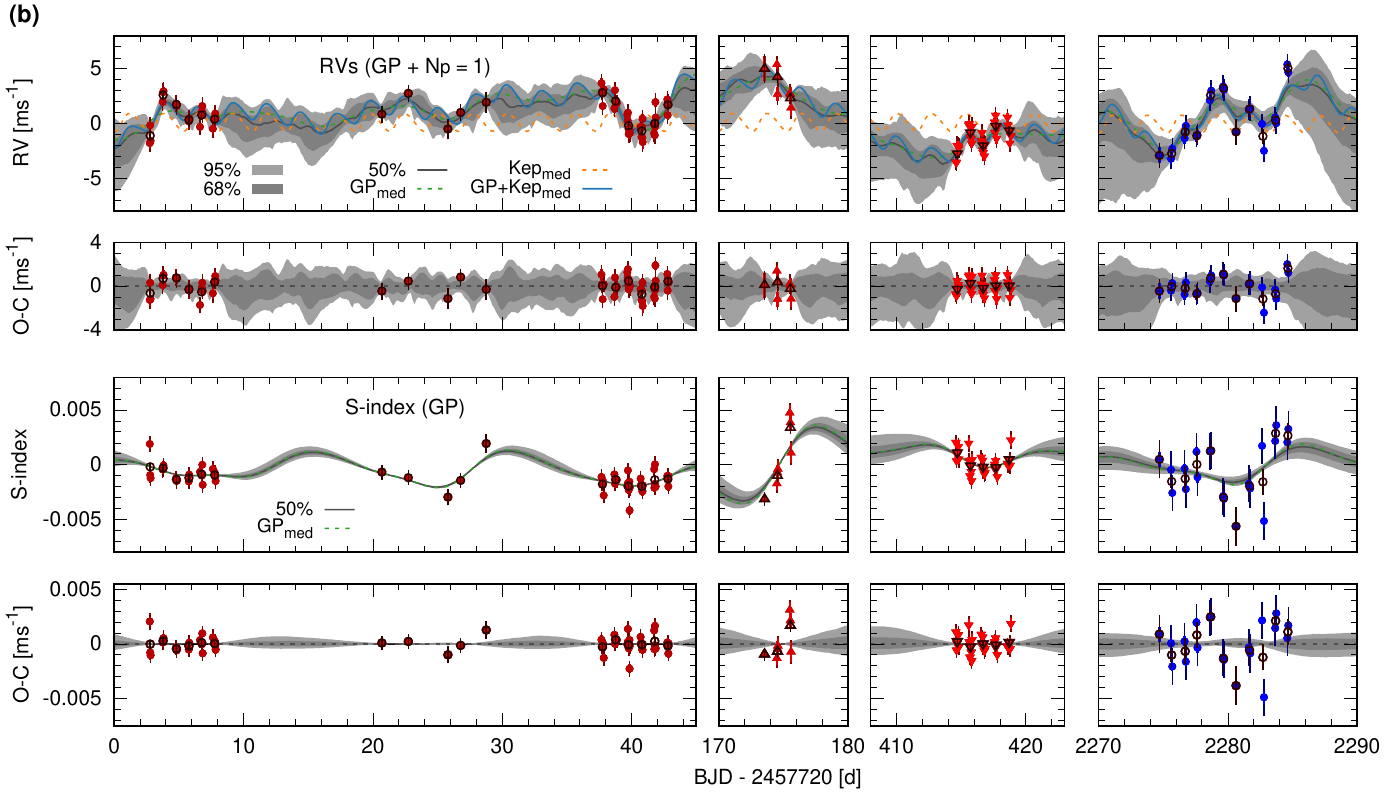}
    \caption{{\hdsix{} solution curves for simultaneous RV + {S$_\textrm{MW}$} models for (a) a pure GP with $N_\textrm{p} = 0$ (Model~3a) and (b) the GP + $N_\textrm{p} = 1$ solution (Model~3b).}}
    %(b) The model with GP included, which prefers $N_\textrm{p} = 0$.}
    \label{fig:hd67200_rvs_activity}
\end{figure*}

\subsubsection{\hdsix{} Activity indicators}
\protect\label{section:hd67200activity}

%We examined the third line moments of the CCF, M$_3$, Bisector Inverse Span (BIS), Full Width at Half Maximum (FWHM) and {S$_\textrm{MW}$} for signs of variability.
With the exception of {S$_\textrm{MW}$}, the likelihood periodograms in \hbox{Fig. \ref{fig:hd67200_periodograms}\,a,b} do not show any evidence for periodic signals (i.e. no power with $\Delta\log{L} > 15$). Significant {S$_\textrm{MW}$} periodicities are present at periods close to $15.99$~d, with a number of closely related significant aliases. There does not appear to be a direct coincidence of the window function peaks (indicated by the vertical dashed lines in {Fig. \ref{fig:hd67200_periodograms}\,a-e}) and significant {S$_\textrm{MW}$} peaks.  A phase fold of {S$_\textrm{MW}$} for the $15.99$\,d period is shown in Fig. \ref{fig:hd67200_sindex_phased} The data points are colour coded to highlight the individual observing epochs.
%P98-P100 (red symbols), P110 HARPS (blue symbols). 
The additional additive noise term from the likelihood model in \textsc{ars} \textrm{(assuming a sinusoidal waveform)} corresponding to this period is 
%respectively $1.4$ and $3.4$ times - 
$3.0$ times the mean {S$_\textrm{MW}$} errors, suggesting significant underestimation of the data uncertainties.

The P98-P100 {(post-2015)} data sets individually span small fractions ($< 31$\%) of the candidate $15.99$\,d rotation period. Although the $10$\,d span of the P110 {(post-2020)} data set covers $60$\% of the  {rotation phase}, there are also too few data in P110 alone to {reliably} identify significant periodicity. It is possible that power at the $15.99$~d period is affected by stochastic changes in chromospheric activity at each observing epoch, so we should not necessarily expect active regions to generate perfectly sinusoidal modulation. While the mean {S$_\textrm{MW}$} levels in P98 and P100 are different, suggesting long term activity changes, these data fall within the {S$_\textrm{MW}$} range spanned by the P110 data. We note however that the mean SNR~$= 33$ ({Ca} \textsc{ii} {H\&K orders}) for P110 is lower than the preceding epochs, where the mean SNR~$=41$. {The P98-100 and P110 data sets yield inconsistent {S$_\textrm{MW}$} vs SNR correlations of $r=-0.20 \pm 0.10$ ($p=0.05$) and $r=0.52_{-0.20}^{+0.16}$ ($p=0.02$)}. Confirmation of a $15.99$\,d stellar rotation period for \hdsix{} would imply active regions contributing significant chromospheric variability are present at relatively fixed longitudes on the $6.25$\,year span of the observations. {Even if active longitudes exist, as has been suggested for over 100 years of solar data by \cite{berdyugina03activelongs}, the long-term coherence of such structures remains unclear.}

%The apparent P110 variability could also be systematically affected by the changing moon illumination and lack of HARPS DRS sky subtraction (\S \ref{section:target_slection}). During P110, \hdsix{} was observed at lunar phases of $0 < \textrm{FLI} < 68$\% and respective moon distances $89$\degs{}~$<\theta_\textrm{dist}$~$101$\degs, but 
%and $\theta_\textrm{sep} \times \textrm{FLI}$, 

%There is however moderate evidence for a correlation between {S$_\textrm{MW}$} and SNR in P110, with $r=0.52_{-0.20}^{+0.16}$.
%For all observations, $r=-0.02 \pm 0.09$. The mean SNR~$= 33$ for P110 is lower than the preceding epochs, where the mean SNR~$=41$).
 
None of the activity indicators show consistent or strong {linear} correlations with the RVs. Each sub-panel of Fig. \ref{fig:hd67200_correlations} gives the Pearson's $r$ linear {RV vs activity} correlation, with upper and lower 68.3\% confidences in parentheses, and the $p$ statistic, with which the null hypothesis of no correlation can be rejected. The P110 data set is probably too small to return meaningful correlations. This is reflected in the large Pearson's $r$ uncertainties for this data set. The RVs are offset by the mean value {for each of the {P98--100} and P110 periods}. The {RV vs BIS} correlation in the {P98--P100} data sets is consistent with no correlation ($r=-0.06$). Similarly, the {RV vs M$_3$} correlation of $r=-0.28$ and {RV vs FWHM} correlation of $r=0.26$ are weak. Despite the tentative {S$_\textrm{MW}$} periodicity at $15.99$\,d, {the P98--P100 data suggests weak RV vs {S$_\textrm{MW}$} negative correlation ($r=-0.16 \pm {0.10}$, $p=0.12$ and there is weak to moderate evidence for a linear correlation with Pearson's $r=0.32_{-0.24}^{+0.20}$ and $p=0.17$ in P110 suggesting the hypothesis of no correlation cannot be reliably ruled out. The RV vs {S$_\textrm{MW}$} distribution shown in Fig. \ref{fig:hd67200_correlations}\,d} appears to show offsets between observing epochs, which may be indicative of activity changes. {Further, it is worth noting that the P98 data, which span $40$\,d, appear to show a correlation. Two data points from P98 sit above the main group of points, both with amongst the highest uncertainties. Hence, while with all P98 data, we obtain $r=0.13_{-0.21}^{+0.14}$, $p=0.32$), removal of the outliers yields $r=0.25_{-0.13}^{+0.12}$, $p=0.06$, a correlation that is much more consistent with the P110 data. The P99 and P100 subsets are too small to yield useful correlations.}

{It is probable that changing activity masks a clean linear correlation. Also, it is not clear that {S$_\textrm{MW}$ or FWHM} should correlate linearly with RV; \cite{barragan22pyaneti2} showed that the RVs correlate with the time derivative of \logrhk{}. This behaviour was also demonstrated between RVs and the second line moment of the RV CCF, $M_2$, in \cite{barnes24moments}.
%Phase shifts between RVs and stellar activity indicators are expected as a result of the geometry and temporal evolution of active regions on the stellar surface. 
Solar observations and time-resolved analyses show that correlations between RVs and activity indicators are not stationary in time, and may vary with activity level and phase coverage (e.g. \citealt{simola22,klein24}). Moreover, quasi-periodic models of stellar activity indicate that these signals arise from evolving surface features, such that no single, fixed phase relationship is expected (e.g. \citealt{nicholson22}). Since for \hdsix{}, {S$_\textrm{MW}$} is the only RV vs activity correlator that shows both a potential linear correlation} \textit{and} {significant periodicity, we performed cross-correlation analysis between the RVs and {S$_\textrm{MW}$}. We used the P98 data (with two outliers removed) and the P110 data (spanning $11$\,d) to search for signs of a phase lag on timescales of the $15.99$\,d periodicity. The cross-correlation function exhibits peaks at lags of  $\sim +0.6$\,d and $+15.7$\,d for P98, implying periodic behaviour that matches the {S$_\textrm{MW}$} $15.99$\,d period. For P110, a lag of  $-5$\,d is seen. These values are not consistent between datasets and correspond to a substantial fraction of the inferred stellar rotation period in P110. Visual inspection of the P110 RVs and {S$_\textrm{MW}$} in Fig. \ref{fig:hd67200_rvs_activity} suggests correlated variability, but the presence of a clear one-to-one relationship is not obvious. This, combined with the incomplete phase coverage means the $-5$\,d lag is unlikely to be robust. The observed behaviour is consistent with quasi-periodic variability and evolving active regions, for which the apparent lag depends on phase coverage and time baseline. We therefore conclude that, although RV and {S$_\textrm{MW}$} variations are probably correlated, there is no robust evidence for a well-defined or stable phase offset between them.}

In summary, \hdsix{} appears to be photometrically very quiet. {Because of probable instrument systematics}, there is no convincing evidence from photometry for periodicity associated with the expected stellar rotation period. {The tentative {S$_\textrm{MW}$} periodicity is probably affected by changes in activity between observing runs. However, the combination of {S$_\textrm{MW}$} periodicity with weak to moderate linear correlation at some epochs, means careful consideration is warranted when inspecting the RVs for evidence of potential Keplerian signals. No clear correlation between the RVs and other activity indicators, nor any corresponding significant periodicity, is readily apparent when examining the entire data set or observing run subsets.}

\begin{figure}
    	\includegraphics[trim=0mm 10mm 0mm 0mm, width=1\columnwidth]{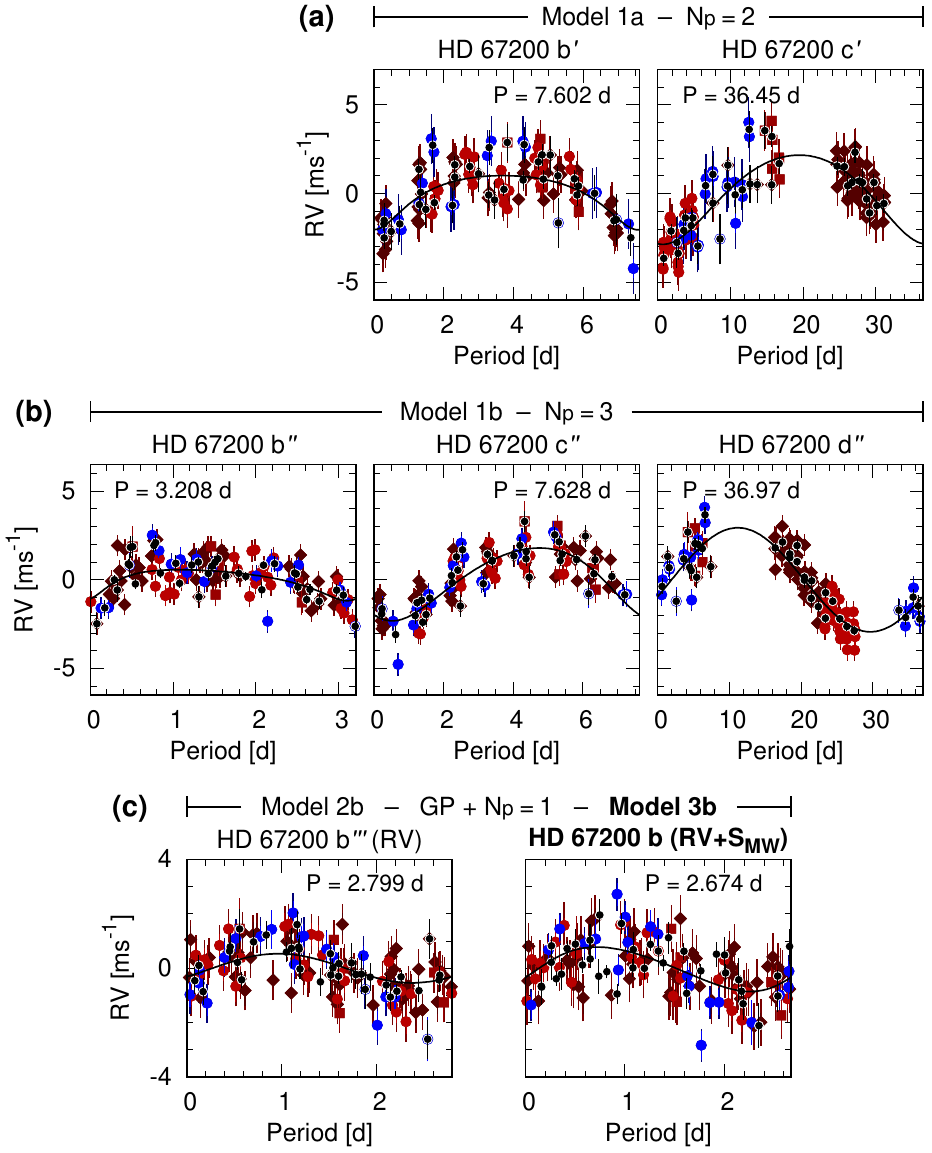}
    \caption{{\hdsix{} phase-folded RVs for the purely Keplerian models with (a) $N_\textrm{p} = 2$ (Model 1a) and (b) $N_\textrm{p} = 3$ (Model 1b). Shown in (c) are the Keplerian phase-folds with $N_\textrm{p} = 1$ (Model~2b and preferred Model~3b, highlighted in bold) after subtracting the GP components. The nightly binned data (as in Figs. \ref{fig:hd67200_rvs} and \ref{fig:hd67200_rvs_activity}) are shown by small filled circles.}}
    \label{fig:hd67200_rvphases}
\end{figure}

\subsubsection{\hdsix{} - Purely Keplerian model}
\protect\label{section:hd67200rvs}

{The recursive maximum likelihood periodograms (Fig. \ref{fig:hd67200_periodograms}\,a,\,b\,and\,c) indicate three persistent periodicities in the $6.25$~yr span of the data, all with $\Delta \log L > 15$}. The third peak has $P=2.83$\,d and $e=0.47$. A likelihood model with fixed eccentricity, $e=0$, yields a different period of $P=3.88$\,d. The distinct period-power distributions yield very similar likelihoods in both instances, indicating the need for a fully Bayesian approach to properly explore posterior space. {RV solutions with} \texttt{kima} {presented by \citetalias{standing26dmpp} found strong evidence for two purely Keplerian signals, identified as DMPP-6\,b and DMPP-6\,c. Further moderate evidence for a $P \sim 3$\,d signal was also found.}

With \texttt{kima}, using a {purely} Keplerian model with $N_\textrm{pmax} = 4$ (i.e. integer prior $N_\textrm{p}: \mathcal{U}[0; 4]$), we {confirm strong evidence for 2 Keplerians ($N_\textrm{p} = 2$) with $\textrm{BF}(N_\textrm{p}=2/N_\textrm{p}=1) > 2213$. Note that the Bayes Factor is a lower limit since no posterior samples (generated by \texttt{kima} from the weighted} \texttt{DNest4} {raw samples) with $N_\textrm{p}<2$ are generated for the \hdsix{} data with this model. In other words, $N_\textrm{p}=0$ and $N_\textrm{p}=1$ are very strongly disfavoured.} The Model~1a parameters, statistics and derived parameters for the candidate signals, \hdsix{}\,b$^\prime$ and \hdsix{}\,c$^\prime$, are listed in Table~\ref{tab:hd67200_solution}. The data and {$N_\textrm{p} = 2$ model curves are shown in Fig.~\ref{fig:hd67200_rvs}\,a; the median 50\% curve and the corresponding 68\% and 95\% envelopes are determined from 1000 randomly selected samples from the $N_\textrm{p} = 2$ posteriors. The single sample most representative of the 50\% median is determined via least squares and shown as the blue curve. The corresponding} phase-folded RVs are shown in the upper panels of Fig.~\ref{fig:hd67200_rvphases}. {Our Model~1a signals correspond to the two maximum posterior \hbox{DMPP-6} solutions presented in \citetalias{standing26dmpp}.} There is moderate evidence for 3 Keplerians ($N_\textrm{p} = 3$), with of {$\textrm{BF}(N_\textrm{p} = 3/N_\textrm{p} = 2) = 23.2$. The parameters for the Model~1b solution are presented} in Table~\ref{tab:hd67200_solution} with candidate planets {\hdsix{}\,b$^{\prime\prime}$, \hdsix{}\,c$^{\prime\prime}$ and \hdsix{}\,d$^{\prime\prime}$. }The apparent improvement in fit ({maximum posterior $\log{L} = -138.85$, compared with $-161.20$} for $N_\textrm{p}=2$) is visually evident in the full RV solution curves shown in Fig. \ref{fig:hd67200_rvs}\,b, particularly in the 2023 (P110) data set (blue points). The white noise {terms, $\sigma_\textrm{p15}$ and $\sigma_\textrm{p20}$ (post-2015 and post-2020 respectively)} in the likelihood model with $N_\textrm{p} = 2$ {accommodate} some of the variability that can be fit with the additional Keplerian signal in the $N_\textrm{p} = 3$ model. {The ratio of $\sigma_\textrm{p15}/\sigma_\textrm{p20}$ in the $N_\textrm{p} = 3$ model is also slightly reduced over $N_\textrm{p} = 2$ model, suggesting that the post-2020 data are better fit with the additional Keplerian.}

Despite imposing AMD stability checks and {using} the Beta distribution eccentricity prior (\S \ref{section:kimamodel}), it is possible that the data sampling has biased the eccentricity distributions of the planet candidates. For the $N_\textrm{p} = 3$ solution, both the inner planets' eccentricities, with $e>0.2$, are greater than the $e=0.031$ of the outer planet candidate. Since the eccentricity of the planet candidates in the $N_\textrm{p} = 2$ scenario are greater than those in the $N_\textrm{p} = 3$ solution, it is possible that more data would provide evidence for more signals with even lower eccentricities. On the other hand, we cannot ignore the possibility that stellar activity {contributes to the RV variability or describes the RV variability better.}

\subsubsection{\hdsix{} - Keplerian model with a GP}
\protect\label{section:hd67200rvskepgp}

{We employed the} \textsc{s+leaf} {ESP kernel to model the RVs (Model~2) and used relatively broad hyperparameter priors to check for simultaneous Keplerian and activity signals (see Table \ref{tab:hd67200_solution}). A broad uniform prior of $\eta_3 :\,\mathcal{U}[10;100]$ (with lower bound informed by the Monte Carlo simulation) was used in conjunction with an $\eta_2 :\,\mathcal{U}[>0.5\,\eta_3; t_\mathrm{span}]$ prior as in \S \ref{section:hd67200photometry}. Although \hdsix{} is more massive than the Sun, its low \vsini{} and low activity most closely resemble a sub-solar minimum to solar minimum state. If a close-orbiting planet is present, as suggested by the original sub-basal activity selection of this target,
%(with even lower apparent \logrhk{}), 
it is possible that the underlying activity level is above solar minimum. The solar min and max models in \citet{barnes24moments}, which model both cool spots and facular regions, indicate the presence of harmonics at $\eta_3/2$ and $\eta_3/3$, in the $M_1$ central line moments (a proxy for RV) typically recovering periodicity at the fundamental characteristic (rotation) $\eta_3$ and $\eta_3/3$ (\citealt{barnes24moments}; Fig. 7). To match these signatures, we allowed for harmonics with $3:2$ and $3:1$ amplitude contributions of $\eta_3/2$ and $\eta_3/3$, by assuming a prior $\eta_4$ of $\mathcal{U}[0.8;10]$ (see Table \ref{tab:harmonics}).}

{We again used a prior of $N_\textrm{pmax} = 4$ for the number of Keplerian signals. For Model~2a (Table \ref{tab:hd67200_solution}), a BF(GP/no-GP)~$> 1271$ (the lower BF limit is determined by the total number of generated posterior samples in this converged model) signals strong evidence to include a GP with no additional Keplerian signals. In other words for Model~2a, all the RV variability is accommodated by the GP. Fig. \ref{fig:hd67200_rvs}\,c shows the resulting GP~+~$N_\textrm{p}=0$ model solution curves. The global evidence, $\log \mathcal{Z} = -207.67$, strongly favours Model~2 with a GP over the purely Keplerian Model~1, with $\Delta\log \mathcal{Z} = 5.58$ (i.e. BF~$\sim 265$). The marginalised characteristic period for Model~2a is $\eta_3~=~10.83_{-0.73}^{+0.43}$\,d.
%with pileup of samples at the lower boundary $10$\,d boundary of the prior. 
With $\eta_2 = 9.130$\,d and a maximum posterior at $\sim\eta_3/2$, it is clear that the GP prefers to fit the short period signals with periodicities that are not optimally matched to the system parameters and Monte Carlo simulation.}

{The evidence for a GP with a single Keplerian (Model~2b; Table \ref{tab:hd67200_solution}) is \hbox{$\textrm{BF}(N_\textrm{p}=1/N_\textrm{p}=0)=10.7$}. The Model~2b solution, plotted in Fig. \ref{fig:hd67200_rvs}\,d, approaches the moderate $\textrm{BF}=12$ threshold on the \cite{trotta08bayes} scale. The Model~2b Keplerian periodicity of $P=2.799$\,d is close to the $P_3 = 2.83$\,d periodogram signal and the $3.208$\,d \hdsix{}\,d in Model~1b. In Model~2b, $\eta_3 = 14.74_{-0.59}^{+3.44}$\,d also closely matches the peak power periodicity in the {S$_\textrm{MW}$}. The GP amplitude dominates over the RV variability with a median $\eta_1 = 2.90\,$\ms{} over a Keplerian amplitude of $0.89$\,\ms{}. The marginalised eccentricity distribution is also skewed to large amplitudes ($e=0.21_{-0.17}^{+0.35}$) with a maximum likelihood of $e=0.65$.}

\subsubsection{\hdsix{} - Keplerian model with a GP using simultaneous RV and {S$_\textrm{MW}$} timeseries}
\protect\label{section:hd67200rvskepgp_actgp}

{We extended the RV-only analysis in \S \ref{section:hd67200rvskepgp} by adopting the {S$_\textrm{MW}$} as a simultaneous activity timeseries modelled in conjunction with the RVs. The GP hyperparameters, $\eta_2, \eta_3$~and~$\eta_4$ are common to the RVs and {S$_\textrm{MW}$} measurements, while separate $\eta_1$ and $\eta_1^\textrm{act}$ are used. Adopting the same priors as in \S \ref{section:hd67200rvskepgp} yields strong evidence for a model with a GP and no additional Keplerian signals. Model~3a (GP~+~$N_\textrm{p}=0$) is shown in Table \ref{tab:hd67200_solution_rv_activity} and plotted in Fig. \ref{fig:hd67200_rvs_activity}\,a. Since Model~3 contains a simultaneous activity timeseries, the global evidence of $\ln \mathcal{Z} = -400.04$ can not be compared directly with Models 1 and 2. The previously identified {S$_\textrm{MW}$} periodicity leads to a posterior $\eta_3 = 15.10_{-0.52}^{+0.75}$\,d. We note that the additive noise term is slightly larger on the RVs than for Model~2a, while the {S$_\textrm{MW}$} post-2020 observations with large additive noise do not contribute significant evidence for this solution. In Model~3b, $\textrm{BF}(N_\textrm{p}=1/N_\textrm{p}=0)=10.3$ is found, again approaching the moderate evidence threshold (see Table \ref{tab:hd67200_solution_rv_activity} and Fig. \ref{fig:hd67200_rvs_activity}b). The posterior period distribution favours a Keplerian with $P = {2.674}_{ -0.195}^{+0.308}$\,d and a moderate eccentricity with posterior that is skewed to higher values ($e = {0.18}_{-0.14}^{+0.35}$\,d). The $K = {0.84}$\,\ms{} and GP $\eta_1 = 2.82$\,\ms amplitude are similar to Model~2b. We refer the Model~3b planet candidate as \hdsix{}\,b.}

\subsubsection{\hdsix{} summary}
\protect\label{section:summaryhd67200}

{Recursive likelihood periodogram analysis of the \hdsix{} RVs reveals long-term coherent periodicity. Posterior sampling with $N_\textrm{p}$ as a free parameter yields strong evidence for at least two Keplerian signals and moderate evidence for three Keplerians. Despite this, there remain systematics in the residuals, which point to either additional Keplerian signals or stellar variability. For the purely Keplerian model, the observation sampling results in phase gaps for the longest period candidates: i.e. the Model~1a ($N_\textrm{p} = 2$), \hdsix{}\,c$^\prime$ phase fold ($P = 36.45$\,d) and the Model~1b ($N_\textrm{p} = 3$), \hdsix{}\,d$^{\prime\prime}$ phase fold ($P = 36.97$\,d), shown respectively in Fig.~\ref{fig:hd67200_rvphases}\,a and Fig.~\ref{fig:hd67200_rvphases}\,b. Apparent clustering of RVs from different runs at specific phases is also seen in Fig.~\ref{fig:hd67200_rvphases}\,b for the Model~1b ($N_\textrm{p} = 3$) \hdsix{}\,c$^{\prime\prime}$ phase fold.}

{Introducing a model with a GP to investigate a stellar component in the RVs is supported by the apparent {S$_\textrm{MW}$} periodicity at $\sim 15$\,d, which is in broad agreement with the expected stellar rotation period. The Keplerian models with a GP for the RV timeseries and the simultaneous RV + {S$_\textrm{MW}$} timeseries both show strong evidence for a pure GP, with $N_\textrm{p} = 0$. In other words, both the RV and RV + {S$_\textrm{MW}$} variabilities are well-modelled by the GP alone. The global model evidence can be compared in the RV-only models, allowing us to select Model~2 with a GP over the purely Keplerian model. We thus consider the Keplerian signals reported as \hbox{DMPP-6\,b} and \hbox{DMPP-6\,c} in our previous study (\citealt{standing26dmpp}) to be superseded by the models with GPs presented here. While the global model evidence favours Model~2 over Model~1 with BF~$=~265$, the GP model shows a preference for the single Keplerian Model~2b (GP + $N_\textrm{p} = 1$)  with evidence approaching the moderate threshold.}

{The phase folds for Model~2b and 3b (GP + $N_\textrm{p} = 1$) are shown in Fig.~\ref{fig:hd67200_rvphases}\,c. Both Model~2b and Model~3b yield broadly similar hyperparameters and Keplerian parameters. Since more data are used as evidence in Model~3b (i.e. the simultaneous {S$_\textrm{MW}$} timeseries), we retain it, with its corresponding tentative \hdsix{}\,b candidate (Fig.~\ref{fig:hd67200_rvphases}\,c, right panel), as our preferred solution. If confirmed with further more extensively sampled observations, \hdsix{}\,b, with  ${2.07}^{+0.52}_{-0.47}$\,M$_\oplus$ is expected to be in a $0.041$\,AU orbit; consistent with the hypothesis underlying DMPP that the target stars host short period mass-losing planets. }

%%%%%%%%%%%%%%%%%%%%%%%%%%%%%%%%%%%%%%%%%%%%%%%%%%%%%%%%%%%%
%%%%%%%%%%%%%%%%%%%%%%% HD118006 %%%%%%%%%%%%%%%%%%%%%%%%%%%
\subsection{\hdone}
\protect\label{section:hd118006}
%%%%%%%%%%%%%%%%%%%%%%%%%%%%%%%%%%%%%%%%%%%%%%%%%%%%%%%%%%%%%
%%%%%%%%%%%%%%%%%%%%%%%%%%%%%%%%%%%%%%%%%%%%%%%%%%%%%%%%%%%%%

%%%%%%%%%%%%%%%%%%%%%%%% FIGURE: HD118006 PERIODOGRAMS and ACTIVITIES
\begin{figure}
% \begin{tabular}{cc}
    \includegraphics[trim=0mm 6mm 0mm 0mm, width=1.0\columnwidth]{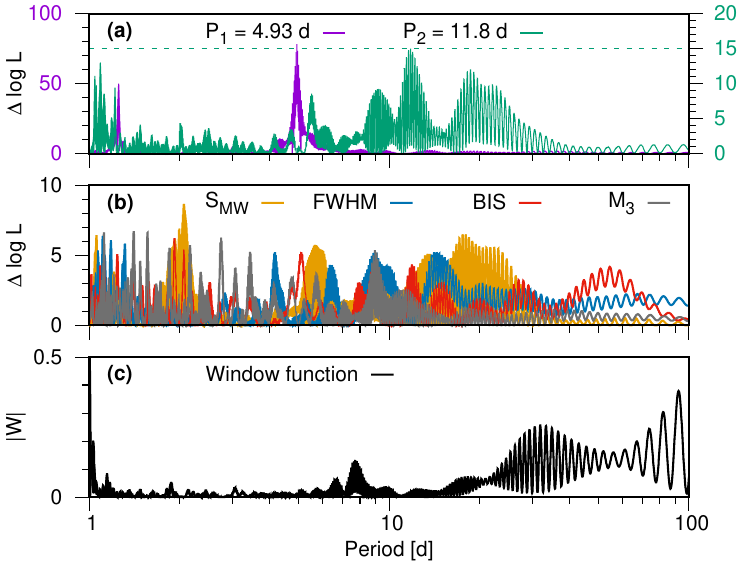}
    %\includegraphics[trim=4mm 5mm 0mm 0mm,  width=1.04\columnwidth]{DMPPmulti/HD118006_activities_v4.pdf} \\
% \end{tabular}
    \caption{\hdone{} RV and activity periodograms as in Fig. \ref{fig:hd67200_periodograms}. The second RV period, $P_2$, in (a) is borderline significant, as indicated by the dashed green line indicating $\Delta\log L = 15$.}
    \label{fig:hd118006_periodograms}
\end{figure}

\begin{figure}
% \begin{tabular}{cc}
    %\includegraphics[trim=3mm -3mm 2mm 0mm, width=0.95\columnwidth]{DMPPmulti/HD118006_periodograms_activities_v2.pdf}
    \includegraphics[trim=0mm 10mm 0mm 0mm,  width=1.0\columnwidth]{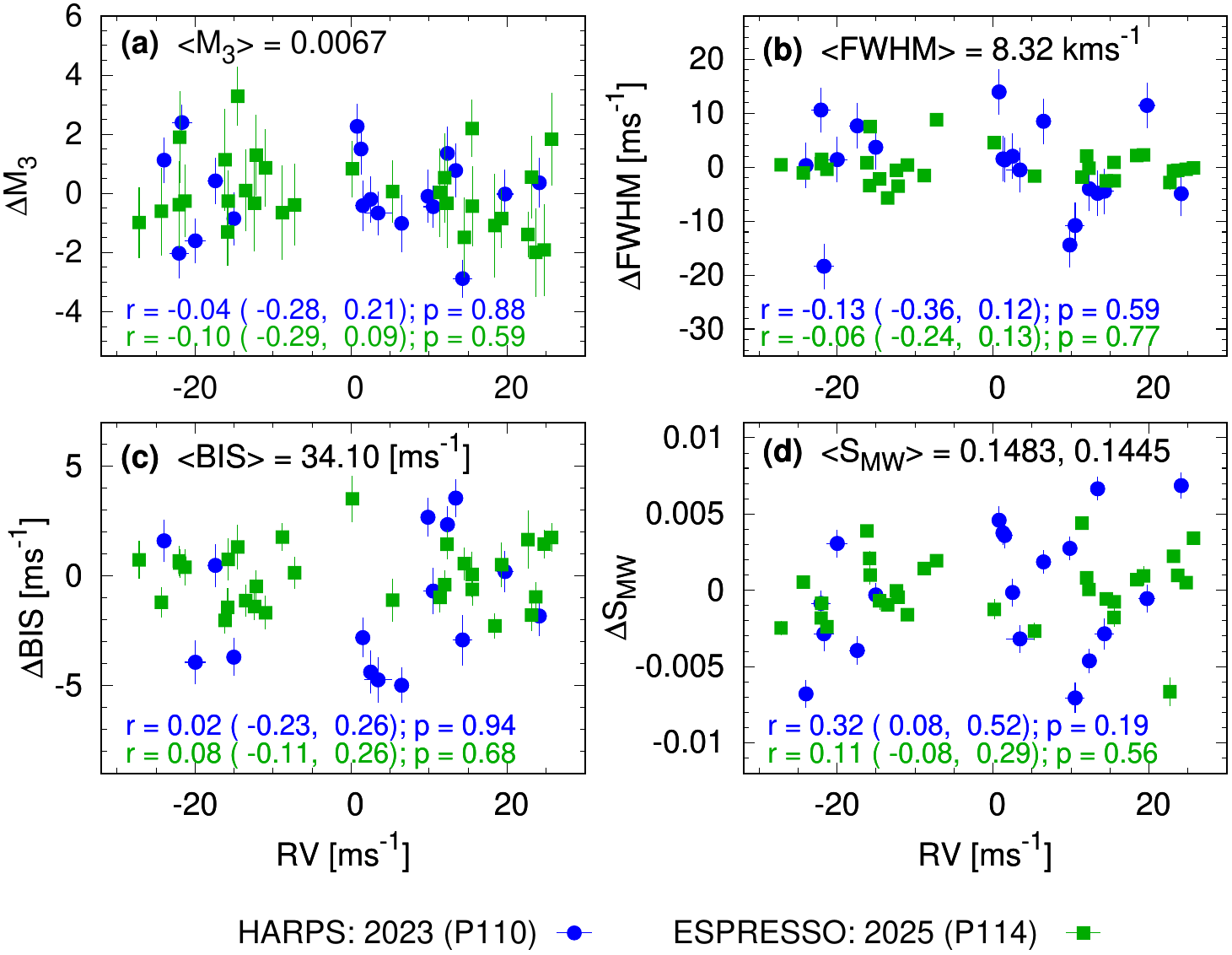}
% \end{tabular}
    \caption{\hdone{} Activity vs RV for the P110 and P114 observations indicating no significant correlations.}
    \label{fig:hd118006_correlations}
\end{figure}

%%%%%%%%%%%%%%%%%%%%%%%% FIGURE: HD118006 RV SOLUTIONS (Keplerian and Keplerian + GP)
\begin{figure*}
\includegraphics[trim=4mm 0mm 5mm 0mm, width=2\columnwidth]{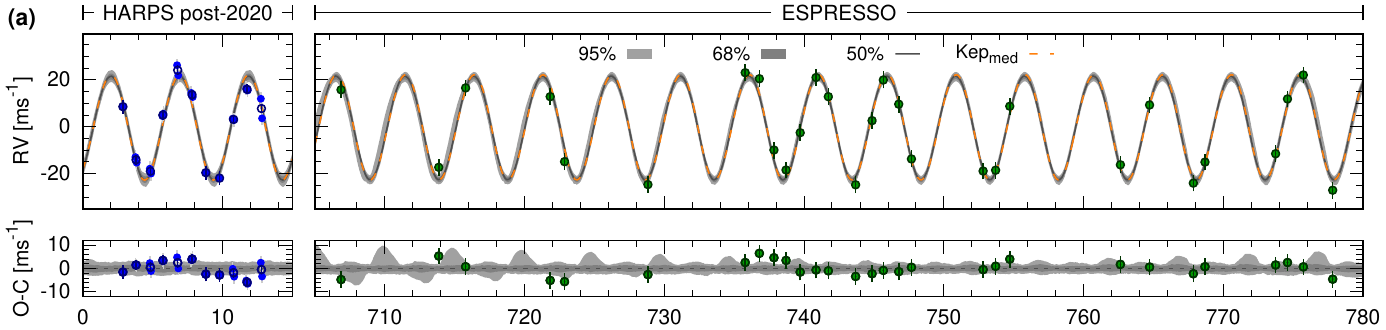}
\includegraphics[trim=4mm 0mm 5mm 0mm, width=2\columnwidth]{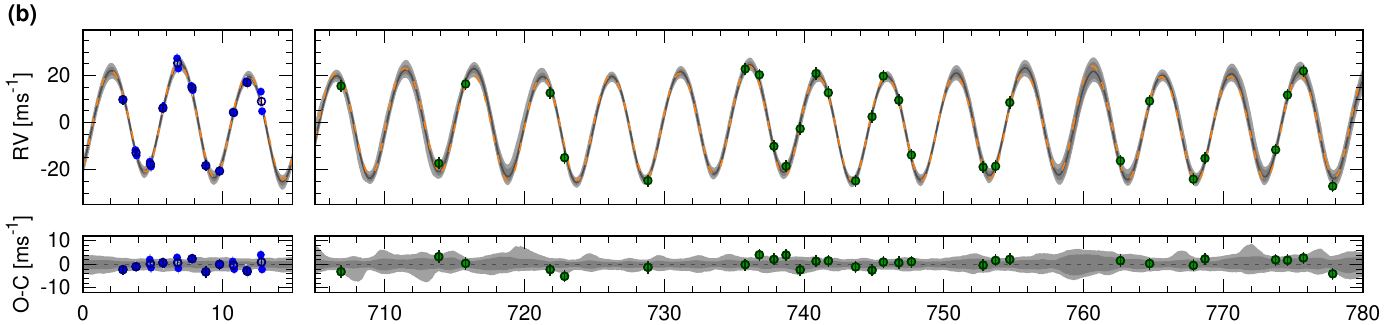}
\includegraphics[trim=4mm 0mm 5mm 0mm, width=2\columnwidth]{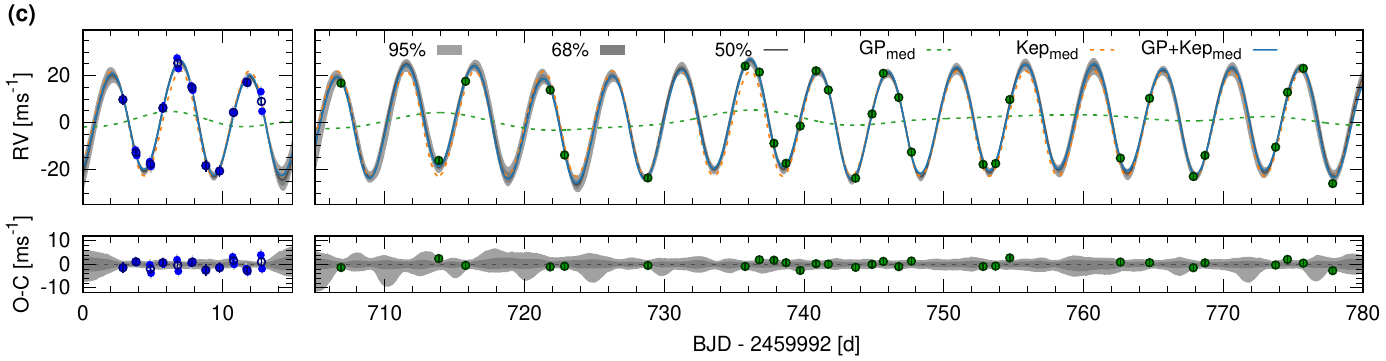}
\caption{{Solution curves for the \hdone{} RVs. The purely Keplerian solutions are shown in (a) for $N_\textrm{p} = 1$ (Model~1a) and in (b) for the $N_\textrm{p} = 2$ (Model~1b). The GP + Keplerian solution is shown in (c) for $N_\textrm{p} = 1$ (Model~2). See Table \ref{tab:hd118006_GPsolution} for model solution details.}}
    \label{fig:hd118006_rvsolutions}
\end{figure*}

%%%%%%%%%%%%%%%%%%%%%%%% FIGURE: HD118006 hyperparams
%\begin{figure}
%\includegraphics[trim=0mm -2mm 0mm 0mm, width=0.99\columnwidth]{DMPPmulti/HD118006_hyperparams_posteriors.pdf}
%\caption{\hdone{} hyperparameter posteriors for Model~$3$ with a Keplerian + GP.}
%    \label{fig:hd118006_hyperparams}
%\end{figure}

\begin{figure}
    \begin{center}
	   \includegraphics[trim=0mm 0mm 0mm 0mm, width=0.90\columnwidth]{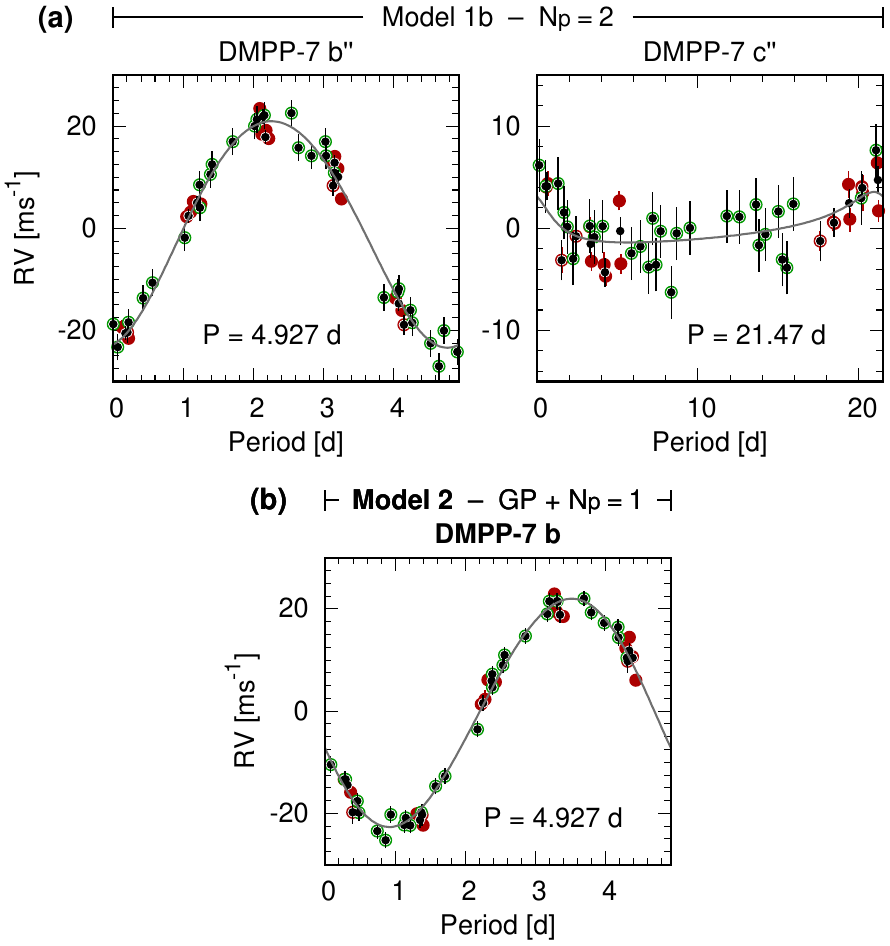}
    \end{center}
    \caption{{\hdone{} phase-folded RVs for the purely Keplerian model with (a) $N_\textrm{p} = 2$ (Model 1b). Shown in (b) is the Keplerian phase-fold with $N_\textrm{p} = 1$ (preferred Model~2, highlighted in bold) after subtracting the GP component.}}

    \label{fig:hd118006_rvphases}
\end{figure}

%---------------------------------------------------------- HD 118006 RESULTS TABLE -------------------------------------------------------
%\begin{adjustbox}{width=0.95\textwidth}
\begin{table*}
    
\renewcommand{\arraystretch}{1.25}
\centering
\setlength\tabcolsep{0.02\columnwidth}

\begin{tabular}{lccc}
%%% RESULTS DIR: /data/kima-venv2/HD118006_kumara/All_HARPS_ESPRESSO_remove6_npmax4

% N.B.  gamma=36651 added back on here as the RVs have it included for HD118006 (original unsubtracted from sBART) - we want without and then
%                   put gamma value in the table as gamma-36651
% Np=1 cluster_hdbscan.py --nkep=1 --csize=100 --nparms=3  --sigma=5.0 --t0=60382.351505 --mstar="1.23 0.02" --gamma=36651.
% Np=2 cluster_hdbscan.py --nkep=2 --csize=1000 --nparms=3  --sigma=5.0 --t0=60382.351505 --mstar="1.23 0.02" --gamma=36651.

% Np=1               Nsamples = 165
% 21.5d cluster Nsamples = 5493 --> 5493 / 165 = 33.3
% 11.8d cluster Nsamples = 4426 --> 4426 / 165 = 26.8

%%% CHECK FILES: lastrun_output_nkep1.dat         and lastrun_output_nkep2.dat         (Kima)
%%%              lastrun_output_nkep1.dat-CLUSTER and lastrun_output_nkep2.dat-CLUSTER (CLUSTER_HDBSCAN)
\hline
& {Model 1a (${N_\textrm{p}=1}$)} & \multicolumn{2}{c}{{Model~1b (${N_\textrm{p}=2}$)}} \\
{Parameter}               &  {\hdone{}\,b$^\prime$}  & {\hdone{}\,b$^{\prime\prime}$} &
                                 {\hdone{}\,c$^{\prime\prime}$} \\
\hline
%{\hdone{}\,b$^{\prime\prime}$}                      & {\hdone{}\,c$^{\prime\prime}$}                                          \\
\multicolumn{4}{c}{\vspace{-3mm}} \\
$P$                           & ${4.92658}^{+0.00042}_{-0.00032} [4.92673]$           & ${4.92662}^{+0.00036}_{-0.00032} [4.92654]$            & ${21.46877}^{+2.10500}_{-3.18984} [23.56424]$          \\
$K$                           & ${22.10}^{+0.66}_{-0.69} [22.13]$                     & ${22.38}^{+0.57}_{-0.56} [22.46]$                      & ${3.37}^{+0.73}_{-0.72} [3.78]$                        \\
$M0$                          & ${2.386}^{+1.684}_{-0.990} [2.284]$                   & ${2.452}^{+0.522}_{-0.393} [2.458]$                    & ${3.032}^{+2.475}_{-1.412} [1.469]$                    \\
$e$                           & ${0.032}^{+0.030}_{-0.025} [0.057]$                   & ${0.056}^{+0.033}_{-0.034} [0.102]$                    & ${0.457}^{+0.151}_{-0.254} [0.556]$                    \\
$\omega$                      & ${2.964}^{+1.091}_{-1.077} [2.892]$                   & ${2.768}^{+0.411}_{-0.379} [2.746]$                    & ${1.009}^{+1.377}_{-0.663} [0.691]$                    \\

$\gamma_\textrm{HARPS}-36651$ [ms$^{-1}$]    & ${0.61}^{+0.93}_{-0.90} [0.38]$            & \multicolumn{2}{c}{${-0.40}^{+1.06}_{-1.09} [-0.67]$} \\
$\gamma_\textrm{ESPRESSO}-36651$ [ms$^{-1}$] & ${-13.17}^{+0.69}_{-0.53} [-12.73]$        & \multicolumn{2}{c}{${-12.97}^{+0.52}_{-0.51} [-12.88]$} \\
$\sigma_\textrm{HARPS}$ [ms$^{-1}$]          & ${3.42}^{+0.79}_{-0.67} [2.92]$            & \multicolumn{2}{c}{${2.17}^{+0.65}_{-0.49} [1.78]$} \\
$\sigma_\textrm{ESPRESSO}$ [ms$^{-1}$]       & ${3.28}^{+0.54}_{-0.35} [3.43]$            & \multicolumn{2}{c}{${2.39}^{+0.47}_{-0.34} [2.08]$} \\
%t0 =  60382.35150
\multicolumn{4}{c}{\vspace{-3mm}} \\
$a$ [AU]                      & ${0.0607}^{+0.0004}_{-0.0003} [0.0605]$               & ${0.0607}^{+0.0003}_{-0.0003} [0.0610]$                & ${0.1614}^{+0.0113}_{-0.0153} [0.1733]$                \\
$m_p$\,sin\,$i$ [M$_\oplus$]  & ${67.335}^{+2.178}_{-2.176} [67.039]$                 & ${68.191}^{+1.906}_{-1.838} [68.938]$                  & ${14.436}^{+3.202}_{-2.840} [16.336]$                  \\
$t_\textrm{c}$ [JD]           & ${60379.537}^{+0.071}_{-0.058}$                       & ${60379.583}^{+0.055}_{-0.058} $                       & ${60374.704}^{+2.890}_{-8.992} [60377.702]$            \\
                              & $[60379.610]$                                         & $[60379.646]$                                          & $[60377.702]$                                          \\
\multicolumn{4}{c}{\vspace{-3mm}} \\
$\ln{\mathcal{L}}$ [MAP]                    & $-123.45$     & \multicolumn{2}{c}{$-100.70$} \\
BF($N_\textrm{p}/(N_\textrm{p}-1))$         & $>165$        & \multicolumn{2}{c}{$60.03$} \\ %(33.30 for this cluster)
%\multicolumn{4}{c}{} \\
Global evidence, $\ln \mathcal{Z}$ ($N_\textrm{p}\leq4$) & \multicolumn{3}{c}{$-153.73$} \\
%ln\,$Z$ ($N_\textrm{p}\leq2$) & \multicolumn{3}{c}{} \\

%%% RESULTS DIR: /data/kima-venv2/HD118006_kumara/All_HARPS_ESPRESSO_GP_remove6_npmax4_wider-eta3-SPLEAF
%%% CHECK FILES: 
%  lastrun_output.dat_Npmax4_eta3-22p3-tol0p333333

% N.B.  gamma=36651 added back on here as the RVs have it included for HD118006 (original unsubtracted from sBART) - we want without and then
%                   put gamma value in the table as gamma-36651
%cluster_hdbscan.py --nkep=1 --csize=1200 --nparms=101  --sigma=5.0 --eta3="22.3 0.333333" --t0=60382.351505 --mstar="1.23 0.02" --gamma=36651.

% awk '{if($11==1 && $7 > 11.8*0.666667 && $7 < 11.8*1.333333)print $0}' posterior_sample.txt | wc -l ---> 1544 samples
% awk '{if($11==1 && $7 > 22.15*0.666667 && $7 < 22.15*1.333333)print $0}' posterior_sample.txt | wc -l ---> 2833 samples
\multicolumn{4}{c}{\vspace{-3mm}} \\
\hline
& \multicolumn{3}{c}{{Model 2} ({GP +} ${N_\textrm{p}=1}$) } \\
{Parameter}             & \multicolumn{3}{c}{{\hdone{}\,b}}  \\
\hline
\multicolumn{4}{c}{\vspace{-3mm}} \\
$P$ [d]                     & \multicolumn{3}{c}{${4.92666}^{+0.00030}_{-0.00030} [4.92661]$}        \\
$K$ [ms$^{-1}$]             & \multicolumn{3}{c}{${22.61}^{+0.49}_{-0.49} [22.62]$          }        \\
$M_0$ [rad]                 & \multicolumn{3}{c}{${2.553}^{+0.501}_{-0.386} [2.741]$        }        \\
e                           & \multicolumn{3}{c}{${0.054}^{+0.025}_{-0.029} [0.059]$        }        \\
$\omega$ [rad]              & \multicolumn{3}{c}{${2.641}^{+0.372}_{-0.445} [2.445]$        }        \\

$\eta_1$ [ms$^{-1}$]          & \multicolumn{3}{c}{${5.13 }^{+6.49 }_{-2.08 } [3.91 ]$} \\
$\eta_2$ [d]                  & \multicolumn{3}{c}{${49.742}^{+119.105}_{-28.149} [34.862]$} \\
$\eta_3$ [d]                  & \multicolumn{3}{c}{${22.152}^{+1.613}_{-3.019} [23.182]$} \\
$\eta_4$                      & \multicolumn{3}{c}{${1.461}^{+2.089}_{-0.705} [0.967]$} \\

$\gamma_\textrm{HARPS}-36651$ [ms$^{-1}$]    & \multicolumn{3}{c}{${-0.36}^{+5.88}_{-6.06} [-2.68]$} \\
$\gamma_\textrm{ESPRESSO}-36651$ [ms$^{-1}$] & \multicolumn{3}{c}{${-14.00}^{+3.26}_{-4.69} [-16.13]$} \\
$\sigma_\textrm{HARPS}$ [ms$^{-1}$]          & \multicolumn{3}{c}{${2.28}^{+0.69}_{-0.51} [1.78]$} \\
$\sigma_\textrm{ESPRESSO}$ [ms$^{-1}$]       & \multicolumn{3}{c}{${1.93}^{+0.62}_{-0.45} [1.28]$} \\

\multicolumn{4}{c}{\vspace{-3mm}} \\
%t0 =  60382.351505
$a$ [AU]                                               & \multicolumn{3}{c}{${0.0607}^{+0.0003}_{-0.0003} [0.0603]$}     \\
$m_\textrm{p}$\,sin\,$i$ [M$_\oplus$]                  & \multicolumn{3}{c}{${68.958}^{+1.692}_{-1.663} [68.066]$  }     \\

$T_\textrm{eq} (A_\textrm{B} = 0, 0.36)$ [K]           & \multicolumn{3}{c}{$1456, 1302$}                                \\
$R_\textrm{p}$ (predicted from \mpsini{}) [R$_\oplus$] & \multicolumn{3}{c}{$9.55 \pm 2.09$}                             \\
$R_\textrm{p}$ (predicted, $i=57.3$\degs) [R$_\oplus$] & \multicolumn{3}{c}{$10.7 \pm 2.4$}                              \\

$t_\textrm{c}$ [JD]                                    & \multicolumn{3}{c}{${60379.591}^{+0.046}_{-0.054} [60379.586]$} \\

\multicolumn{4}{c}{\vspace{-3mm}} \\
$\ln{\mathcal{L}}$ [MAP]                             & \multicolumn{3}{c}{$-114.033$} \\
BF(GP + $N_\textrm{p}=1 / N_\textrm{p}=0$)           & \multicolumn{3}{c}{$> 4564$} \\ % (>2833 for the cluster at 21d +/- 33.333% )
Global evidence, $\ln \mathcal{Z}$                   & \multicolumn{3}{c}{$-154.59$} \\

\end{tabular}
\caption{{\hdone{} parameters for purely Keplerian models, $N_\textrm{p}=1$ (Model~1a) and $N_\textrm{p}=2$ (Model~1b) and the GP model with $N_\textrm{p}=1$ (Model~2).}}
\label{tab:hd118006_GPsolution}
%\end{adjustbox}
\end{table*}

\subsubsection{\hdone{} stellar parameters and evolution status}
\protect\label{section:HD118006params}

HD\,118006 (hereafter \hdone{}) was classified as a G1V star by \citet{Houk1999}. It was selected as a DMPP target from our measurement of log~$R^\prime_\textrm{HK} = -5.11$ based on an single spectrum from the California Planet Search (CPS) program reported by \citet{isaacson10activity}. No information on sky subtraction is given by the CPS, though we assume that inter-order scattered light subtraction was performed. Our subsequent HARPS spectra show log~$R^\prime_\textrm{HK} = -5.05 \pm 0.03$ (Table \ref{tab:stellar_params}), with two of the 20 HARPS spectra taken in P110 indicating sub-basal values. The $M_*$ and $T_\textrm{eff}$ and $B-V$ colour estimate (Table \ref{tab:stellar_params}) are indicative of a late-F star or early-G star, while our age estimate of $3.87^{+0.28}_{-0.31}$\,Gyr and \hbox{$\textrm{log}\,g = 4.17 \pm 0.09$} indicate \hdone{} is a main sequence star. These parameters are at tension with the corresponding radius estimate of $R_* = 1.49 \pm 0.02$~R$_\odot$, which implies slightly evolved status. Further, an independent radius estimate using parameters derived from the Gaia mission \citep{gaia23dr3} returns $R_* = 1.64\pm0.04$~R$_\odot$. With $\Delta M_\textrm{v} = +0.44$ above the default main-sequence, it only just meets our DMPP selection criterion (\S \ref{section:target_slection}). The renormalised unit weight error (RUWE) value of 0.844 from Gaia \citep{lindegren21dr3,gaia23dr3} is below the 1.25 threshold, which would potentially indicate a dynamical companion \citep{penoyre22,castro-ginard24}. A fainter background star, Gaia~DR3~3663052713405023616, at 1.25 kpc, with $G=14.56$ \citep{gaia23dr3} is located $\sim 19^{\prime\prime}$~NW of \hdone{} and visible in a number of optical and infrared survey images such as the Sloan Digital Sky Survey Data Release 9 \citep{ahn2012sdss9}. The Gaia DR3 estimate of $T_\textrm{eff} = 5278$\,K indicates it is likely an early K star. A second, closer companion at $\sim 12^{\prime\prime}$ NNE of \hdone{} is only seen in the Two Micron All Sky Survey (2MASS) J band image \citep{skrutskie062mass}. Neither of these companions will contribute to our \hdone{} HARPS spectra. They are also too faint (i.e. Gaia DR3 3663052713405023616 is 5.9 mag fainter than \hdone{}) to contribute to elevated absolute magnitude measurements that would indicate evolved status of \hdone{}.

We find 
%\vsini{}~$= 3.04 \pm 0.74$~\kms{} 
\vsini{}~$= 3.0 \pm 0.7$~\kms{} 
from \textsc{species}, yielding (Fig.~\ref{fig:all_monte}) most probable $\hat{P}_\textrm{rot} = 19.8$\,d, with median, upper and lower 68.3\% confidences of 
\hbox{$\tilde{P}_\textrm{rot} = 20.4~(12.9; 28.3)$\,d}.
The rotational velocity estimate from \citetalias{murphy16}, gives \vsini{} $= 2.6^{+2.3}_{-2.6}$~\kms{}, with 
$\hat{P}_\textrm{rot} = 12.8$\,d and
\hbox{$\tilde{P}_\textrm{rot} = 16.8 (13.4; 46.1)$\,d}.

\subsubsection{\hdone{} recursive RV solution}
\protect\label{section:HD118006RV}

Recursive likelihood periodograms of the RVs are presented in Fig.~\ref{fig:hd118006_periodograms}\,a. The RVs show very strong evidence for a single Keplerian with $P_1=4.93$\,d. A second Keplerian with $P_2 = 11.8$\,d and significance of $\Delta\log{L} = 15.0$ is found, with a wider, less significant double-peaked cluster at longer periods of $18.7$\,d and $20.3$\,d. The phase folds nevertheless reveal significant outliers in both the HARPS and ESPRESSO RVs; with respective additive white noise of $2.1$\,\ms{} and $1.8$\,\ms{}, the residual r.m.s. values are $2.4$\,\ms{} and $1.8$\,\ms{}. These values are particularly significant for the ESPRESSO RVs. While the $P=4.93$\,d signal is persistent on the timescale of the observations, with $K = 22.5$\,\ms{}, it is not clear that the second signal, with $K = 4.8$\,\ms{}, is dynamically induced by an orbiting body.

\subsubsection{\hdone{} activity indicators and TESS photometry}
\protect\label{section:HD118006activity}

The activity periodograms (Fig.~\ref{fig:hd118006_periodograms}\,b) do not show evidence for significant periodicities. Further, there is no convincing evidence for a {linear} correlation between the activity indicators and the RVs (Fig.~\ref{fig:hd118006_correlations}). \hdone{} has been observed in TESS Sectors 23, 46, 50 and 91. {The lightcurves obtained with \texttt{unpopular} are shown in Fig.~\ref{fig:hd118006_lightcurve} revealing the large mid-sector gaps in the data, and clipping of systematics. $R_\textrm{var} = 546$\,ppm is somewhat higher than was found for \hdsix{}. There is a significant $P = 21.5$\,d power, with a sinusoidal semi-amplitude of $A = 111$\,ppm, while a lower significance $9.4$\,d peak is a suspected harmonic of this period. Although the ratio of $R_\textrm{var} / A$ is lower than we see for \hdsix{}, we we were unable to recover reliable periodicities using the} \textsc{s+leaf} {GP model.}

\subsubsection{{\hdone{} - Purely Keplerian solution}}
\protect\label{section:dmpp7kep}

{Assuming a prior in \texttt{kima} of $N_\textrm{p}: \mathcal{U}[0; 4]$, returned strong evidence for a single Keplerian. Table \ref{tab:hd118006_GPsolution} presents the $N_\textrm{p} = 1$ solution (Model~$1a$) with $P=4.93$\,d, confirming the high amplitude signal identified through the likelihood periodogram search. The RVs, and model solution are plotted in Fig. \ref{fig:hd118006_rvsolutions}\,a. The residuals reveal additional, potentially correlated scatter, which is partially accommodated by the white noise terms ($\sigma_\textrm{HARPS} = 3.4$\,\ms{} and $\sigma_\textrm{ESPRESSO} = 3.3$\,\ms{}.}

{Moderate evidence is found for a second signal, with $\textrm{BF}(N_\textrm{p} = 2 / N_\textrm{p} = 1)=60.0$ (Fig. \ref{fig:hd118006_rvsolutions}\,b). The scatter and white noise terms are reduced with respect to Model~1a. However, the posterior period distribution reveals a multimodal solution, with periods at $P \sim 11.8$\,d and a second group at $\sim 21$\,d, in general agreement with the periodogram. The second group is split into two clusters centred on $\sim 19$\,d and $\sim 23$\,d. We tabulate the solution as Model~1b in Table \ref{tab:hd118006_GPsolution} with $P = 21.47^{+2.10}_{-3.19}$\,d. The eccentricity of \hdone{}\,c$^{\prime\prime}$ of \hbox{$e={0.46}^{+0.15}_{-0.25}$} is high and the phase fold (Fig. \ref{fig:hd118006_rvphases}\,a, right panel) reveals this to be driven by only a few data points. Despite higher evidence, this period is not conclusively favoured ($\textrm{BF} = 1.2$) over the $P = {11.79}^{+0.22}_{-0.20}$\,d alternative second period, which has a lower eccentricity ($e = {0.30}^{+0.20}_{-0.16}$).}

\subsubsection{{\hdone{} - Keplerian solution with a GP}}
\protect\label{section:dmpp7kepgp}
 
{With a GP, we again adopted a prior of $N_\textrm{p}: \mathcal{U}[0; 4]$ and the same GP priors as the \hdsix{} analysis. We find strong evidence for a single Keplerian with with $\textrm{BF}(N_\textrm{p} = 1 / N_\textrm{p} = 0) > 4564$ (Table \ref{tab:hd118006_GPsolution}, Model~2). The Model~1 and Model~2 $P=4.93$\,d Keplerian periods are essentially identical, while the preferred $\eta_3 = 22.15^{+1.61}_{-3.02}$\,d in Model~$2$ is again inconclusively preferred over the posterior cluster of $\eta_3 = {11.81}^{+0.50}_{-0.40}$\,d samples with $\textrm{BF} = 1.8$. These periodicities are consistent with the $22.5$\,d Keplerian period from Model~1b and are also consistent with the Monte Carlo period distribution and photometric $21.5$\,d power. The correlated noise in the residuals is visibly reduced in Fig. \ref{fig:hd118006_rvsolutions}\,c, while $\sigma_\textrm{ESPRESSO}$ is lower but not distinct from the distribution in Model~1b (see Table \ref{tab:hd118006_GPsolution}). The \hdone{}\,b (Model~2) phase fold in Fig. \ref{fig:hd118006_rvphases}\,b shows notably less scatter than the Model~1b phase fold for \hdone{}\,b$^{\prime\prime}$ in Fig. \ref{fig:hd118006_rvphases}\,a. There is only weak evidence for an additional Keplerian in this model, where we find BF($N_\textrm{p} = 2/N_\textrm{p} = 1$) = 4.0.}

{Despite a preference for Model~$1$ over Model~$2$, the relative global evidence of $\Delta\log\mathcal{Z} = 0.86$ (BF~$\sim2.4$) is inconclusive. Since the two models yield essentially the same global evidence, we also performed additional comparisons with fixed $N_\textrm{p}$ to compare models at equal dimensionality. Variants of Models~1a and 1b, resulting from fixed $N_\textrm{p}$ priors of $N_\textrm{p} : \mathcal{F}[1]$ or $\mathcal{F}[2]$), give respective $\log \mathcal{Z} = -157.74$ and $-153.68$.
%and $-155.21$.
Within this purely Keplerian framework, $N_\textrm{p} = 2$ is again moderately preferred ($\Delta\log\mathcal{Z} = 4.1$; BF~$\sim60.0$), in agreement with the default Model~1 run with prior $N_\textrm{p}: \mathcal{U}[0; 4]$. A variant of Model~2 with a GP using fixed $N_\textrm{p}$ priors of $N_\textrm{p} : \mathcal{F}[1]$ yields global evidence of $\log\mathcal{Z} = -155.21$. Comparing the variants of Model~2 and Model~1b  confirms only a weak preference for Model~1b over Model~2 (i.e. $\Delta\log\mathcal{Z} = 1.53$; BF~$\sim4.6$), confirming a degeneracy between modelling the additional variability as either a second Keplerian or correlated noise within this class of variant models.}

% ORIGIN OF THE NUMBERS FOR THE ABOVE ARGUMENT:
%cluster_hdbscan.py --nkep=1 --csize=1000 --nparms=3  --sigma=5.0 --t0=60382.351505 ---> 
% ---> $\eta_1$ [ms$^{-1}$]          & \multicolumn{1}{c}{${5.144}^{+6.428}_{-2.174} [2.840]$} \\
%cluster_hdbscan.py --nkep=2 --csize=2000 --nparms=3  --sigma=5.0 --t0=60382.351505 --scluster="0 1"
% ---> $\eta_1$ [ms$^{-1}$]          & \multicolumn{2}{c}{${2.163}^{+5.361}_{-1.843} [1.494]$} \\

\subsubsection{{A solar-like spot dynamo and starspot pattern?}}
\protect\label{section:dmpp7spots}

{A compact system architecture with interior hot Jupiter and exterior Neptune is not likely from both a theoretical and observational evidence perspective \citep{steffen12,dawson18,zink23}. Although the AMD checks suggest that Model~1b with two Keplerian orbits represents a stable configuration, there is insufficient evidence to accept this scenario over Model~2. The high posterior eccentricity of \hdone{}\,c$^{\prime\prime}$ may be evidence the Model~1b solution is driven by the data sampling.  We suspect that the $11.8$\,d period cluster found in the posteriors of both the purely Keplerian and GP models and the photometric $9.4$\,d periodogram peak  are stellar $P_\textrm{rot}/2$ aliases. The split of the characteristic $P_\textrm{rot} \sim 21$\,d in the Keplerian posteriors in Model~1 and of $\eta_3$ in Model~2 into two clusters with median posterior periods of $\eta_3\sim 18.9$~and~$\sim 22.5$\,d is indicative of differential rotation. This scenario may also indicate the presence of two active regions at different latitudes, possibly separated by $\sim 180$\degs{} \citep{berdyugina03activelongs}. Assuming solar-like latitude dependent rotation and shear magnitude and scaling relationships from \citet{barnes17mdwarfs}, an equatorial rotation period of $\sim 22.5$\,d would lead us to expect a $\sim 18.9$\,d rotation at a latitude of $32$\degs. This picture is entirely consistent with a solar-like dynamo mechanism in \hdone{} with cool spots restricted to low latitudes. The activity pattern is clearly evolving between the HARPS observations and the ESPRESSO observations; a more clearly defined periodicity at $P_\textrm{rot}/2$ is evident in the HARPS data set (see Fig. \ref{fig:hd118006_rvsolutions}\,c, dashed GP$_\textrm{med}$ curve). In addition to possible differential rotation, the implied activity evolution is a likely contributor to the multimodal periodicity in the secondary periodic signal in the posteriors.}

\subsubsection{\hdone{} Summary}
\protect\label{section:dmpp7summary}

{For \hdone{}, activity-related periodicities appear to be linked to stellar rotation. Further, the high eccentricity of the purely Keplerian} $N_{\textrm{p}}=2$ {solution leads us to a preferred interpretation of an active host star that is slightly evolved and which harbours a single close-orbiting giant planet (i.e. we prefer Model~2 with a GP and single Keplerian). Nevertheless, modelling the lower amplitude second signal in the RVs as either a Keplerian or as activity via a GP has little impact on the $4.297$\,d planet candidate {parameters, which are consistent between Models~1a, 1b and 2. With the addition of ESPRESSO data, the planet parameters are more precisely determined than the HARPS-only solution presented in \citetalias{standing26dmpp}. Our Model~2 minimum mass of \mpsini{} $= 69.0 \pm 1.7$\,M$_\oplus$ ($\Delta M/M = 2.4$\%) is 11\% higher than the \citetalias{standing26dmpp} solution, implying \hdone{} is a $0.72$ Saturn mass planet.}}

%%%%%%%%%%%%%%%%%%%%%%%%%%%%%%%%%%%%%%%%%%%%%%%%%%%%%%%%%%%%%%%%%%%%%%%%%%%%%%%%%%%%%%%%%%%%%%%%%%%%%%%%%%%%
%%%%%%%%%%%%%%%%%%%%%%%% HD2134 %%%%%%%%%%%%%%%%%%%%%%%%%%%%%%%%%%%%%%%%%%%%%%%%%%%%%%%%%%%%%%%%%%%%%%%%%%%%
\subsection{\hdtwo}
\protect\label{section:hd2134}
%%%%%%%%%%%%%%%%%%%%%%%%%%%%%%%%%%%%%%%%%%%%%%%%%%%%%%%%%%%%%%%%%%%%%%%%%%%%%%%%%%%%%%%%%%%%%%%%%%%%%%%%%%%%
%%%%%%%%%%%%%%%%%%%%%%%%%%%%%%%%%%%%%%%%%%%%%%%%%%%%%%%%%%%%%%%%%%%%%%%%%%%%%%%%%%%%%%%%%%%%%%%%%%%%%%%%%%%%

%\subsubsection{\hdtwo{} stellar parameters and evolution status}
\protect\label{section:hd2134params}

%%%%%%%%%%%%%%%%%%%%%%%% HD2134 GP analysis of TESS lightcurves 

%\begin{figure*}
%%%%\includegraphics[trim=4mm 0mm 6mm 0mm, width=2.0\columnwidth]{DMPPmulti/HD2134_unpopular-lightcurve-kimasolution.png}
%\includegraphics[trim=4mm 0mm 6mm 0mm, width=2.0\columnwidth]{DMPPmulti/HD2134_unpopular-lightcurve-kimasolution_dec25.pdf}
%\caption{\hdtwo{} TESS photometry and GP fit. Shown are the unbinned data (light  grey), the $0.5$\,d binned points (black), the maximum posterior GP curve (black) and the 68.3\% and 95.4\% intervals from 1000 posterior samples (dark and light pink).}
%    \label{fig:hd2134_lightcurve}
%\end{figure*}

%%%%%%%%%%%%%%%%%%%%%%%% HD2134 activity periodograms and correlations
\begin{figure}
    \includegraphics[trim=0mm 6mm 0mm 0mm, width=1.0\columnwidth]{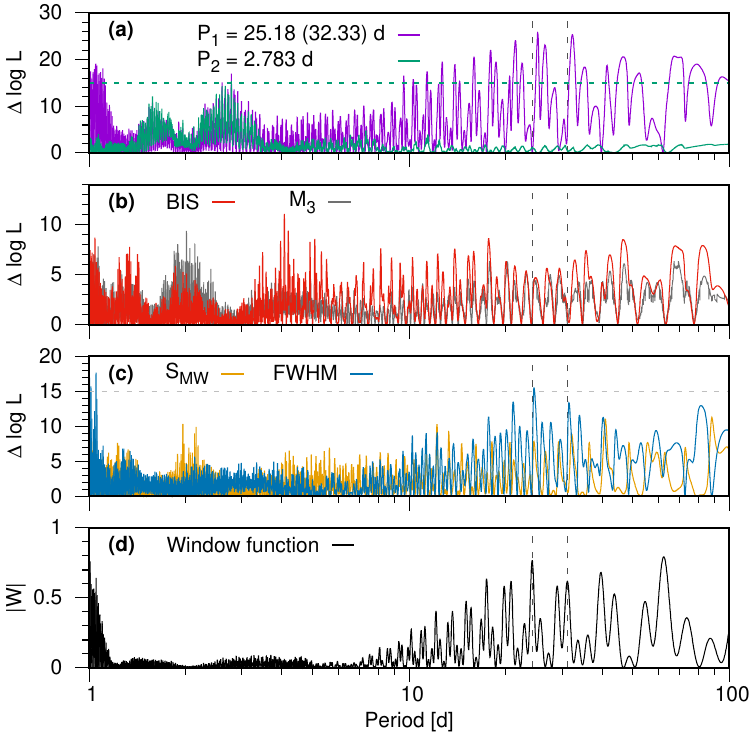}
    \caption{\hdtwo{} RV and activity periodograms showing (a) recursive log-likelihood periodograms, {(b,c)} activity periodograms and (d) the corresponding Window function with periods at 24.2\,d and 31.2\,d plotted as vertical dashed lines in all panels.}
    \label{fig:hd2134_periodograms}
\end{figure}

\begin{figure}
    \includegraphics[trim=0mm 10mm 0mm 0mm,  width=1.0\columnwidth]{{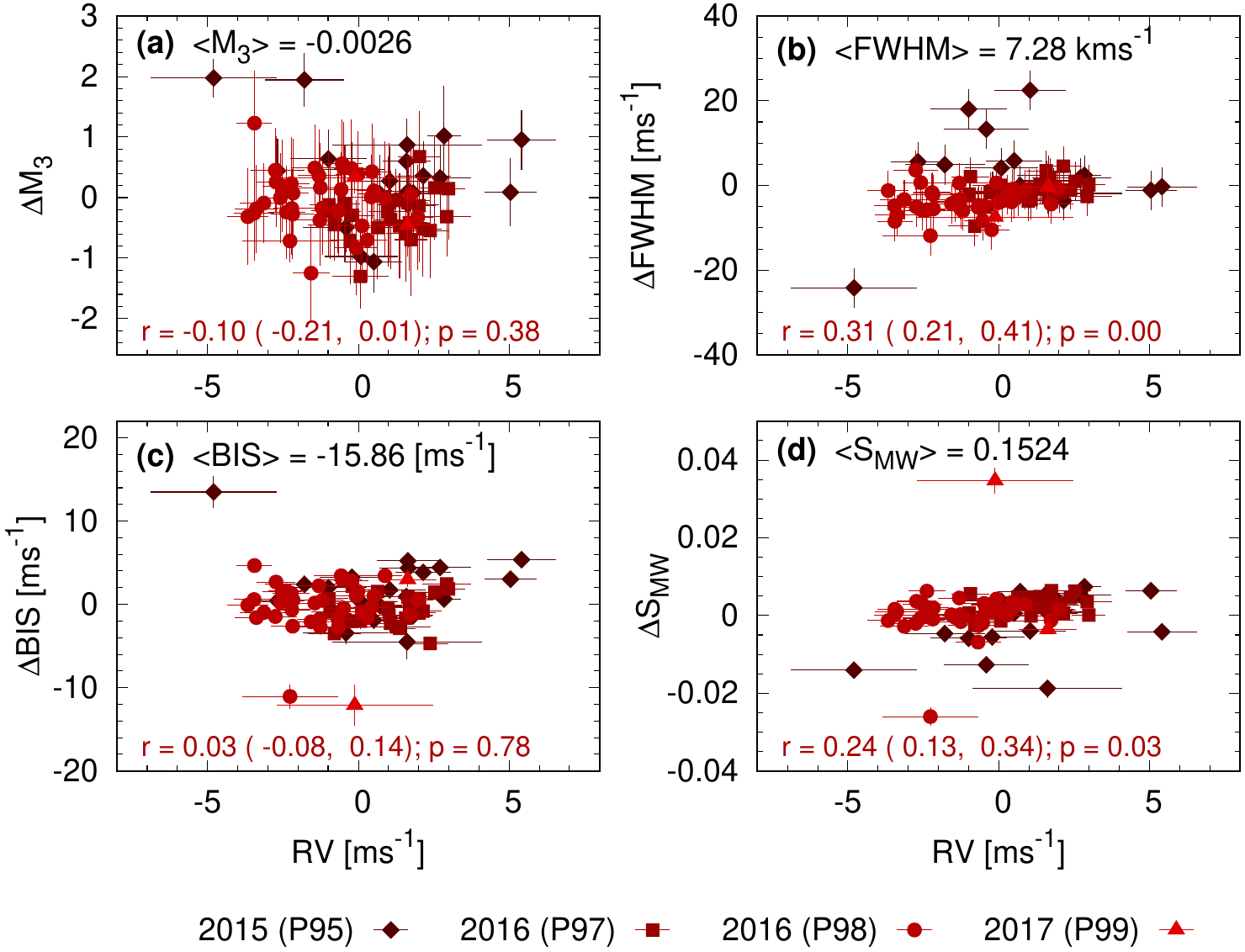}}
    \caption{\hdtwo{} activity vs RV indicators correlations. The data are colour coded according to observing runs in ESO Periods P95, P97, P98 and P99.}
    \label{fig:hd2134_correlations}
\end{figure}

%%%%%%%%%%%%%%%%%%%%%%%% HD2134 Keplerian solutions Np=1 and Np=2

\begin{figure*}
    \includegraphics[width=2\columnwidth]{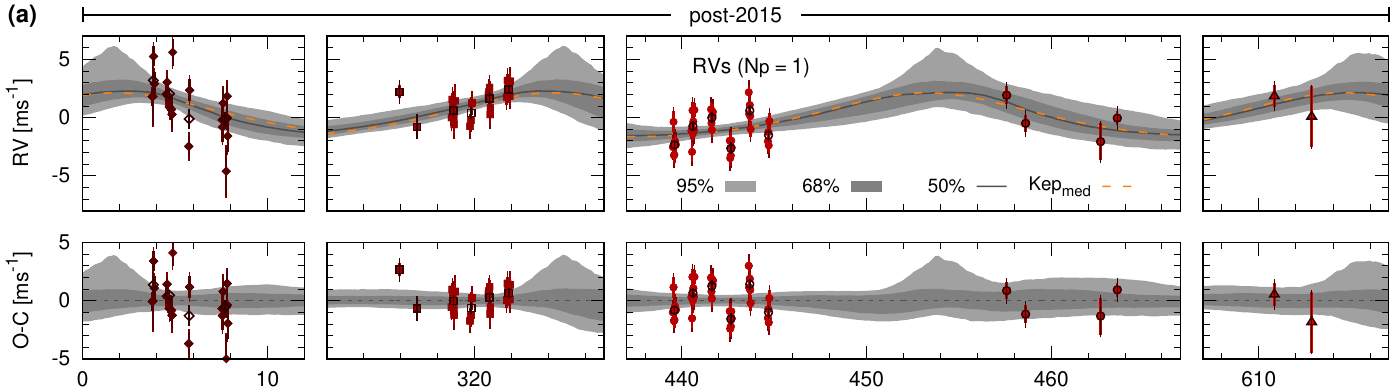}
    \includegraphics[width=2\columnwidth]{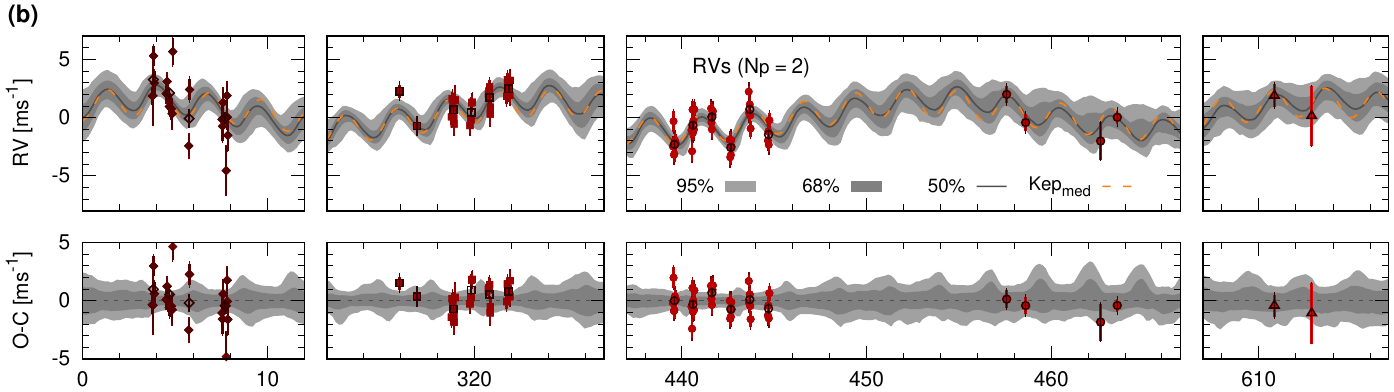}
    %\vspace{2mm} \\
    %\includegraphics[width=2\columnwidth]{DMPPmulti/HD2134_solution_GP_np-0kep.pdf}
    \includegraphics[width=2\columnwidth]{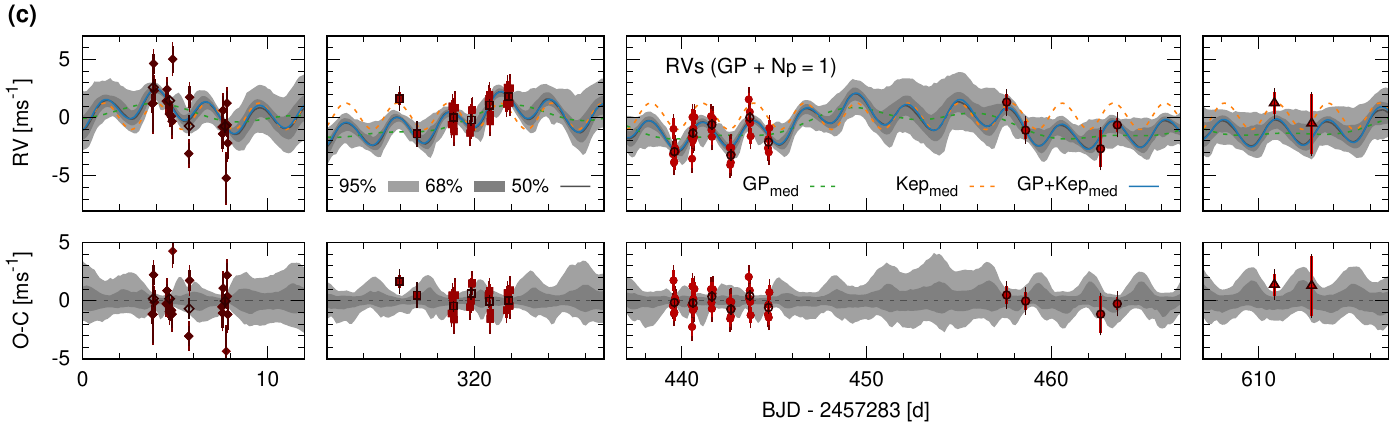}
    \caption{{Solution curves for the \hdtwo{} RVs. The purely Keplerian solutions are shown in (a) for $N_\textrm{p} = 1$ (Model~1a) and in (b) for the $N_\textrm{p} = 2$ (Model~1b). The moderately significant GP + Keplerian solution is shown in (c) for $N_\textrm{p} = 1$ (Model~2). See Table \ref{tab:hd2134_GPsolution} for \hdtwo{} solution details.}}
    \label{fig:hd2134_rvs}
\end{figure*}

\begin{figure*}
    \includegraphics[width=2\columnwidth]{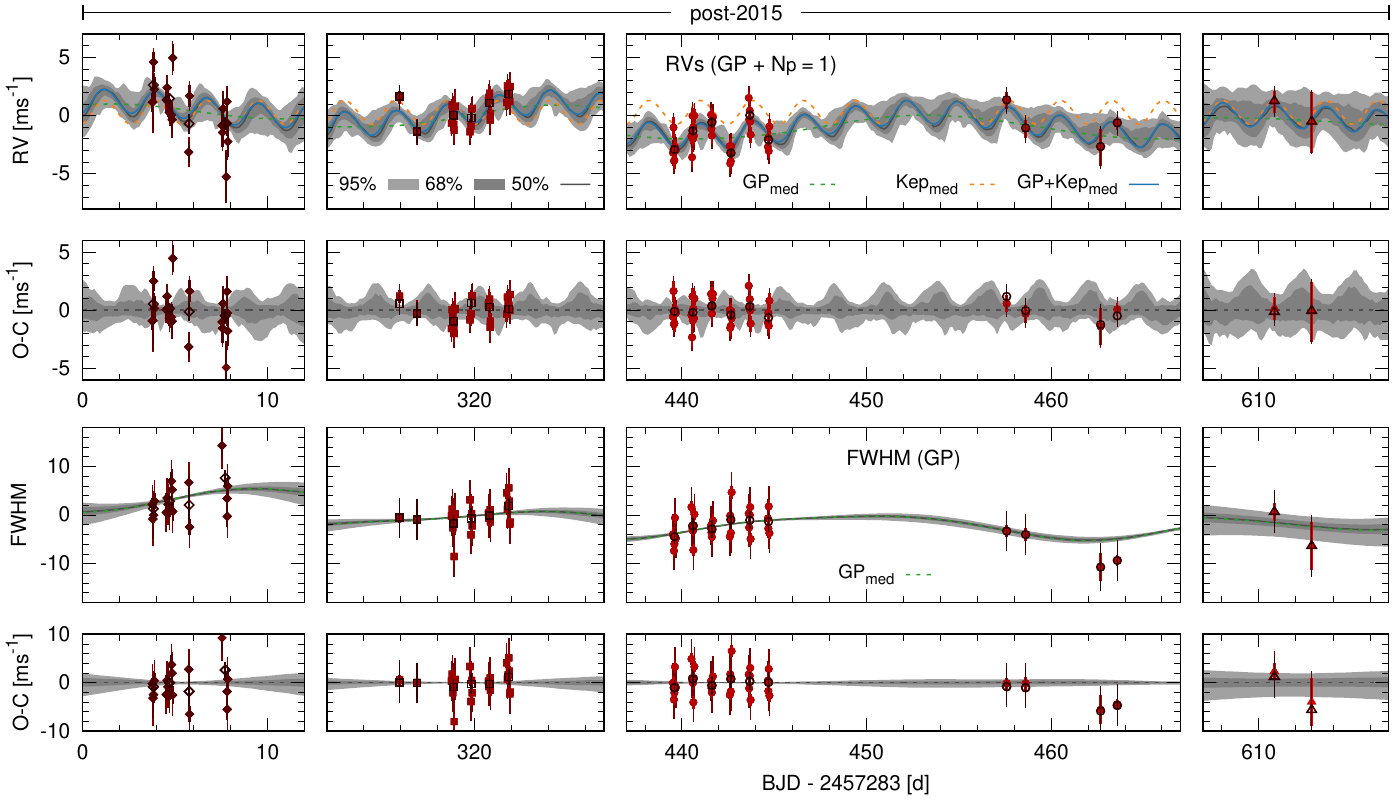}
    \caption{{The \hdtwo{} moderately significant GP + Keplerian solution for $N_\textrm{p} = 1$ (Model~3) using simultaneous RV (upper panels) and FWHM data (lower panels). See Table \ref{tab:hd2134_GPsolution} for \hdtwo{} solution details.}}
    \label{fig:hd2134_rvsfwhm}
\end{figure*}

\begin{figure}
    \centering
	\includegraphics[trim=0mm 2mm 0mm 0mm, width=0.9\columnwidth]{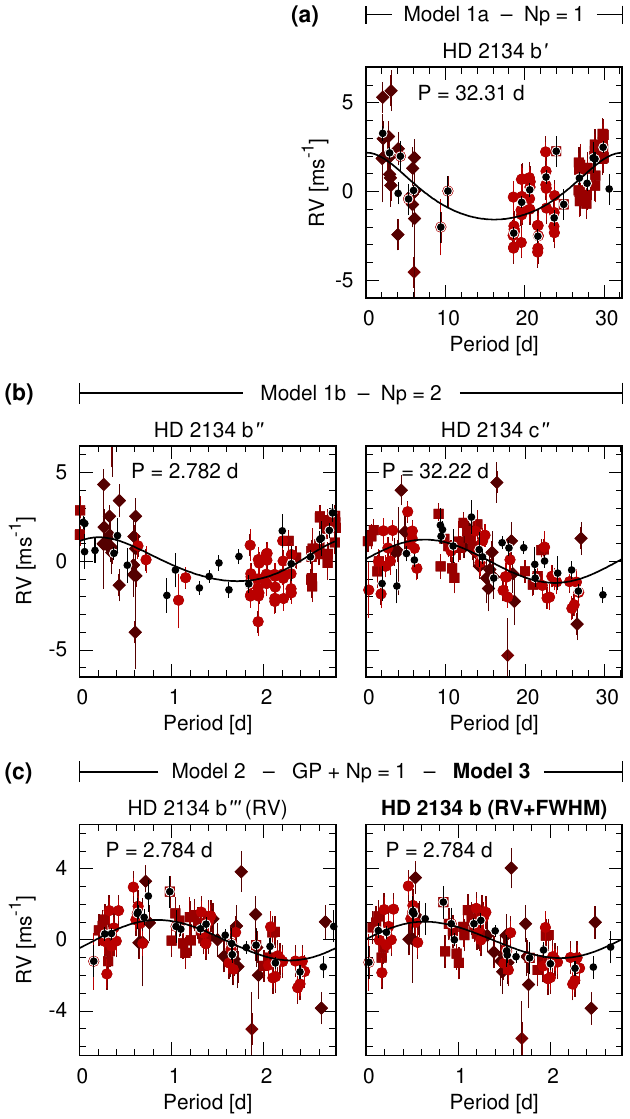}
    \caption{{\hdtwo{} phase-folded RVs for the purely Keplerian models with (a) $N_\textrm{p} = 1$ (Model 1a) and (b) $N_\textrm{p} = 2$ (Model 1b). Shown in (c) are the Keplerian phase-folds with $N_\textrm{p} = 1$ (Model 2 and preferred Model~3, highlighted in bold) after subtracting the GP components.}}
    \label{fig:hd2134_rvphases}
\end{figure}

%------------------------------ HD 2134 RESULTS TABLE ----------------------------------------------
%%%%%%%% KEPLERIAN MODELS
% Re-extracted on 21/05/2026:
%cluster_hdbscan.py --nkep=1 --csize=250 --nparms=3  --sigma=5.0 --t0=57591.311214 --binsz=0.1 --gamma=0. --mstar="1.05 0.02"
%cluster_hdbscan.py --nkep=2 --csize=2500 --nparms=3  --sigma=5.0 --t0=57591.311214 --binsz=0.1 --gamma=0. --mstar="1.05 0.02"

\begin{table*}
    
\renewcommand{\arraystretch}{1.25}
\centering
\setlength\tabcolsep{0.02\columnwidth}

\begin{tabular}{lccc}
\hline
%Keplerian model               & ${N_\textrm{p}=1}$ & \multicolumn{2}{c}{{${N_\textrm{p}=2}$}} \\
& {Model 1a} (${N_\textrm{p}=1}$) & \multicolumn{2}{c}{{Model 1b} (${N_\textrm{p}=2}$)}\\
{Parameter} & {\hdtwo{}\,b$^\prime$} & {\hdtwo{}\,b$^{\prime\prime}$} & {\hdtwo{}\,c$^{\prime\prime}$} \\
\hline
%Parameter                               &  HD\,21334\,b                                        & HD\,2134\,b                                          & HD\,2134\,c                                          \\
$P$ [d]                       & ${32.31289}^{+0.13035}_{-0.13736} [32.30419]$          & ${2.78247}^{+0.00275}_{-0.18088} [2.78488]$            & ${32.22380}^{+0.17118}_{-0.20701} [31.59403]$          \\
$K$ [ms$^{-1}$]               & ${2.28}^{+0.58}_{-0.40} [3.50]$                        & ${1.22}^{+0.28}_{-0.26} [1.79]$                        & ${1.77}^{+0.35}_{-0.29} [1.44]$                        \\
$M_0$ [rad]                   & ${2.890}^{+0.546}_{-0.700} [2.926]$                    & ${3.343}^{+1.556}_{-1.818} [3.506]$                    & ${3.271}^{+0.675}_{-0.875} [4.032]$                    \\
e                             & ${0.254}^{+0.176}_{-0.170} [0.526]$                    & ${0.143}^{+0.198}_{-0.109} [0.082]$                    & ${0.225}^{+0.146}_{-0.158} [0.734]$                    \\
$\omega$ [rad]                & ${0.887}^{+4.897}_{-0.640} [0.326]$                    & ${3.891}^{+1.156}_{-1.927} [4.486]$                    & ${4.562}^{+1.386}_{-4.194} [5.185]$                    \\

\multicolumn{4}{c}{\vspace{-3mm}} \\
$a$ [AU]                                & ${0.2018}^{+0.0014}_{-0.0015} [0.2041]$      & ${0.0391}^{+0.0005}_{-0.0015} [0.0396]$                & ${0.2014}^{+0.0015}_{-0.0016} [0.1996]$                \\
$m_p$\,sin\,$i$ [M$_\oplus$]            & ${11.325}^{+2.412}_{-1.920} [15.685]$        & ${2.676}^{+0.569}_{-0.586} [4.104]$                    & ${8.842}^{+1.652}_{-1.594} [5.042]$                    \\
$t_\textrm{c}$ [JD]                     & ${57580.664}^{+2.041}_{-2.058} [57578.315]$  & ${57589.521}^{+1.646}_{-1.036} [57588.483]$            & ${57581.232}^{+2.076}_{-2.169} [57577.997]$            \\

\multicolumn{4}{c}{\vspace{-3mm}} \\

$\gamma_\textrm{HARPS} - 4263$ [ms$^{-1}$] & ${-0.16}^{+0.29}_{-0.29} [0.01]$                  & \multicolumn{2}{c}{${-0.20}^{+0.27}_{-0.27} [-0.39]$} \\
$\sigma_\textrm{HARPS}$ [ms$^{-1}$]     & ${1.16}^{+0.16}_{-0.14} [1.14]$                      & \multicolumn{2}{c}{${0.85}^{+0.16}_{-0.15} [0.70]$} \\

\multicolumn{4}{c}{\vspace{-3mm}} \\
$\ln{\mathcal{L}}$ [MAP]    & $-144.74$     & \multicolumn{2}{c}{$-126.96$} \\

BF($N_\textrm{p}/(N_\textrm{p} - 1$)) & $>2822$        & \multicolumn{2}{c}{$31.10$} \\ %(7.93)
%\multicolumn{4}{c}{} \\
Global evidence, $\ln \mathcal{Z}$  & \multicolumn{3}{c}{$-166.24$} \\
\multicolumn{4}{c}{\vspace{-3mm}} \\
\hline
\end{tabular}
\setlength\tabcolsep{0.109\columnwidth}
\begin{tabular}{lcc}
                             & {Model~2} ({GP + }${N_\textrm{p}=1}$ )                                        & {Model 3} ({GP +} ${N_\textrm{p}=1}$ )                         \\
{Parameters}             & {\hdtwo{}\,b$^{\prime\prime\prime}$ (RV)}  & {\hdtwo{}\,b (RV+FWHM)} \\
%                             & ${N_\textrm{p}=1}$ + GP         & ${N_\textrm{p}=1}$ + GP & \\
\hline

% RVs
% cluster_hdbscan.py --nkep=1 --csize=1000 --nparms=101  --sigma=5.0 --eta3="25.0 0.4" --t0=57591.31121383 --mstar="1.05 0.02" --gamma=0.
% RVs + FWHM
% cluster_hdbscan.py --nkep=1 --csize=2000 --nparms=101  --sigma=5.0 --eta3="25 0.4" --t0=57591.31121383 --mstar="1.05 0.02" --gamma=0.

$P$ [d]                      & ${2.78379}^{+0.00221}_{-0.11880} [2.34923]$          & ${2.78368}^{+0.00223}_{-0.12077} [2.78366]$            \\
$K$ [ms$^{-1}$]              & ${1.33}^{+0.31}_{-0.27} [2.21]$                      & ${1.31}^{+0.31}_{-0.27} [1.74]$                        \\
$M_0$ [rad]                  & ${3.567}^{+1.498}_{-1.683} [4.964]$                  & ${3.489}^{+1.469}_{-1.619} [3.266]$                    \\
e                            & ${0.153}^{+0.211}_{-0.118} [0.840]$                  & ${0.156}^{+0.216}_{-0.119} [0.320]$                    \\
$\omega$ [rad]               & ${3.894}^{+1.193}_{-2.011} [4.248]$                  & ${3.919}^{+1.225}_{-1.973} [4.654]$                    \\

$\eta_1$ [ms$^{-1}$]         & {${1.74}^{+1.82}_{-0.70} [1.38]$}                    & {${2.03}^{+2.12}_{-0.89} [1.10]$} \\
$\eta_1^\textrm{act}$(FWHM) [ms$^{-1}$]   &                                         & {${4.57}^{+4.10}_{-1.86} [3.95]$} \\
$\eta_2$ [d]                 & {${82.740}^{+262.302}_{-58.480} [574.590]$}          & {${156.316}^{+262.976}_{-120.185} [133.863]$} \\
$\eta_3$ [d]                 & {${24.272}^{+7.379}_{-6.089} [21.174]$}              & {${25.610}^{+6.236}_{-4.940} [25.704]$} \\
$\eta_4$                     & {${2.731}^{+3.923}_{-1.665} [0.567]$}                & {${2.613}^{+3.342}_{-1.388} [0.725]$} \\

$\gamma - 4263$ [ms$^{-1}$]  & ${0.32}^{+1.22}_{-1.07} [0.02]$                      & ${0.37}^{+1.59}_{-1.27} [0.24]$ \\
$\gamma_\textrm{FWHM}$ [ms$^{-1}$]  &                                               & ${-1.12}^{+3.50}_{-3.00} [-1.60]$ \\
$\sigma$ [ms$^{-1}$]         & ${0.91}^{+0.17}_{-0.15} [0.72]$                      & ${0.92}^{+0.17}_{-0.14} [0.71]$   \\
$\sigma$(FWHM) [ms$^{-1}$]   &                                                      & ${4.06}^{+0.45}_{-0.42} [3.76]$   \\

\multicolumn{3}{c}{\vspace{-3mm}} \\

$a$ [AU]                     & ${0.0393}^{+0.0004}_{-0.0010} [0.0356]$              & ${0.0393}^{+0.0004}_{-0.0013} [0.0395]$            \\
$m_p$\,sin\,$i$ [M$_\oplus$] & ${2.919}^{+0.648}_{-0.604} [2.633]$                  & ${2.862}^{+0.668}_{-0.650} [3.760]$                \\
$T_\textrm{eq} (A_\textrm{B} = 0, 0.36)$ [K]  &                                     & $1487, 1330$                                          \\
$R_\textrm{p}$ (predicted from \mpsini{}) [R$_\oplus$]       &                      & $1.35 \pm 0.11$                                       \\
$R_\textrm{p}$ (predicted, $i=57.3$\degs) [R$_\oplus$]       &                      & $1.42 \pm 0.12$                                       \\
$t_\textrm{c}$ [JD]          & ${57589.079}^{+2.075}_{-0.645} [57589.094]$          & ${57589.426}^{+1.713}_{-0.987} [57588.520]$    \\

\multicolumn{3}{c}{\vspace{-3mm}} \\

$\ln{\mathcal{L}}$ [MAP]    & $-138.01$                                             & $-390.309$ \\
%BF($N_\textrm{p}=1/N_\textrm{p}=0$)   & $(28.55)12.09$                              & $18.39(8.91)$ \\
BF($N_\textrm{p}=1/N_\textrm{p}=0$)   & $28.55$                                     & $18.39$ \\
Global evidence, $\ln \mathcal{Z}$    & $-165.19$                                   & $-423.88$ \\
\end{tabular}
\caption{{\hdtwo{} parameters. Top: the purely Keplerian models with $N_\textrm{p}=1$ (Model~1a) and $N_\textrm{p}=2$ (Model~1b). Bottom: GP + $N_\textrm{p}=1$ parameters using only the RV timeseries  (Model~2) and the corresponding simultaneous RV and FWHM timeseries (Model~3).}}
\label{tab:hd2134_GPsolution}
%\end{adjustbox}
\end{table*}

In ESO periods P95, P97, P98 and P99 (2015--2017) $83$ observations of \hdtwo{} were obtained with $1-7$ observations per night. The last two observations of the four obtained in P110 (2023) at high airmasses of $1.97 - 2.01$ had to be cut short due {to a tip-tilt guiding system fault}, leaving only two observations. Because of the need for separate RV offsets between the P95--P99 and P110 data sets, we did not include the P110 data.
%In the absence of strong evidence for a Keplerian in the following analysis, we continue to refer to this system as \hdtwo{}.

\subsubsection{\hdtwo{} Stellar parameters}
\protect\label{section:stellarparamshd2134}

The effective temperature, $T_\textrm{eff} = 5675 \pm 50$\,K, and other stellar parameters for \hdtwo{} (Table \ref{tab:stellar_params}) are consistent with those of a main-sequence early-mid G dwarf. It is {just over $71$\,pc away} and slightly older than the Sun, with an estimated age of $6.64^{+0.49}_{-0.86}$\,Gyr. \hdtwo{} was selected for DMPP {through \logrhk{}~$= -5.23 \pm 0.03$ from the S-index} reported by \citet{jenkins11activities}. The low \logrhk{} is likely to be robust against systematic effects of order a few percent, meaning that even sky background subtraction would probably not yield activity above the basal flux level. Table \ref{tab:stellar_params} shows higher \logrhk{} during the HARPS observations; we find 10/87 observations (11\%) {have} \logrhk{}~$< -5.1$.
%\vsini{} = $4.0 \pm 0.3$\,\kms{} is consistent with the \textsc{species}-derived estimate of \citet{perdelwitz24}. 
The \textsc{species} estimate of \vsini{}~$= 2.6 \pm 0.6$\,\kms{} (after applying the $v_\textrm{mac}$ as in \S \ref{section:stellarparamshd2134}) is given in Table~\ref{tab:stellar_params}. 
For \hdtwo{}, as with \hdsix{}, the method of \citetalias{murphy16} is not able to obtain a finite positive estimate for \vsini{}, but yields \vsini{} = 0 \kms{} and an upper (68.3\%) limit of 3 \kms{}. Monte Carlo simulations (see Fig. \ref{fig:all_monte}) using $R = 1.16 \pm 0.01$\,R$_\odot$, randomly oriented stellar axial inclinations and the \textsc{species} \vsini{} estimate suggest a {most probable} stellar rotation period of $\hat{P}_\textrm{rot}$~$ = 18.4$~d, with a median and $68.3$\% ranges of 
\hbox{$\tilde{P}_\textrm{rot} = 18.5~(11.8; 24.8)$\,d}.
Assuming \vsini~$= 0$\,\kms{} and a $68.3\%$ upper limit of 3\,\kms{}, we find  $\hat{P}_\textrm{rot}$~$= 10.8$~d {and a period} distribution that is skewed to longer rotation rates due to small \vsini{} values, with median
$\tilde{P}_\textrm{rot} = 22.4 (9.9; 76.7)$\,d. An age estimate of $6.2 \pm 2.9$ Gyr from \citet{mamajek08age} is consistent with our \textsc{species} estimate and leads us to expect $P_\textrm{rot} = 32.9 \pm 3.4$~d, based on our \logrhk{} measurements.

\subsubsection{\hdtwo{} TESS photometry}
\protect\label{section:photometryhd2134}

{TESS observed \hdtwo{} in} Years 3, 5 and 7 in three consecutive pairs of TESS sectors: Sectors 27 and 28, Sectors 67 and 68 and Sectors 94 and 95. {The lightcurves (Fig. \ref{fig:hd2134_lightcurve}) reveal peak period power at $P = 26.4$\,d, sinusoidal semi-amplitude $A=105$\,ppm and a similar $R_{var}/A$ to \hdone{} (Table \ref{tab:TESS_GP_params}). As reported in \citealt{standing26dmpp}, analysis restricted to Sectors 27 and 28 indicates peak power at $13.4$\,d with $99$\,ppm sinusoidal amplitude while Sectors 67 and 68 yielded $27.2$\,d with $115$\,ppm. For Sectors 94 and 95, peaks at $6.2$\,d ($110$\,ppm) and $18.0$\,d ($113$\,ppm) were found along with a peak at $27.5$\,d ($114$\,ppm). Application of the} \textsc{s+leaf} {ESP kernel to investigate periodicities again reveals a degeneracy between the hyperparameters due to instrumental systematics. Recovering a consistent reliable characteristic GP period was not possible for the available \hdtwo{} TESS sectors.}

\subsubsection{\hdtwo{} Recursive RV solution and activity correlations}
\protect\label{section:activityhd2134}

Recursive periodogram analysis reveals two significant periodicities in the RVs, with $P_1 = 25.2$\,d and $2.78$\,d (Fig. \ref{fig:hd2134_periodograms}\,a). The $25.2$\,d {signal is not} significantly preferred over the second $32.3$\,d peak ($\Delta \textrm{log}\,L = 0.4$). Both peaks are potentially either caused, or influenced, by the {nearby} window function peaks at $24.2$\,d and $31.2$\,d, {though these coincide with regions of low power in the RV periodograms}.

{The correlation between activity indicators and RV are shown in Fig. \ref{fig:hd2134_correlations} for the observing post-2015 runs in P95, P97, P98 and P99. There does not appear to be any correlation between BIS or $M_3$ and RV. Similarly only weak correlation is seen between {S$_\textrm{MW}$} and RV. The FWHM in Fig. \ref{fig:hd2134_correlations}\,b is the only indicator showing a correlation, albeit weak to moderate, with $r > 0.3$, and $p < 0.05$. The three FWHM peaks with most power {at $21.1$\,d, $24.5$\,d and $31.6$\,d (Fig. \ref{fig:hd2134_periodograms}\,c)} again show potential influence of the window function (Fig. \ref{fig:hd2134_periodograms}\,d). Cross-correlation of the RV and FWHM timeseries does not indicate a significant phase lag.}

\subsubsection{{\hdtwo{} - Purely Keplerian model}}
\protect\label{section:kimahd2134kep}

% (i.e. $\textrm{BF} > 150$
%\hdtwo{} shows tentative evidence for RV power at $\sim 32.3$\,d. 
%As with the FWHM periodogram, there appears to be potential influence from the window function, though we note $31.2$\,d peak in the latter coincides with very low power in the RV periodogram where $\sim 32.3$\,d lies. The RV power may thus also be related to the periodicities revealed in the photometry and the simultaneously recovered FWHM. 

With \texttt{kima}, {we find strong evidence of $\textrm{BF}(N_\textrm{p} = 1/N_\textrm{p}=0) > 2822$ for a single Keplerian (Model~1a) and moderate evidence of $\textrm{BF}(N_\textrm{p} = 2/N_\textrm{p}=1) = 31.1$ for a model with two Keplerians (Model~1b). The RV solutions are shown in Table~\ref{tab:hd2134_GPsolution} and Fig.~\ref{fig:hd2134_rvs}\,a,b, which reveal a dominant RV signal at $P \sim 32.3$\,d and $32.2$\,d for Models~1a and 1b. This period is favoured over posterior samples that contain periods at $P = 21.50$\,d and $26.54$\,d with BF~$= 5.2$~and~$1.8$. The $32.31$\,d period is thus inconclusively favoured over the $26.54$\,d period. The second Keplerian in Model~1b has a period of $P = 2.782^{+0.003}_{-0.181}$\,d and $K=1.2$\,\ms{}, which is revealed through the high cadence sampling of the observations. For this model, three competing posterior solutions again reveal essentially the same periods, at $P = 21.43$\,d, $26.50$\,d and $32.33$\,d. The Bayes factors, $\textrm{BF}(32.33\,\textrm{d}\,/\,21.43\,\textrm{d}) = 1.0$~and~$\textrm{BF}(32.33\,\textrm{d}\,/\,26.50\,\textrm{d}) = 1.1$, mean none of these three periods are preferred.}

\subsubsection{{\hdtwo{} - Keplerian model with a GP}}
\protect\label{section:kimahd2134gp}

{Since the TESS photometry and FWHM provide evidence for periodicities matching the $21$\,d\,--\,$32$\,d periods found in the RVs and there is tentative evidence for a weak to moderate FWHM correlation ($r = 0.31$), we modelled the data with an activity component. We first modelled only the RV data by adding a GP to the Keplerian model. The same hyperparameter priors as used for \hdsix{} and \hdone{} were adopted.
% Get BFs for the eta3 clusters for Np=0 in /home/jbarnes/data/kima-venv2/HD2134/HD2134_kumara/Npmax4_no2023_GP_SPLEAF  (!!N.B. only 459 samples for Np=0!!)
%awk '{if($8==0 && $4>20.25 && $4<22.75)print $0}' posterior_sample.txt | wc -l --> 30
%awk '{if($8==0 && $4>24.25 && $4<28.75)print $0}' posterior_sample.txt | wc -l --> 55
%awk '{if($8==0 && $4>29.75 && $4<33.75)print $0}' posterior_sample.txt | wc -l --> 39
Inconclusive evidence for Model~2 over Model~1 is found, with $\Delta\ln \mathcal{Z} = 1.06$ or $\textrm{BF}=2.9$ (see Table \ref{tab:hd2134_GPsolution}). For Model~2, the evidence for a pure GP and no additional Keplerians (GP + $N_\textrm{p} = 0$) is strong, with $\textrm{BF}(\textrm{GP}/\textrm{no-GP}) > 457$. We have not tabulated this model for \hdtwo{}, but note that it has an amplitude $\eta_1 = {1.95}_{-0.60}^{+1.63}$\,\ms{} that is consistent with the Keplerian semi-amplitude, $K = {2.28}^{+0.58}_{-0.40}$\,\ms{}, in Model~1a. Clustering of samples in the $\eta_3 = {25.26}_{-5.90}^{+7.31}$\,d posterior range is again concentrated at $\eta_3 \sim 22$\,d, $25$\,d and $32$\,d with relative Bayes Factors of $\textrm{BF}(25.5\,\textrm{d}\,/\,22.0\textrm\,{d})=1.83$~and~$\textrm{BF}(25.5\,\textrm{d}\,/\,32.0\,\textrm{d})=1.41$.}

{Model~2 shows moderate evidence for a single Keplerian, with $\textrm{BF}(N_\textrm{p} = 1/N_\textrm{p}=0) =28.6$. The solution is detailed in Table \ref{tab:hd2134_GPsolution} (\hbox{GP + $N_\textrm{p} = 1$}) and illustrated in Fig. \ref{fig:hd2134_rvs}\,c. The samples comprise Keplerians with a median orbital period of $P=2.784$\,d and $\eta_3 = {24.27}^{+7.38}_{-6.09}$\,d, which are consistent with the purely Keplerian periods found in Model~1b.}

\subsubsection{{\hdtwo{} - Keplerian model with a GP using simultaneous RV and FWHM timeseries}}
\protect\label{section:kimahd2134gpfwhm}

{We modelled the RV and simultaneous FWHM timeseries with the same broad prior on $\eta_3$. Strong evidence is found for a pure GP model, with $\textrm{BF}(\textrm{GP}/\textrm{no-GP}) > 949$ (as in \S \ref{section:kimahd2134gp}, for the sake of brevity, we do not tabulate the parameters). Additional moderate evidence for a single Keplerian is indicated with $\textrm{BF}(N_\textrm{p} = 1/N_\textrm{p}=0) = 18.39$. We find again find an orbital period of $P=2.784$\,d and the same three longer periodicities, which are described by the GP characteristic period distribution of $\eta_3 = {25.61}^{+6.24}_{-4.94}$\,d. The corresponding complete $N_\textrm{p} = 1$ parameters are presented in Table \ref{tab:hd2134_GPsolution} as Model~3 and shown in Fig. \ref{fig:hd2134_rvsfwhm}.}
%The $\textrm{BF}(N_\textrm{p} = 2/N_\textrm{p}=1) = 1.04$ in favour of $N_\textrm{p} = 2$.}
%Since the evidence cannot be compared directly with Models~1, and 2, we inspected the Model~$3$ posteriors, to compare the GP + $N_\textrm{p} = 1$ model samples with the $N_\textrm{p} = 2$ (no GP) samples. For $N_\textrm{p} = 2$, we use all samples with $\eta_1 < 0.25$\,\ms{}, which is half the height of the posterior $\eta_1$ frequency peak (i.e. samples with essentially no GP contribution). We find $\textrm{BF}$($N_\textrm{p} = 1, \eta_3 = 25.61^{+6.24}_{-4.94}$\,d / $N_\textrm{p} = 2,\eta_1 \leq 0.25\,\textrm{ms}^{-1}$)~$=1.8$. Even for $\textrm{BF}$($N_\textrm{p} = 1, \eta_3 <35$\,d / $N_\textrm{p} = 2,\eta_1 \leq 0.25\,\textrm{ms}^{-1}$)~$=3.6$; hence, the evidence in favour of the GP model with a Keplerian is inconclusive or weak.}
% Last comparison above for 1Kep+GP (eta<0.25) vs 2Kep (eta3 as returned)
% awk '\{if(\$10==1 \&& \$6>25.61-4.94 \&& \$6<25.61+6.24)print \$0\}' posterior\_sample.txt | wc -l --> 5777
% awk '\{if(\$10==2 \&& \$3<0.25)print \$0\}' posterior\_sample.txt | wc -l --> 3138
%

\subsubsection{\hdtwo{} summary}
\protect\label{section:summaryhd2134}

{The $21$\,d\,--\,$32$\,d periodicities seen in the \hdtwo{} RVs are either astrophysical or arise from window function contamination. Since the same periods also appear in} \textit{only} {the simultaneous CCF FWHM timeseries with no strong indications in the other activity timeseries, and there is tentative evidence of a weak to moderate RV-FWHM correlation, we considered both purely Keplerian and models with a GP. Resolution of this issue requires further precise and carefully sampled radial velocities. All the models with an additional periodicity (i.e. Models 1b, 2 and 3) show moderate evidence of BF $= 18.4$\,--\,$31.1$ for a short period candidate signal at $2.782$\,d or $2.784$\,d. This short period, candidate is also close to, but marginally longer than periods} typically expected from supergranulation \citep{rincon18sg}. We discuss this possibility further in \S \ref{section:discussion}, but note here the persistent periodicity implies coherence on the $1.7$~yr timespan of the observations. Furthermore, a short-period planet is expected a priori under the DMPP hypothesis. The Keplerian signal appears well sampled throughout the orbital phase, as shown in Fig. \ref{fig:hd2134_rvphases}.
%Moderate evidence for a $2.78$\,d planet is revealed in the posterior distributions for a purely Keplerian model using the RVs and for a model that also includes a GP. 
{
%Adding the simultaneous FWHM timeseries (Model~3) reduces the evidence for a single Keplerian with a GP when comparing with the RV-only timeseries (Model~2).
%The evidence for the $2.78$\,d periodicity is slightly lower for the joint GP fit to both RV and simultaneous FWHM timeseries in Model~3 when compared with the RV timeseries in Model~2. 
%This may be a consequence of the weak-moderate RV--FWHM correlation; 
Joint modelling of RVs and FWHM may ultimately be improved by considering the FWHM timeseries derivative if better-sampled data becomes available \citep{barragan22pyaneti2,barnes24moments}.}
%The flexibility of the quasi-periodic GP means that the GP model that only uses the RVs does not require any Keplerians. 
For all models presented in Table \ref{tab:hd2134_GPsolution}, where the $2.78$\,d signal is present, the derived orbital radii and minimum masses of the putative planet are consistent within the uncertainties. {Model~3, using the RV and FWHM data, indicates a planet candidate with mass ${2.86}^{+0.67}_{-0.65}$\,M$_\oplus$ in a ${0.0393}^{+0.0004}_{-0.0013}$ AU orbit. We consider this further in \S \ref{section:discussion}, though emphasise that more data are needed for signal confirmation.}

\subsection{Transit searches}
\protect\label{section:transits}

Transiting planets identified in TESS data with signatures of around $100-400$\,ppm \citep{jones20dmpp1,serrano22hd93963,silverstein24lhs} are not common as they often do not pass the S/N criterion for TESS Object of Interest (TOI) identification \citep{twicken18keplervalidation,guerrero21toi}.
%The M2 dwarf, LHS 1678 (TOI-696) contains the smallest reported transit in a multiplanet system, with LHS~1678~b yielding a transit depth of $363$\,ppm \citep{silverstein24lhs}.
For instance, HD~93965~A~b, with a transit depth of $131.5$\,ppm was only identified through further analysis, following the $812.5$\,ppm transit candidate previously identified as TOI-1797.01, and subsequently designated HD~93965~A~c \citep{serrano22hd93963}. Individual searches for shallow transits, where planet candidates have been identified by other means, are thus prudent.

For each target, we searched for transits using TESS photometry.
For all target stars, the $120$\,s cadence light curves files were downloaded from the Mikulski Archive for Space Telescopes (MAST)\footnote{https://mast.stsci.edu/portal/Mashup/Clients/\ Mast/Portal.html}. We used the PDCSAP data with all non-zero quality cadences removed {and performed $3\sigma$ clipping of outlier fluxes.
%We removed cadences with flux greater than three times the median value.
We also supplemented the dataset with the $200$\,s cadence data from the FFIs from Sectors $67$ and $68$ (since PDCSAP lightcurves were not available).}
%but this did not reveal any other signals.
We then further detrended the data using a biweight filter from the \textsc{$\textrm{w\=otan}$} Python package \citep{hippke_wotan_2019}. Detrending window lengths of $1.0$\,d or $1.5$\,d were chosen such that they were at least three times the approximate transit duration of the largest period from the RV models calculated using \textsc{$\textrm{w\=otan}$}'s duration approximation function.
We removed cadences within $2.5$ hours of mid-sector and inter-sector data gaps.
For \hdsix{} specifically, we binned the data in $30$ minute bins prior to searching for transits.
We then searched for transits using the TransitLeastSquares (TLS) algorithm \citep{hippke_transit_2019} for periods between $0.5$ and $40$\,d and intensively in small windows around the periods from the RV solutions.
The data were phase-folded using the best-fit ephemerides from the RV models.

With the largest predicted radius amongst our planets (Table \ref{tab:hd118006_GPsolution} and \S \ref{section:discussion} discussion below), \hdone{} might be expected to most readily reveal a transit signature, with a depth of around $3400$\,ppm. However, no significant transit signals were found. 
{From the \citetalias{muller24} predicted radii, the \hdsix{}\,b candidate (Table \ref{tab:hd67200_solution_rv_activity}, Model~3b) should exhibit a transit depths of $\sim 80$\,--\,$90$\,ppm, while the putative $2.783$\,d \hdtwo{}\,b candidate (Table \ref{tab:hd2134_GPsolution}; Model~3) could show a $\sim 114$\,--\,$126$\,ppm transit}. Despite the large number of sectors and multiple RV planets, no convincing transit candidates were found for \hdsix{}. We found evidence for a $25.995$\,d transit candidate about \hdtwo{}, {matching the central $21$\,d\,--\,$35$\,d candidate periodicities}, but the transits for this signal coincided with noisy regions of data from sectors $27$, $94$ and $95$. The transit candidates thus probably arise from improperly corrected systematics.

\section{Discussion}
\protect\label{section:discussion}

\begin{figure}
    \centering
    \includegraphics[trim=0mm 0mm 1mm 0mm, width=1\columnwidth]{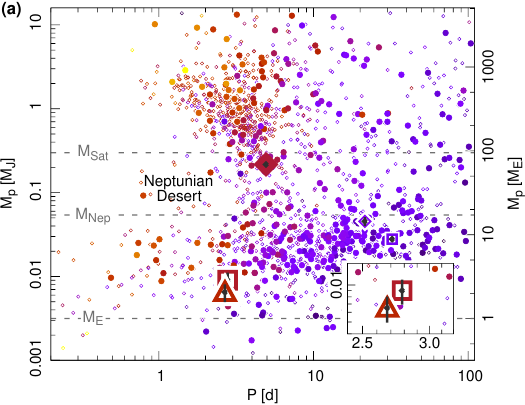}
    \vspace{-2mm} \\
    \includegraphics[trim=0mm 0mm 1mm 0mm, width=1\columnwidth]{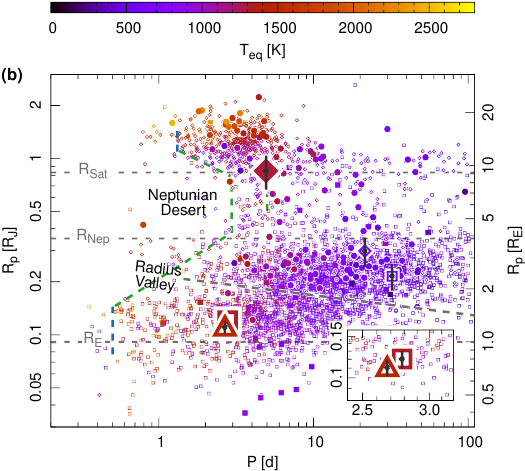} \\
    \caption{(a) Mass-Period and (b) Radius-Period diagrams for planet candidates. Masses and predicted radii for \hdone{} (large solid diamond). The moderate evidence candidates \hdsix{}\,b and \hdtwo{}\,b are also shown in the main panels and inset plots (large open triangles and squares). Planets corresponding to the other candidate Keplerian signals for each host star are shown by the corresponding smaller open symbols. Archival planets with either measured minimum mass, or true mass (where inclination is known) are plotted for bright $V < 10$ planets (filled circles) and for $V \geq 10$ planets (open circles) in both (a) and (b). Planets with measured radii but no mass are similarly plotted with small filled and open squares in (b). The colour scale indicates the equilibrium temperature, $T_\textrm{eq}$. The location of the radius valley (grey dashed curve) and the boundaries of the Neptunian desert features are shown in (b).}
    \label{fig:demographics}
\end{figure}

Fig. \ref{fig:demographics} presents exoplanet mass-period ($M_\textrm{p}$-$P$) and radius-period ($R_\textrm{p}$-$P$) demographic plots and illustrates some of the statistics discussed in \S \ref{section:intro}. {Assuming stellar activity periodicities are present in all three stellar hosts, \hdone{} is the only planet candidate with strong evidence in this study. The $M_\textrm{p}$-$P$ panel in Fig. \ref{fig:demographics}\,a (large symbols) shows  the location of \hdone{}\,b (Model~3) and the tentative \hdsix{}\,b (Model~3b) and \hdtwo{}\,b (Model~3) planet candidates identified with moderate evidence in \S \ref{section:results}. Because we cannot definitively rule out the purely Keplerian models for \hdone{} and \hdtwo{}, we also plot \hdone{}\,c$^{\prime\prime}$ and \hdtwo{}\,b$^{\prime\prime}$ from the Model~1b solutions (smaller symbols).
%For completeness, we \hdsix{}\,b$^{\prime\prime}$,\,c$^{\prime\prime}$\,d$^{\prime\prime}$. 
Fig. \ref{fig:demographics}\,a also shows} all confirmed planets\footref{note1} for which an estimated equilibrium temperature, $T_\textrm{eq}$, and either a mass or minimum mass, $m_\textrm{p}\,\textrm{sin}\,i$, are known. The \hdsix{} and \hdtwo{} planet candidates belong to the population of less massive planets in the Mass-Period diagram, whereas the 0.72 Saturn mass of \hbox{\hdone{}\,b}  places it in the upper region of the Neptunian savannah. \hdsix{}\,b is the coolest candidate planet presented in this paper, with equilibrium temperature of $T_\textrm{eff} = 616$\,K, assuming an average solar-system Bond albedo of $A_\textrm{B} = 0.36$. The remaining planets receive considerably higher instellation, indicating equilibrium temperatures of $1041\,\textrm{K} <  T_\textrm{eff} < 1393\,\textrm{K}$.

Fig. \ref{fig:demographics}\,b shows the $R_\textrm{p}$-$P$ diagram for planets with an estimated equilibrium temperature and either a measured transit radius, {or in the case of our candidates, a predicted radius (\citetalias{muller24})}. The location of the radius valley is shown, where $R_\textrm{p}(P) = (1.88/11.2)(P/10)^{-0.11}$ $M_\textrm{J}$ \citep{affolter2023}. The boundaries related to the Neptunian desert, as defined by \citealt{castro-gonzalez24} (\citetalias{castro-gonzalez24}), are also plotted. The small open squares in Fig. \ref{fig:demographics}\,b illustrate the sheer number of $V>10$ planets with measured radius, but no mass measurement. In addition to radii, the radius uncertainties for our planets are also obtained from \citetalias{muller24} and represent the ranges arising from the population spread of radius for a given mass. The predicted radii and their uncertainties are given for the {main} solutions for each system in Tables \ref{tab:hd67200_solution}, \ref{tab:hd118006_GPsolution} and \ref{tab:hd2134_GPsolution}.
%Assuming a $\sin i$ distribution of exoplanet orbital planes, 
{Assuming an isotropic distribution of orbital orientations,}
true exoplanet masses derived from RVs should on average be 18.8\% higher than the measured minimum mass. The corresponding adjusted radii for the inclination distribution mean of $i=57.3$\degs{} are given for each target in Tables \ref{tab:hd67200_solution}, \ref{tab:hd118006_GPsolution} \& \ref{tab:hd2134_GPsolution} to enable an assessment of potential systematic radius bias. The difference between the radii predicted from \mpsini{} and the adjusted radii are probably pessimistic as the DMPP hypothesis leads us to expect edge-on or near edge-on systems. Assuming the lack of transits is not limited by data quality, orbital inclinations of $i < 82$\degs~--~$83$\degs{} are implied for our planets. Orbits with these marginal non-detection inclinations yield true masses that are $<1$\% greater than \mpsini{}.

DMPP-7\,b has a well defined period and small minimum mass uncertainty of $\Delta M/M \sim 2.6$\%. The main uncertainty in the corresponding {predicted} radius ($\Delta R_\textrm{p} / R$ = 22\%) arises from the relatively larger spread in measured radii at higher masses (\citetalias{muller24}). {At $69.0$\,M$_\oplus$, equivalently $0.22$\,M$_\textrm{J}$, $0.72$\,M$_\textrm{Sat}$ or $4.0$\,M$_\textrm{Nept}$, \hdone{}\,b} lies at the upper end of the mass range where radii increase rapidly; hence the \textrm{apparently large predicted $R_\textrm{p} = 9.6 \pm 2.1$\,R$_\oplus$ or $0.87 \pm 0.19$\,R$_\textrm{J}$}. It lies at the $5$\,d boundary of the Neptunian ridge and Neptunian savannah regions recently identified by \citetalias{castro-gonzalez24}. Since such planets are expected to form beyond the ice line, at periods longer than shown in Fig. \ref{fig:demographics}, and rocky bodies with sufficient mass are generally expected to rapidly accrete to become Jupiter-mass/radius planets, the Neptunes in the ridge and savannah are likely to have been disrupted in some way. Disc migration is expected to result in low-eccentricity planets as is found in the savannah region ($P>5$\,d), whereas the Kozai-Lidov mechanism would lead to inclined orbits and the higher eccentricities that are found in the over-dense Neptunian ridge (\citetalias{castro-gonzalez24}). The low eccentricity of \hdone{}\,b favours the disc migration mechanism. A smaller radius would intriguingly place \hdone{}\,b closer to the Ridge over-density. However, with no transit signal apparent in the four TESS sectors at the $\sim 5$\,d orbital period of the planet, we are unable to obtain a direct radius estimate.

{Since the Bayesian evidence does not enable the purely Keplerian solution to be formally rejected, we consider the \hdone{}\,c$^{\prime\prime}$ candidate, corresponding to Model~1b. \citet{burn24steam} updated the Bern planet population synthesis model \citep{emsenhuber21bernmodel} to include mixed H/He + H$_2$O envelopes. By simulating migrating ``steam worlds'', they are able to explain the radius valley at $\sim 1.7$\,R$_\oplus$ and also the radius cliff at $\sim 3.0$\,R$_\oplus$, which is clear in Fig. \ref{fig:demographics}\,b. The radius cliff marks the transition from planets with H$_2$O rich atmospheres to exoplanets that contain several tens of per cent of H/He by mass. This corresponds to the drop in occurrence rate at around the mass of Neptune, where rapid gas accretion then leads to the more massive gas giant population.  The location of \hdone{}\,c$^{\prime\prime}$ with a predicted mass of ${14.4}^{+3.2}_{-2.8}$ \,M$_\oplus$ and radius of $3.4^{+0.6}_{-0.7}$\,R$_\oplus$ (respectively $19$\% and $12$\% higher if $i=57.3$\degs{}) is thus intriguing as it is potentially represents one of the relatively few (13-16\%) Neptune-mass planets simulated by \citet{burn24steam} with some H/He in their atmospheres.}

{The Keplerian orbital periods and amplitudes for the moderately significant $N_\textrm{p} = 1$ models for \hdsix{} (Model~3b) and \hdtwo{} (Model 3) are both small.} The data sampling for most of the observing runs, with multiple observations per night, means that we are particularly sensitive to $1$\,d\,--\,$2$\,d signals, regardless of whether they are stellar or dynamical in origin. Solar supergranulation effects are generally seen on $0.5$\,d\,--\,$2$\,d timescales \citep{rincon18sg} {with amplitudes of $0.7-\gtrsim 2$\,\ms{} \citep{osullivan25sg}}. While modelling of the Sun has been carried out by \citet{meunier15sg}, stellar `jitter' that includes convective granulation and supergranulation effects is typically expected to be $\sim 2$\,\ms{} for the lowest activity stars with \logrhk{}~$\sim -5.0$ \citep{meunier19activity}. Further investigation of these effects by \citet{meunier20sg} found potentially significant impact on the recovery of habitable zone planet masses, with the effect being most prominent {in} F stars; up to 100\% mass uncertainties are possible for $1$\,M$_\oplus$ and $2$\,M$_\oplus$ planets. Analysis of Solar Cycle 24 and 25 observations by \citet{osullivan25sg} reveals that supergranulation is highest, at around $2$\,\ms{} during the solar minimum period. Our two shortest period planet candidates, \hdsix{}\,b and \hdtwo{}\,b are revealed from data taken on multiple year timescales. Coherent periodicities cannot be attributed to stochastic convective jitter, but we cannot rule out the possibility that the RVs are contaminated by stellar convection effects. Depending on exact periodicity interplay, one may expect masses to  be overestimated if short period signals comprise both stochastic and Keplerian components.

{Fig. \ref{fig:demographics} shows that the \hdsix{}\,b and \hdtwo{}\,b candidates, which lie below the radius valley with respective $T_\textrm{eff} = 1650$ and $1487$\,K ($A_\textrm{B} = 0$), are likely to be atmosphereless planets that could be responsible for mass-loss in these systems.} Given their implied equilibrium temperatures, exospheres and mass loss may result from the presence of liquid magma; as noted by  \citet{staab20dmpp1}, $1100$\,K--\,$1500$\,K is typical for liquid magma on the Earth \citep{sigurdsson2015encyclopedia}. Even for exoplanets such as HD\,219134\,b, that are not hot enough to suggest liquid magma oceans might be present, \citet{vidotto18} showed that an exosphere of up to several planetary radii can be formed by sputtering of surface particles from a dense stellar wind. This explanation was also posited by \citet{barnes20dmpp3} as a possible explanation for mass loss from DMPP-3A\,b and may be important for the slightly cooler equilibrium temperatures expected when $A_\textrm{B} > 0$.

{Because the  $N_\textrm{p} = 2$ model for \hdtwo{} is not conclusively rejected, we plot \hdtwo{}\,c$^{\prime\prime}$ in Fig. \ref{fig:demographics} and note the implied planet would likely be a steam world with a significant atmosphere.
%To highlight the potential pitfalls of using standard periodogram analyses with recursive addition of purely Keplerian signals, we plot the location of the implied planets from the \hdsix{} Model~1 solutions. \hdsix{}\,b$^{\prime\prime}$,\,c$^{\prime\prime}$\,d$^{\prime\prime}$ would imply planets spanning the entire low-mass and low-radius populations of Fig. \ref{fig:demographics}, with \hdsix{}\,c$^{\prime\prime}$ lying inside the radius valley (a likely evaporating steam world) and \hdsix{}\,d$^{\prime\prime}$ representing another H/He world, similar to \hdone{}\,c$^{\prime\prime}$. 
It is clear that high precision RV instruments and more intensive monitoring campaigns are needed to efficiently monitor \hdsix{} and \hdtwo{} to confirm and improve solutions or remove inconclusive candidates. }

\section{Conclusions}

{We have found strong Bayesian evidence in support of a $0.72$ Saturn mass planet, \hdone{}\,b, and moderate evidence for two low mass planet candidates orbiting \hdsix{} and \hdtwo{}. The existence of ultra-short orbiting exoplanets \citep{rappaport12,rappaport14,sanchis-ojeda15,hon25} and the lack of previously reported massive exoplanets for the bright systems in this study led us to expect the presence of predominantly low-mass, short-period planets. Even for the lowest activity stars, the burden of proof for dynamically induced radial velocity signals is a significant obstacle for low amplitude signals.
%It is easy to overstate the case for multiplanet systems when purely Keplerian models are assumed. 
The evidence for multiplanet systems can be overstated when purely Keplerian models are assumed.
Models that include activity generally yield different configurations with fewer planet candidates, as we have demonstrated for \hdsix{}. Although short period signals are easier to identify with short observing campaigns, stochastic stellar activity with similar timescales and amplitudes can also easily contaminate or obscure true Keplerian signals.}

{Modelling the number of Keplerians as a free parameter with nested sampling in conjunction with GPs has facilitated an objective way to present competing models and intra-modal signals with similar significance. Adopting a Gaussian Process such as the } \textsc{s+leaf} { ESP kernel, with relatively physical hyperparameters, can prove problematic without careful consideration of appropriate priors. This further compounds the difficulties of correctly modelling sparsely sampled data. Nevertheless, we have demonstrated that the ESP kernel is remarkably stable if the number of harmonics and their relative contributions are carefully controlled (\S \ref{section:data_analysis} and Appendix \ref{section:appendixGP}). }

In the wider context, the future recovery of masses for the under-represented fainter, low-mass planetary systems, from the Kepler and K2 missions in particular, must adopt methods {offering} a full exploration of posterior space. This approach offers the best chance of obtaining the least biased outcomes with the optimal combination of signals for which strong evidence is present while enabling an assessment of more tentative signals with moderate evidence. This will be especially important when the small, fainter host stars targeted by missions such as Ariel \citep{edwards22ariel} are adopted. The considerable observing time needed to obtain reliable Ariel mission target masses \citep{barnes22ariel}, will be further optimised with Bayesian analysis tools such as \texttt{kima}. To put this in context, approximately 73\% of known systems with measured radii and $V>10$ do not currently have measured masses and will require {time on the largest optical and near infrared facilities, which are under high demand.}

The {planet candidates} we report here belong to quite different parts of the mass-period and radius-period diagrams. Despite their {likely} differing formation mechanisms, they all have high instellation and projected equilibrium temperatures, that could lead to mass-loss via different mechanisms. The dominant species that are `evaporated' may differ, depending on the presence or absence of significant atmospheres. \citet{lammer22exosphere}, for instance, have reviewed the various mechanisms that can lead to mass loss of refractory elements from close orbiting rocky exoplanets with liquid magma {ocean}. Studies of the signatures of multiple species expected from ongoing mass loss are needed and may shed further light on the bulk compositions of the components of exoplanets in the systems studied here.

\section*{Acknowledgements}
This paper is based on data taken with the European Southern Observatory (ESO) 3.6 m telescope under programme IDs: 095.C-0799(A), 097.C-0390(B), 098.C-0269(A), 096.C0499(A), 098.C-0269(B), 098.C-0269(B), 099.C-0798(A), 0100.C-0836(A) and 0110.248C.001 and the VLT facility under programme ID: 114.27LM.001. This paper includes data collected by the TESS mission and obtained from the MAST data archive at the Space Telescope Science Institute (STScI). STScI is operated by the Association of Universities for Research in Astronomy, Inc., under NASA contract NAS 5–26555.

JRB and CAH were funded by STFC under consolidated grant ST/T000295/1 and ST/X001164/1. MRS acknowledges support from the European Space Agency as an ESA Research Fellow, and was supported by grant ST/T000295/1 from the Science and Technology Facilities Council (STFC) for the early part of this work. ATS, ZOBR and ER were supported by Science and Technology Facilities Council (STFC) studentships. 

\section*{Data Availability}
The data underlying this article will be shared on reasonable request to the corresponding author.



%\bibliographystyle{mnras}
\bibliographystyle{mnras}
\bibliography{master,ownrefs}

\newcommand{\noop}[1]{}
\begin{thebibliography}{}
\makeatletter
\relax
\def\mn@urlcharsother{\let\do\@makeother \do\$\do\&\do\#\do\^\do\_\do\%\do\~}
\def\mn@doi{\begingroup\mn@urlcharsother \@ifnextchar [ {\mn@doi@}
  {\mn@doi@[]}}
\def\mn@doi@[#1]#2{\def\@tempa{#1}\ifx\@tempa\@empty \href
  {http://dx.doi.org/#2} {doi:#2}\else \href {http://dx.doi.org/#2} {#1}\fi
  \endgroup}
\def\mn@eprint#1#2{\mn@eprint@#1:#2::\@nil}
\def\mn@eprint@arXiv#1{\href {http://arxiv.org/abs/#1} {{\tt arXiv:#1}}}
\def\mn@eprint@dblp#1{\href {http://dblp.uni-trier.de/rec/bibtex/#1.xml}
  {dblp:#1}}
\def\mn@eprint@#1:#2:#3:#4\@nil{\def\@tempa {#1}\def\@tempb {#2}\def\@tempc
  {#3}\ifx \@tempc \@empty \let \@tempc \@tempb \let \@tempb \@tempa \fi \ifx
  \@tempb \@empty \def\@tempb {arXiv}\fi \@ifundefined
  {mn@eprint@\@tempb}{\@tempb:\@tempc}{\expandafter \expandafter \csname
  mn@eprint@\@tempb\endcsname \expandafter{\@tempc}}}

\bibitem[\protect\citeauthoryear{{Affolter}, {Mordasini}, {Oza}, {Kubyshkina}
  \& {Fossati}}{{Affolter} et~al.}{2023}]{affolter2023}
{Affolter} L.,  {Mordasini} C.,  {Oza} A.~V.,  {Kubyshkina} D.,   {Fossati} L.,
   2023, \mn@doi [\aap] {10.1051/0004-6361/202142205}, \href
  {https://ui.adsabs.harvard.edu/abs/2023A&A...676A.119A} {676, A119}

\bibitem[\protect\citeauthoryear{{Ahn} et~al.,}{{Ahn}
  et~al.}{2012}]{ahn2012sdss9}
{Ahn} C.~P.,  et~al., 2012, \mn@doi [\apjs] {10.1088/0067-0049/203/2/21}, \href
  {https://ui.adsabs.harvard.edu/abs/2012ApJS..203...21A} {203, 21}

\bibitem[\protect\citeauthoryear{Ambikasaran}{Ambikasaran}{2015}]{ambikasaran15}
Ambikasaran S.,  2015, \mn@doi [Numerical Linear Algebra with Applications]
  {https://doi.org/10.1002/nla.2003}, 22, 1102

\bibitem[\protect\citeauthoryear{{Anglada-Escud{\'e}}
  et~al.,}{{Anglada-Escud{\'e}} et~al.}{2013}]{anglada13}
{Anglada-Escud{\'e}} G.,  et~al., 2013, \mn@doi [\aap]
  {10.1051/0004-6361/201321331}, \href
  {http://cdsads.u-strasbg.fr/abs/2013A%26A...556A.126A} {556, A126}

\bibitem[\protect\citeauthoryear{{Anglada-Escud{\'e}}
  et~al.,}{{Anglada-Escud{\'e}} et~al.}{2016}]{anglada16proxima}
{Anglada-Escud{\'e}} G.,  et~al., 2016, \mn@doi [\nat] {10.1038/nature19106},
  \href {https://ui.adsabs.harvard.edu/abs/2016Natur.536..437A} {536, 437}

\bibitem[\protect\citeauthoryear{{Arriagada}}{{Arriagada}}{2011}]{arriagada11activities}
{Arriagada} P.,  2011, \mn@doi [\apj] {10.1088/0004-637X/734/1/70}, \href
  {https://ui.adsabs.harvard.edu/abs/2011ApJ...734...70A} {734, 70}

\bibitem[\protect\citeauthoryear{{Barnes} \& {Haswell}}{{Barnes} \&
  {Haswell}}{2022}]{barnes22ariel}
{Barnes} J.~R.,  {Haswell} C.~A.,  2022, \mn@doi [Experimental Astronomy]
  {10.1007/s10686-021-09758-0}, \href
  {https://ui.adsabs.harvard.edu/abs/2022ExA....53..589B} {53, 589}

\bibitem[\protect\citeauthoryear{{Barnes}, {Jeffers}, {Haswell}, {Jones},
  {Shulyak}, {Pavlenko}  \& {Jenkins}}{{Barnes} et~al.}{2017}]{barnes17mdwarfs}
{Barnes} J.~R.,  {Jeffers} S.~V.,  {Haswell} C.~A.,  {Jones} H.~R.~A.,
  {Shulyak} D.,  {Pavlenko} Y.~V.,   {Jenkins} J.~S.,  2017, \mn@doi [\mnras]
  {10.1093/mnras/stx1482}, \href
  {https://ui.adsabs.harvard.edu/abs/2017MNRAS.471..811B} {471, 811}

\bibitem[\protect\citeauthoryear{{Barnes} et~al.,}{{Barnes}
  et~al.}{2020}]{barnes20dmpp3}
{Barnes} J.~R.,  et~al., 2020, \mn@doi [Nature Astronomy]
  {10.1038/s41550-019-0972-z}, \href
  {https://ui.adsabs.harvard.edu/abs/2020NatAs...4..419B} {4, 419}

\bibitem[\protect\citeauthoryear{{Barnes}, {Jeffers}, {Haswell}, {Damasso},
  {Del Sordo}, {Liebing}, {Perger}  \& {Anglada-Escud{\'e}}}{{Barnes}
  et~al.}{2024}]{barnes24moments}
{Barnes} J.~R.,  {Jeffers} S.~V.,  {Haswell} C.~A.,  {Damasso} M.,  {Del Sordo}
  F.,  {Liebing} F.,  {Perger} M.,   {Anglada-Escud{\'e}} G.,  2024, \mn@doi
  [\mnras] {10.1093/mnras/stae2125}, \href
  {https://ui.adsabs.harvard.edu/abs/2024MNRAS.534.1257B} {534, 1257}

\bibitem[\protect\citeauthoryear{{Barrag{\'a}n}, {Aigrain}, {Rajpaul}  \&
  {Zicher}}{{Barrag{\'a}n} et~al.}{2022}]{barragan22pyaneti2}
{Barrag{\'a}n} O.,  {Aigrain} S.,  {Rajpaul} V.~M.,   {Zicher} N.,  2022,
  \mn@doi [\mnras] {10.1093/mnras/stab2889}, \href
  {https://ui.adsabs.harvard.edu/abs/2022MNRAS.509..866B} {509, 866}

\bibitem[\protect\citeauthoryear{{Basri} \& {Nguyen}}{{Basri} \&
  {Nguyen}}{2018}]{basri18doubledip}
{Basri} G.,  {Nguyen} H.~T.,  2018, \mn@doi [\apj] {10.3847/1538-4357/aad3b6},
  \href {https://ui.adsabs.harvard.edu/abs/2018ApJ...863..190B} {863, 190}

\bibitem[\protect\citeauthoryear{{Basri} et~al.,}{{Basri}
  et~al.}{2011}]{basri11}
{Basri} G.,  et~al., 2011, \mn@doi [\aj] {10.1088/0004-6256/141/1/20}, \href
  {https://ui.adsabs.harvard.edu/abs/2011AJ....141...20B} {141, 20}

\bibitem[\protect\citeauthoryear{{Berdyugina} \& {Usoskin}}{{Berdyugina} \&
  {Usoskin}}{2003}]{berdyugina03activelongs}
{Berdyugina} S.~V.,  {Usoskin} I.~G.,  2003, \mn@doi [\aap]
  {10.1051/0004-6361:20030748}, \href
  {https://ui.adsabs.harvard.edu/abs/2003A&A...405.1121B} {405, 1121}

\bibitem[\protect\citeauthoryear{{Boyle}, {Mann}  \& {Bush}}{{Boyle}
  et~al.}{2025}]{boyle25}
{Boyle} A.~W.,  {Mann} A.~W.,   {Bush} J.,  2025, \mn@doi [\apj]
  {10.3847/1538-4357/adcecc}, \href
  {https://ui.adsabs.harvard.edu/abs/2025ApJ...985..233B} {985, 233}

\bibitem[\protect\citeauthoryear{{Brewer} \& {Foreman-Mackey}}{{Brewer} \&
  {Foreman-Mackey}}{2016}]{brewer16dnest4}
{Brewer} B.~J.,  {Foreman-Mackey} D.,  2016, \mn@doi [arXiv e-prints]
  {10.48550/arXiv.1606.03757}, \href
  {https://ui.adsabs.harvard.edu/abs/2016arXiv160603757B} {p. arXiv:1606.03757}

\bibitem[\protect\citeauthoryear{{Burn}, {Mordasini}, {Mishra}, {Haldemann},
  {Venturini}, {Emsenhuber}  \& {Henning}}{{Burn} et~al.}{2024}]{burn24steam}
{Burn} R.,  {Mordasini} C.,  {Mishra} L.,  {Haldemann} J.,  {Venturini} J.,
  {Emsenhuber} A.,   {Henning} T.,  2024, \mn@doi [Nature Astronomy]
  {10.1038/s41550-023-02183-7}, \href
  {https://ui.adsabs.harvard.edu/abs/2024NatAs...8..463B} {8, 463}

\bibitem[\protect\citeauthoryear{{Castro-Ginard} et~al.,}{{Castro-Ginard}
  et~al.}{2024}]{castro-ginard24}
{Castro-Ginard} A.,  et~al., 2024, \mn@doi [\aap]
  {10.1051/0004-6361/202450172}, \href
  {https://ui.adsabs.harvard.edu/abs/2024A&A...688A...1C} {688, A1}

\bibitem[\protect\citeauthoryear{{Castro-Gonz{\'a}lez}, {Bourrier},
  {Lillo-Box}, {Delisle}, {Armstrong}, {Barrado}  \&
  {Correia}}{{Castro-Gonz{\'a}lez} et~al.}{2024}]{castro-gonzalez24}
{Castro-Gonz{\'a}lez} A.,  {Bourrier} V.,  {Lillo-Box} J.,  {Delisle} J.-B.,
  {Armstrong} D.~J.,  {Barrado} D.,   {Correia} A.~C.~M.,  2024, \mn@doi [\aap]
  {10.1051/0004-6361/202450957}, \href
  {https://ui.adsabs.harvard.edu/abs/2024A&A...689A.250C} {689, A250}

\bibitem[\protect\citeauthoryear{Christiansen}{Christiansen}{2025}]{christiansen2025}
Christiansen J. L. e.~a.,  2025, \mn@doi [Planetary Science Journal]
  {10.3847/PSJ/ade3c2}, 6, 186

\bibitem[\protect\citeauthoryear{{Claytor}, {van Saders}, {Llama}, {Sadowski},
  {Quach}  \& {Avallone}}{{Claytor} et~al.}{2022}]{claytor22TESSdeeplearning}
{Claytor} Z.~R.,  {van Saders} J.~L.,  {Llama} J.,  {Sadowski} P.,  {Quach} B.,
    {Avallone} E.~A.,  2022, \mn@doi [\apj] {10.3847/1538-4357/ac498f}, \href
  {https://ui.adsabs.harvard.edu/abs/2022ApJ...927..219C} {927, 219}

\bibitem[\protect\citeauthoryear{{Colman}, {Angus}, {David}, {Curtis},
  {Hattori}  \& {Lu}}{{Colman} et~al.}{2024}]{colman24tessrotation}
{Colman} I.~L.,  {Angus} R.,  {David} T.,  {Curtis} J.,  {Hattori} S.,   {Lu}
  Y.~L.,  2024, \mn@doi [\aj] {10.3847/1538-3881/ad2c86}, \href
  {https://ui.adsabs.harvard.edu/abs/2024AJ....167..189C} {167, 189}

\bibitem[\protect\citeauthoryear{{Dawson} \& {Johnson}}{{Dawson} \&
  {Johnson}}{2018}]{dawson18}
{Dawson} R.~I.,  {Johnson} J.~A.,  2018, \mn@doi [\araa]
  {10.1146/annurev-astro-081817-051853}, \href
  {https://ui.adsabs.harvard.edu/abs/2018ARA&A..56..175D} {56, 175}

\bibitem[\protect\citeauthoryear{{Delisle}, {Hara}  \&
  {S{\'e}gransan}}{{Delisle} et~al.}{2020}]{delisle20spleaf}
{Delisle} J.-B.,  {Hara} N.,   {S{\'e}gransan} D.,  2020, \mn@doi [\aap]
  {10.1051/0004-6361/201936906}, \href
  {https://ui.adsabs.harvard.edu/abs/2020A&A...638A..95D} {638, A95}

\bibitem[\protect\citeauthoryear{{Edwards} \& {Tinetti}}{{Edwards} \&
  {Tinetti}}{2022}]{edwards22ariel}
{Edwards} B.,  {Tinetti} G.,  2022, \mn@doi [\aj] {10.3847/1538-3881/ac6bf9},
  \href {https://ui.adsabs.harvard.edu/abs/2022AJ....164...15E} {164, 15}

\bibitem[\protect\citeauthoryear{{Emsenhuber}, {Mordasini}, {Burn}, {Alibert},
  {Benz}  \& {Asphaug}}{{Emsenhuber} et~al.}{2021}]{emsenhuber21bernmodel}
{Emsenhuber} A.,  {Mordasini} C.,  {Burn} R.,  {Alibert} Y.,  {Benz} W.,
  {Asphaug} E.,  2021, \mn@doi [\aap] {10.1051/0004-6361/202038553}, \href
  {https://ui.adsabs.harvard.edu/abs/2021A&A...656A..69E} {656, A69}

\bibitem[\protect\citeauthoryear{{Faria}, {Santos}, {Figueira}  \&
  {Brewer}}{{Faria} et~al.}{2018}]{faria18kima}
{Faria} J.~P.,  {Santos} N.~C.,  {Figueira} P.,   {Brewer} B.~J.,  2018,
  \mn@doi [The Journal of Open Source Software] {10.21105/joss.00487}, \href
  {https://ui.adsabs.harvard.edu/abs/2018JOSS....3..487F} {3, 487}

\bibitem[\protect\citeauthoryear{{Faria}, {Santos}, {Figueira}  \&
  {Brewer}}{{Faria} et~al.}{2023}]{faria23kimanote}
{Faria} J.~P.,  {Santos} N.~C.,  {Figueira} P.,   {Brewer} B.~J.,  2023, kima:
  Exoplanet detection in RVs with DNest4 and GPs, \url
  {https://ascl.net/2302.014}

\bibitem[\protect\citeauthoryear{{Foreman-Mackey}, {Agol}, {Ambikasaran}  \&
  {Angus}}{{Foreman-Mackey} et~al.}{2017}]{foremanmackey17celerite}
{Foreman-Mackey} D.,  {Agol} E.,  {Ambikasaran} S.,   {Angus} R.,  2017,
  \mn@doi [\aj] {10.3847/1538-3881/aa9332}, \href
  {https://ui.adsabs.harvard.edu/abs/2017AJ....154..220F} {154, 220}

\bibitem[\protect\citeauthoryear{{Gaia Collaboration} et~al.,}{{Gaia
  Collaboration} et~al.}{2023}]{gaia23dr3}
{Gaia Collaboration} et~al., 2023, \mn@doi [\aap]
  {10.1051/0004-6361/202243940}, \href
  {https://ui.adsabs.harvard.edu/abs/2023A&A...674A...1G} {674, A1}

\bibitem[\protect\citeauthoryear{{Gomes da Silva}, {Figueira}, {Santos}  \&
  {Faria}}{{Gomes da Silva} et~al.}{2018}]{gomesdasilva18actin}
{Gomes da Silva} J.,  {Figueira} P.,  {Santos} N.,   {Faria} J.,  2018, \mn@doi
  [The Journal of Open Source Software] {10.21105/joss.00667}, \href
  {https://ui.adsabs.harvard.edu/abs/2018JOSS....3..667G} {3, 667}

\bibitem[\protect\citeauthoryear{{Gomes da Silva} et~al.,}{{Gomes da Silva}
  et~al.}{2021}]{gomesdasilva2021activities}
{Gomes da Silva} J.,  et~al., 2021, \mn@doi [\aap]
  {10.1051/0004-6361/202039765}, \href
  {https://ui.adsabs.harvard.edu/abs/2021A&A...646A..77G} {646, A77}

\bibitem[\protect\citeauthoryear{{Gomes da Silva}, {Bensabat}, {Monteiro}  \&
  {Santos}}{{Gomes da Silva} et~al.}{2022}]{gomesdasilva2022activities}
{Gomes da Silva} J.,  {Bensabat} A.,  {Monteiro} T.,   {Santos} N.~C.,  2022,
  \mn@doi [\aap] {10.1051/0004-6361/202244595}, \href
  {https://ui.adsabs.harvard.edu/abs/2022A&A...668A.174G} {668, A174}

\bibitem[\protect\citeauthoryear{{Guerrero} et~al.,}{{Guerrero}
  et~al.}{2021}]{guerrero21toi}
{Guerrero} N.~M.,  et~al., 2021, \mn@doi [\apjs] {10.3847/1538-4365/abefe1},
  \href {https://ui.adsabs.harvard.edu/abs/2021ApJS..254...39G} {254, 39}

\bibitem[\protect\citeauthoryear{{Haswell} et~al.,}{{Haswell}
  et~al.}{2020}]{haswell20dmpp}
{Haswell} C.~A.,  et~al., 2020, \mn@doi [Nature Astronomy]
  {10.1038/s41550-019-0973-y}, \href
  {https://ui.adsabs.harvard.edu/abs/2020NatAs...4..408H} {4, 408}

\bibitem[\protect\citeauthoryear{{Hattori}, {Foreman-Mackey}, {Hogg}, {Montet},
  {Angus}, {Pritchard}, {Curtis}  \& {Sch{\"o}lkopf}}{{Hattori}
  et~al.}{2022}]{hattori2022unpopular}
{Hattori} S.,  {Foreman-Mackey} D.,  {Hogg} D.~W.,  {Montet} B.~T.,  {Angus}
  R.,  {Pritchard} T.~A.,  {Curtis} J.~L.,   {Sch{\"o}lkopf} B.,  2022, \mn@doi
  [\aj] {10.3847/1538-3881/ac625a}, \href
  {https://ui.adsabs.harvard.edu/abs/2022AJ....163..284H} {163, 284}

\bibitem[\protect\citeauthoryear{{Hattori}, {Angus}, {Foreman-Mackey}, {Lu}  \&
  {Colman}}{{Hattori} et~al.}{2025}]{hattori25tess}
{Hattori} S.,  {Angus} R.,  {Foreman-Mackey} D.,  {Lu} Y.~L.,   {Colman} I.,
  2025, \mn@doi [\aj] {10.3847/1538-3881/add0ab}, \href
  {https://ui.adsabs.harvard.edu/abs/2025AJ....170...15H} {170, 15}

\bibitem[\protect\citeauthoryear{Hippke \& Heller}{Hippke \&
  Heller}{2019}]{hippke_transit_2019}
Hippke M.,  Heller R.,  2019, \mn@doi [Astronomy \& Astrophysics]
  {10.1051/0004-6361/201834672}, 623, A39

\bibitem[\protect\citeauthoryear{Hippke, David, Mulders  \& Heller}{Hippke
  et~al.}{2019}]{hippke_wotan_2019}
Hippke M.,  David T.~J.,  Mulders G.~D.,   Heller R.,  2019, \mn@doi [The
  Astronomical Journal] {10.3847/1538-3881/ab3984}, 158, 143

\bibitem[\protect\citeauthoryear{{H{\o}g} et~al.,}{{H{\o}g}
  et~al.}{2000}]{hog2000tycho}
{H{\o}g} E.,  et~al., 2000, \aap, \href
  {https://ui.adsabs.harvard.edu/abs/2000A&A...355L..27H} {355, L27}

\bibitem[\protect\citeauthoryear{{Hon} et~al.,}{{Hon} et~al.}{2025}]{hon25}
{Hon} M.,  et~al., 2025, \mn@doi [\apjl] {10.3847/2041-8213/adbf21}, \href
  {https://ui.adsabs.harvard.edu/abs/2025ApJ...984L...3H} {984, L3}

\bibitem[\protect\citeauthoryear{{Houk} \& {Swift}}{{Houk} \&
  {Swift}}{1999}]{Houk1999}
{Houk} N.,  {Swift} C.,  1999, Michigan Spectral Survey, \href
  {https://ui.adsabs.harvard.edu/abs/1999MSS...C05....0H} {5, 0}

\bibitem[\protect\citeauthoryear{{Isaacson} \& {Fischer}}{{Isaacson} \&
  {Fischer}}{2010}]{isaacson10activity}
{Isaacson} H.,  {Fischer} D.,  2010, \mn@doi [\apj]
  {10.1088/0004-637X/725/1/875}, \href
  {https://ui.adsabs.harvard.edu/abs/2010ApJ...725..875I} {725, 875}

\bibitem[\protect\citeauthoryear{{Jenkins} et~al.,}{{Jenkins}
  et~al.}{2011}]{jenkins11activities}
{Jenkins} J.~S.,  et~al., 2011, \mn@doi [\aap] {10.1051/0004-6361/201016333},
  \href {https://ui.adsabs.harvard.edu/abs/2011A&A...531A...8J} {531, A8}

\bibitem[\protect\citeauthoryear{{Jones}, {Haswell}, {Barnes}, {Staab}  \&
  {Heller}}{{Jones} et~al.}{2020}]{jones20dmpp1}
{Jones} M.~H.,  {Haswell} C.~A.,  {Barnes} J.~R.,  {Staab} D.,   {Heller} R.,
  2020, \mn@doi [\apjl] {10.3847/2041-8213/ab8f2b}, \href
  {https://ui.adsabs.harvard.edu/abs/2020ApJ...895L..17J} {895, L17}

\bibitem[\protect\citeauthoryear{{Kipping}}{{Kipping}}{2013}]{kipping13}
{Kipping} D.~M.,  2013, \mn@doi [\mnras] {10.1093/mnrasl/slt075}, \href
  {https://ui.adsabs.harvard.edu/abs/2013MNRAS.434L..51K} {434, L51}

\bibitem[\protect\citeauthoryear{{Klein} et~al.,}{{Klein}
  et~al.}{2024}]{klein24}
{Klein} B.,  et~al., 2024, \mn@doi [\mnras] {10.1093/mnras/stae1313}, \href
  {https://ui.adsabs.harvard.edu/abs/2024MNRAS.531.4238K} {531, 4238}

\bibitem[\protect\citeauthoryear{{Lammer} et~al.,}{{Lammer}
  et~al.}{2022}]{lammer22exosphere}
{Lammer} H.,  et~al., 2022, \mn@doi [\ssr] {10.1007/s11214-022-00876-5}, \href
  {https://ui.adsabs.harvard.edu/abs/2022SSRv..218...15L} {218, 15}

\bibitem[\protect\citeauthoryear{{Laskar}}{{Laskar}}{1997}]{Laskar1997}
{Laskar} J.,  1997, \aap, \href
  {https://ui.adsabs.harvard.edu/abs/1997A&A...317L..75L} {317, L75}

\bibitem[\protect\citeauthoryear{{Laskar}}{{Laskar}}{2000}]{laskar2000}
{Laskar} J.,  2000, \mn@doi [\prl] {10.1103/PhysRevLett.84.3240}, \href
  {https://ui.adsabs.harvard.edu/abs/2000PhRvL..84.3240L} {84, 3240}

\bibitem[\protect\citeauthoryear{{Lindegren} et~al.,}{{Lindegren}
  et~al.}{2021}]{lindegren21dr3}
{Lindegren} L.,  et~al., 2021, \mn@doi [\aap] {10.1051/0004-6361/202039709},
  \href {https://ui.adsabs.harvard.edu/abs/2021A&A...649A...2L} {649, A2}

\bibitem[\protect\citeauthoryear{{Mamajek} \& {Hillenbrand}}{{Mamajek} \&
  {Hillenbrand}}{2008}]{mamajek08age}
{Mamajek} E.~E.,  {Hillenbrand} L.~A.,  2008, \mn@doi [\apj] {10.1086/591785},
  \href {https://ui.adsabs.harvard.edu/abs/2008ApJ...687.1264M} {687, 1264}

\bibitem[\protect\citeauthoryear{{Meunier} \& {Lagrange}}{{Meunier} \&
  {Lagrange}}{2020}]{meunier20sg}
{Meunier} N.,  {Lagrange} A.-M.,  2020, \mn@doi [\aap]
  {10.1051/0004-6361/202038376}, \href
  {https://ui.adsabs.harvard.edu/abs/2020A&A...642A.157M} {642, A157}

\bibitem[\protect\citeauthoryear{{Meunier}, {Lagrange}, {Borgniet}  \&
  {Rieutord}}{{Meunier} et~al.}{2015}]{meunier15sg}
{Meunier} N.,  {Lagrange} A.-M.,  {Borgniet} S.,   {Rieutord} M.,  2015,
  \mn@doi [\aap] {10.1051/0004-6361/201525721}, \href
  {https://ui.adsabs.harvard.edu/abs/2015A&A...583A.118M} {583, A118}

\bibitem[\protect\citeauthoryear{{Meunier}, {Lagrange}  \& {Cuzacq}}{{Meunier}
  et~al.}{2019}]{meunier19activity}
{Meunier} N.,  {Lagrange} A.-M.,   {Cuzacq} S.,  2019, \mn@doi [\aap]
  {10.1051/0004-6361/201935348}, \href
  {https://ui.adsabs.harvard.edu/abs/2019A&A...632A..81M} {632, A81}

\bibitem[\protect\citeauthoryear{{M{\"u}ller}, {Baron}, {Helled}, {Bouchy}  \&
  {Parc}}{{M{\"u}ller} et~al.}{2024}]{muller24}
{M{\"u}ller} S.,  {Baron} J.,  {Helled} R.,  {Bouchy} F.,   {Parc} L.,  2024,
  \mn@doi [\aap] {10.1051/0004-6361/202348690}, \href
  {https://ui.adsabs.harvard.edu/abs/2024A&A...686A.296M} {686, A296}

\bibitem[\protect\citeauthoryear{{Murphy}, {Fossati}, {Bedding}, {Saio},
  {Kurtz}, {Grassitelli}  \& {Wang}}{{Murphy} et~al.}{2016}]{murphy16}
{Murphy} S.~J.,  {Fossati} L.,  {Bedding} T.~R.,  {Saio} H.,  {Kurtz} D.~W.,
  {Grassitelli} L.,   {Wang} E.~S.,  2016, \mn@doi [\mnras]
  {10.1093/mnras/stw705}, \href
  {https://ui.adsabs.harvard.edu/abs/2016MNRAS.459.1201M} {459, 1201}

\bibitem[\protect\citeauthoryear{{Nicholson} \& {Aigrain}}{{Nicholson} \&
  {Aigrain}}{2022}]{nicholson22}
{Nicholson} B.~A.,  {Aigrain} S.,  2022, \mn@doi [\mnras]
  {10.1093/mnras/stac2097}, \href
  {https://ui.adsabs.harvard.edu/abs/2022MNRAS.515.5251N} {515, 5251}

\bibitem[\protect\citeauthoryear{{O'Sullivan} et~al.,}{{O'Sullivan}
  et~al.}{2025}]{osullivan25sg}
{O'Sullivan} N.~K.,  et~al., 2025, \mn@doi [\mnras] {10.1093/mnras/staf1168},
  \href {https://ui.adsabs.harvard.edu/abs/2025MNRAS.541.3942O} {541, 3942}

\bibitem[\protect\citeauthoryear{{Pace}}{{Pace}}{2013}]{pace13}
{Pace} G.,  2013, \mn@doi [\aap] {10.1051/0004-6361/201220364}, \href
  {https://ui.adsabs.harvard.edu/abs/2013A&A...551L...8P} {551, L8}

\bibitem[\protect\citeauthoryear{{Penoyre}, {Belokurov}  \& {Evans}}{{Penoyre}
  et~al.}{2022}]{penoyre22}
{Penoyre} Z.,  {Belokurov} V.,   {Evans} N.~W.,  2022, \mn@doi [\mnras]
  {10.1093/mnras/stac959}, \href
  {https://ui.adsabs.harvard.edu/abs/2022MNRAS.513.2437P} {513, 2437}

\bibitem[\protect\citeauthoryear{{Pepe} et~al.,}{{Pepe}
  et~al.}{2013}]{pepe-espresso13}
{Pepe} F.,  et~al., 2013, The Messenger, \href
  {https://ui.adsabs.harvard.edu/abs/2013Msngr.153....6P} {153, 6}

\bibitem[\protect\citeauthoryear{{Perdelwitz}, {Trifonov}, {Teklu}, {Sreenivas}
   \& {Tal-Or}}{{Perdelwitz} et~al.}{2024}]{perdelwitz24}
{Perdelwitz} V.,  {Trifonov} T.,  {Teklu} J.~T.,  {Sreenivas} K.~R.,   {Tal-Or}
  L.,  2024, \mn@doi [\aap] {10.1051/0004-6361/202348263}, \href
  {https://ui.adsabs.harvard.edu/abs/2024A&A...683A.125P} {683, A125}

\bibitem[\protect\citeauthoryear{{Petit}, {Laskar}  \& {Bou{\'e}}}{{Petit}
  et~al.}{2017}]{petit17amd}
{Petit} A.~C.,  {Laskar} J.,   {Bou{\'e}} G.,  2017, \mn@doi [\aap]
  {10.1051/0004-6361/201731196}, \href
  {https://ui.adsabs.harvard.edu/abs/2017A&A...607A..35P} {607, A35}

\bibitem[\protect\citeauthoryear{{Rappaport} et~al.,}{{Rappaport}
  et~al.}{2012}]{rappaport12}
{Rappaport} S.,  et~al., 2012, \mn@doi [\apj] {10.1088/0004-637X/752/1/1},
  \href {https://ui.adsabs.harvard.edu/abs/2012ApJ...752....1R} {752, 1}

\bibitem[\protect\citeauthoryear{{Rappaport}, {Barclay}, {DeVore}, {Rowe},
  {Sanchis-Ojeda}  \& {Still}}{{Rappaport} et~al.}{2014}]{rappaport14}
{Rappaport} S.,  {Barclay} T.,  {DeVore} J.,  {Rowe} J.,  {Sanchis-Ojeda} R.,
  {Still} M.,  2014, \mn@doi [\apj] {10.1088/0004-637X/784/1/40}, \href
  {https://ui.adsabs.harvard.edu/abs/2014ApJ...784...40R} {784, 40}

\bibitem[\protect\citeauthoryear{{Reinhold} \& {Hekker}}{{Reinhold} \&
  {Hekker}}{2020}]{reinhold20}
{Reinhold} T.,  {Hekker} S.,  2020, \mn@doi [\aap]
  {10.1051/0004-6361/201936887}, \href
  {https://ui.adsabs.harvard.edu/abs/2020A&A...635A..43R} {635, A43}

\bibitem[\protect\citeauthoryear{{Rincon} \& {Rieutord}}{{Rincon} \&
  {Rieutord}}{2018}]{rincon18sg}
{Rincon} F.,  {Rieutord} M.,  2018, \mn@doi [Living Reviews in Solar Physics]
  {10.1007/s41116-018-0013-5}, \href
  {https://ui.adsabs.harvard.edu/abs/2018LRSP...15....6R} {15, 6}

\bibitem[\protect\citeauthoryear{{Sanchis-Ojeda} et~al.,}{{Sanchis-Ojeda}
  et~al.}{2015}]{sanchis-ojeda15}
{Sanchis-Ojeda} R.,  et~al., 2015, \mn@doi [\apj]
  {10.1088/0004-637X/812/2/112}, \href
  {https://ui.adsabs.harvard.edu/abs/2015ApJ...812..112S} {812, 112}

\bibitem[\protect\citeauthoryear{{Serrano} et~al.,}{{Serrano}
  et~al.}{2022}]{serrano22hd93963}
{Serrano} L.~M.,  et~al., 2022, \mn@doi [\aap] {10.1051/0004-6361/202243093},
  \href {https://ui.adsabs.harvard.edu/abs/2022A&A...667A...1S} {667, A1}

\bibitem[\protect\citeauthoryear{Sigurdsson, Houghton, McNutt, Rymer  \&
  Stix}{Sigurdsson et~al.}{2015}]{sigurdsson2015encyclopedia}
Sigurdsson H.,  Houghton B.,  McNutt S.,  Rymer H.,   Stix J.,  eds, 2015, {The
  Encyclopedia of Volcanoes}, 2nd edn.
Elsevier Science \& Technology, Amsterdam, \mn@doi{10.1016/C2011-0-00438-6}

\bibitem[\protect\citeauthoryear{{Silva} et~al.,}{{Silva}
  et~al.}{2022}]{silva22sbart}
{Silva} A.~M.,  et~al., 2022, \mn@doi [\aap] {10.1051/0004-6361/202142262},
  \href {https://ui.adsabs.harvard.edu/abs/2022A&A...663A.143S} {663, A143}

\bibitem[\protect\citeauthoryear{{Silverstein} et~al.,}{{Silverstein}
  et~al.}{2024}]{silverstein24lhs}
{Silverstein} M.~L.,  et~al., 2024, \mn@doi [\aj] {10.3847/1538-3881/ad3040},
  \href {https://ui.adsabs.harvard.edu/abs/2024AJ....167..255S} {167, 255}

\bibitem[\protect\citeauthoryear{{Simola}, {Bonfanti}, {Dumusque},
  {Cisewski-Kehe}, {Kaski}  \& {Corander}}{{Simola} et~al.}{2022}]{simola22}
{Simola} U.,  {Bonfanti} A.,  {Dumusque} X.,  {Cisewski-Kehe} J.,  {Kaski} S.,
   {Corander} J.,  2022, \mn@doi [\aap] {10.1051/0004-6361/202142941}, \href
  {https://ui.adsabs.harvard.edu/abs/2022A&A...664A.127S} {664, A127}

\bibitem[\protect\citeauthoryear{{Skrutskie} et~al.,}{{Skrutskie}
  et~al.}{2006}]{skrutskie062mass}
{Skrutskie} M.~F.,  et~al., 2006, \mn@doi [\aj] {10.1086/498708}, \href
  {https://ui.adsabs.harvard.edu/abs/2006AJ....131.1163S} {131, 1163}

\bibitem[\protect\citeauthoryear{{Soto} \& {Jenkins}}{{Soto} \&
  {Jenkins}}{2018}]{soto18species}
{Soto} M.~G.,  {Jenkins} J.~S.,  2018, \mn@doi [\aap]
  {10.1051/0004-6361/201731533}, \href
  {https://ui.adsabs.harvard.edu/abs/2018A&A...615A..76S} {615, A76}

\bibitem[\protect\citeauthoryear{{Soto}, {Jones}  \& {Jenkins}}{{Soto}
  et~al.}{2021}]{soto21species}
{Soto} M.~G.,  {Jones} M.~I.,   {Jenkins} J.~S.,  2021, \mn@doi [\aap]
  {10.1051/0004-6361/202039357}, \href
  {https://ui.adsabs.harvard.edu/abs/2021A&A...647A.157S} {647, A157}

\bibitem[\protect\citeauthoryear{{Staab} et~al.,}{{Staab}
  et~al.}{2020}]{staab20dmpp1}
{Staab} D.,  et~al., 2020, \mn@doi [Nature Astronomy]
  {10.1038/s41550-020-1074-7}, \href
  {https://ui.adsabs.harvard.edu/abs/2020NatAs...4..427S} {4, 427}

\bibitem[\protect\citeauthoryear{{Standing} et~al.,}{{Standing}
  et~al.}{2022}]{standing22bebpo2}
{Standing} M.~R.,  et~al., 2022, \mn@doi [\mnras] {10.1093/mnras/stac113},
  \href {https://ui.adsabs.harvard.edu/abs/2022MNRAS.511.3571S} {511, 3571}

\bibitem[\protect\citeauthoryear{{Standing}, {Barnes}, {Haswell}  \& {et
  al.}}{{Standing} et~al.}{2026}]{standing26dmpp}
{Standing} M.,  {Barnes} J.~R.,  {Haswell} C.~A.~H.,   {et al.}
  \noop{2026}submitted 2026, MNRAS

\bibitem[\protect\citeauthoryear{{Steffen} et~al.,}{{Steffen}
  et~al.}{2012}]{steffen12}
{Steffen} J.~H.,  et~al., 2012, \mn@doi [Proceedings of the National Academy of
  Science] {10.1073/pnas.1120970109}, \href
  {https://ui.adsabs.harvard.edu/abs/2012PNAS..109.7982S} {109, 7982}

\bibitem[\protect\citeauthoryear{{Stevenson}, {Haswell}, {Faria}, {Barnes},
  {Barstow}, {Dickinson}  \& {Standing}}{{Stevenson}
  et~al.}{2025a}]{stevenson25eccentricities}
{Stevenson} A.~T.,  {Haswell} C.~A.,  {Faria} J.~P.,  {Barnes} J.~R.,
  {Barstow} J.~K.,  {Dickinson} H.,   {Standing} M.~R.,  2025a, \mn@doi
  [\mnras] {10.1093/mnras/staf502}, \href
  {https://ui.adsabs.harvard.edu/abs/2025MNRAS.539..727S} {539, 727}

\bibitem[\protect\citeauthoryear{{Stevenson}, {Haswell}, {Barnes}, {Standing},
  {Barstow}, {Ross}, {Freckelton}  \& {Staab}}{{Stevenson}
  et~al.}{2025b}]{stevenson25hd28471}
{Stevenson} A.~T.,  {Haswell} C.~A.,  {Barnes} J.~R.,  {Standing} M.~R.,
  {Barstow} J.~K.,  {Ross} Z.~O.~B.,  {Freckelton} A.~V.,   {Staab} D.,  2025b,
  \mn@doi [\mnras] {10.1093/mnras/staf1405}, \href
  {https://ui.adsabs.harvard.edu/abs/2025MNRAS.543...28S} {543, 28}

\bibitem[\protect\citeauthoryear{{Trotta}}{{Trotta}}{2008}]{trotta08bayes}
{Trotta} R.,  2008, \mn@doi [Contemporary Physics] {10.1080/00107510802066753},
  \href {https://ui.adsabs.harvard.edu/abs/2008ConPh..49...71T} {49, 71}

\bibitem[\protect\citeauthoryear{{Twicken} et~al.,}{{Twicken}
  et~al.}{2018}]{twicken18keplervalidation}
{Twicken} J.~D.,  et~al., 2018, \mn@doi [\pasp] {10.1088/1538-3873/aab694},
  \href {https://ui.adsabs.harvard.edu/abs/2018PASP..130f4502T} {130, 064502}

\bibitem[\protect\citeauthoryear{Twicken et~al.,}{Twicken
  et~al.}{2020}]{Twicken2020TESS}
Twicken J.~D.,  et~al., 2020, NASA Technical Memorandum NASA/TM-20205008729,
  TESS Science Data Products Description Document, \url
  {https://archive.stsci.edu/files/live/sites/mast/files/home/missions-and-data/active-missions/tess/_documents/EXP-TESS-ARC-ICD-TM-0014-Rev-F.pdf}.
NASA Ames Research Center, \url
  {https://archive.stsci.edu/files/live/sites/mast/files/home/missions-and-data/active-missions/tess/_documents/EXP-TESS-ARC-ICD-TM-0014-Rev-F.pdf}

\bibitem[\protect\citeauthoryear{{Vidotto} et~al.,}{{Vidotto}
  et~al.}{2018}]{vidotto18}
{Vidotto} A.~A.,  et~al., 2018, \mn@doi [\mnras] {10.1093/mnras/sty2130}, \href
  {https://ui.adsabs.harvard.edu/abs/2018MNRAS.481.5296V} {481, 5296}

\bibitem[\protect\citeauthoryear{{Zechmeister} \& {K{\"u}rster}}{{Zechmeister}
  \& {K{\"u}rster}}{2009}]{zechmeister09gls}
{Zechmeister} M.,  {K{\"u}rster} M.,  2009, \mn@doi [\aap]
  {10.1051/0004-6361:200811296}, \href
  {https://ui.adsabs.harvard.edu/abs/2009A&A...496..577Z} {496, 577}

\bibitem[\protect\citeauthoryear{{Zink} \& {Howard}}{{Zink} \&
  {Howard}}{2023}]{zink23}
{Zink} J.~K.,  {Howard} A.~W.,  2023, \mn@doi [\apjl]
  {10.3847/2041-8213/acfdab}, \href
  {https://ui.adsabs.harvard.edu/abs/2023ApJ...956L..29Z} {956, L29}

\bibitem[\protect\citeauthoryear{{dos Santos} et~al.,}{{dos Santos}
  et~al.}{2016}]{dossantos16}
{dos Santos} L.~A.,  et~al., 2016, \mn@doi [\aap]
  {10.1051/0004-6361/201628558}, \href
  {https://ui.adsabs.harvard.edu/abs/2016A&A...592A.156D} {592, A156}

\makeatother
\end{thebibliography}

%\begin{thebibliography}{99}
%\end{thebibliography}

%%%%%%%%%%%%%%%%%%%%%%%%%%%%%%%%%%%%%%%%%%%%%%%%%%

%%%%%%%%%%%%%%%%% APPENDICES %%%%%%%%%%%%%%%%%%%%%

\appendix

\section{Gaussian Process implementation in Kima}
\protect\label{section:appendixGP}

The standard  squared-exponential periodic (SEP) kernel in \texttt{kima} is defined as

\begin{equation}
k_{\rm SEP}(\tau)
=
\eta_1^{\,2}
\exp \left[
 -\frac{\tau^{2}}{2\eta_2^{\,2}}
 -\frac{2\sin^{2} \left(\pi\tau/\eta_3\right)}{\eta_4^{\,2}}
\right].
\label{eqn:qpgp}
\end{equation}

\noindent
where $\tau = t_i - t_j$ is the time lag between observations,
$\eta_1$ is the GP amplitude,
$\eta_2$ is the exponential decay timescale, and
$\eta_3$ is the characteristic period.
The amount of sub-structure within each cycle is controlled by $\eta_4$,
which is related to the ``harmonic complexity'' $\Gamma$.
\texttt{Kima} internally uses $\Gamma = 2/\eta_4^2$
whereas a more common convention in the literature is
$\Gamma = 1/(2\eta_4^2)$.

\medskip
\noindent
The SEP kernel requires a full Cholesky decomposition of the
covariance matrix, with computational cost scaling as
$\mathcal{O}(N_\textrm{data}^3)$.
A more efficient semiseparable (\texttt{celerite}-like) representation providing a linear scaling of $\sim \mathcal{O}(N_\textrm{data})$
\citep{foremanmackey17celerite} is provided by the
\textsc{s+leaf} framework \citep{delisle20spleaf}, which includes the
Exponential--Sine Periodic (ESP) kernel.
The ESP kernel used in \texttt{kima} has the form

\begin{multline}
k_{\mathrm{ESP}}(\tau)
=
\left\{
\eta_1^{\,2}\,
\exp \left(
 -\frac{|\tau|}{\eta_2}
\right)
\exp \left[
 -\frac{\bigl(2\sin(\pi \tau / \eta_3)\bigr)^{2}}{\eta_4^{\,2}}
\right]
\right\}
\\[8pt]
\shoveleft{
\times\,
\left\{
  \sum_{h=0}^{n_{\mathrm{harm}}}
  \tilde I_h \left(\frac{1}{\eta_4^{\,2}}\right)
  \cos \left(h\,\frac{2\pi}{\eta_3}\,\tau\right)
\right\},
}
\label{eqn:esp2}
\end{multline}

\noindent
with the normalised coefficients
\begin{equation}
\tilde I_h(f)
=
\frac{I_h(f)}{\displaystyle\sum_{k=0}^{n_{\mathrm{harm}}} I_k(f)},
\qquad
f = \frac{1}{\eta_4^{2}}.
\label{eqn:modbessel}
\end{equation}

\noindent
where $n_{\mathrm{harm}}$ is the number of harmonics.
The first term in Eqn.~\ref{eqn:esp2} is the
Exponential--Sine (ES) envelope, which combines an exponential decay
($\eta_2$) with a periodic smoothing term ($\eta_3$, $\eta_4$).
The second part of Eqn.~\ref{eqn:esp2} is a truncated Fourier--Bessel
series that captures periodic structure through the normalised modified Bessel
functions $\tilde I_h(f)$ in Eqn. \ref{eqn:modbessel}. By default \texttt{kima} restricts this to
$n_{\mathrm{harm}}=3$, corresponding to the fundamental period
$\eta_3$ and its first two harmonics ($\eta_3/2$ and $\eta_3/3$). Table \ref{tab:harmonics}  shows how choice of $\eta_4$ controls the modified Bessel function harmonic weights.

\begin{table*}
\centering
\caption{Correspondence between $\eta_4$ and harmonic complexity in \texttt{Kima}'s implementation of the S+LEAF ESP kernel \citep{delisle20spleaf}. Normalised harmonic weights for the ESP kernel using $f = 1/\eta_4^2$ as the modified Bessel argument. The harmonic complexity is implemented in \texttt{kima} as $\Gamma = 2/\eta_4^2$. 
Weights $w_h = I_h(f)\big/\sum_{k=0}^{3} I_k(f)$ for $h=1,2,3$ 
correspond to the fundamental characteristic period $\eta_3$, the first harmonic $\eta_3/2$, 
and second harmonic $\eta_3/3$, respectively ($I_0$ is a constant term).
The weighting ratios $w_2/w_1$, $w_3/w_2$ and 
$w_3/w_1$ respectively corresponding to $(\eta_3/2)/\eta_3$, $(\eta_3/3)/(\eta_3/2)$ and $(\eta_3/3)/\eta_3$ quantify the relative harmonic strengths.}
\begin{tabular}{cccccccc}
\hline
$\eta_4$ & $f=1/\eta_4^2$ & $\Gamma=2f$
%& $w_1$ (P) & $w_2$ (P/2) & $w_3$ (P/3) 
& $w_2/w_1$ & $w_3/w_2$ & $w_3/w_1$ 
& Comments \\
\hline
0.1  & 100  & 200   
%& 0.2531 & 0.2493 & 0.2431 
& 0.9850 & 0.9752 & 0.9606 
& Extremely harmonic-rich \\

0.2  & 25 & 50      
%& 0.2624 & 0.2468 & 0.2229 
& 0.9406 & 0.9031 & 0.8495 
& Very high harmonic complexity \\

0.5  & 4.0 & 8.0     
%& 0.3167 & 0.2084 & 0.1083 
& 0.6580 & 0.5196 & 0.3420 
& Strong $\eta_3/2$ and $\eta_3/3$; {well-suited to RV activity signals} \\

0.8  & 1.5625 & 3.125  
%& 0.3246 & 0.1156 & 0.0287 
& 0.3561 & 0.2482 & 0.0884 
& {Classic double-dip morphology} ($\eta_3/2$ $\approx 36$\,\%) \\

1.0  & 1.0  & 2.0
%& 0.2841 & 0.06824 & 0.01114 
& 0.2402 & 0.1633 & 0.03923 
& Weak double-dip; modest harmonic content \\

%1.2  & 0.6944   
%%& 0.2359 & 0.04015 & 0.00460 
%& 0.1702 & 0.1146 & 0.01951 
%& Very weak double-dip; nearing purely single-dip behaviour \\

1.5  & 0.4444 & 0.8889 
%& 0.1746 & 0.01924 & 0.00142 
& 0.1102 & 0.07377 & 0.00813 
& $\eta_3/2$ $\approx 11$\%; approaching sinusoidal \\

%1.7  & 0.3460  & 0.6920
%%& 0.1437 & 0.01237 & 0.000712 
%& 0.08608 & 0.05753 & 0.00495 
%& Very weak harmonics; effectively sinusoidal \\

2.0  & 0.25 & 0.50
%& 0.1096 & 0.006830 & 0.000284 
& 0.06234 & 0.04161 & 0.00259 
& Shallow $\eta_3/2$ $\approx 6$\% \\

%3.0  & 0.1111 & 0.2222
%%& 0.05248 & 0.001457 & 0.00002697 
%& 0.02776 & 0.01851 & 0.000514 
%& Very weak harmonics; near-perfect sinusoid \\

5.0  & 0.04 & 0.08  
%& 0.01960 & 0.0001960 & 0.00000131 
& 0.009999 & 0.006666 & $6.7\times 10^{-5}$ 
& Almost pure sinusoid \\

%8.0  & 0.015625 
%%& 0.007751 & 0.00003028 & $7.9\times10^{-8}$ 
%& 0.003906 & 0.002604 & $1.0\times 10^{-5}$ 
%& Pure sinusoid; harmonics negligible \\

10.0 & 0.01 & 0.02     
%%& 0.004975 & 0.00001244 & $2.1\times10^{-8}$ 
& 0.002500 & 0.001667 & $4.2\times 10^{-6}$ 
& Essentially pure sinusoid \\ % (matches \texttt{kima} behaviour) \\

%20.0 & 0.0025   
%%& 0.001248 & $7.80\times10^{-7}$ & $3.25\times10^{-10}$ 
%& 0.000625 & 0.000417 & $2.6\times10^{-7}$ 
%& Harmonics vanish completely \\
\hline
\end{tabular}
\label{tab:harmonics}
\end{table*}

\section{TESS lightcurves}
\protect\label{section:appendixTESS}

\begin{figure*}
\includegraphics[trim=0mm 0mm 0mm 0mm, width=1.98\columnwidth]{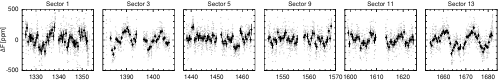} \\
\includegraphics[trim=0mm 0mm 0mm 0mm, width=1.98\columnwidth]{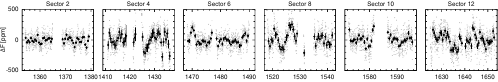} \\
\includegraphics[trim=0mm 0mm 0mm 0mm, width=1.98\columnwidth]{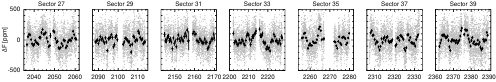} \\
\includegraphics[trim=0mm 0mm 0mm 0mm, width=1.98\columnwidth]{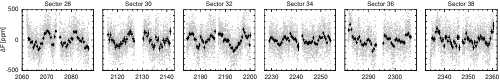} \\
\includegraphics[trim=0mm 0mm 0mm 0mm, width=1.98\columnwidth]{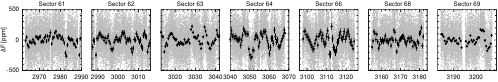} \\
\includegraphics[trim=0mm 0mm 0mm 0mm, width=1.98\columnwidth]{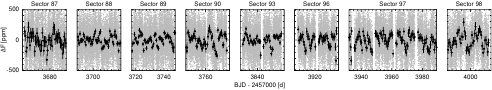} \\
%\vspace{1mm} \\
%%\includegraphics[trim=4mm 0mm 0mm 0mm, height=0.41\columnwidth]{DMPPmulti/HD67200_unpopular-lightcurve-kimasolution-27-39.png}
%\vspace{1mm} \\
%\includegraphics[trim=4mm 0mm 0mm 0mm, height=0.41\columnwidth]{DMPPmulti/HD67200_unpopular-lightcurve-kimasolution-61-69.png}
%\vspace{1mm} \\
%\includegraphics[trim=4mm 0mm 0mm 0mm, height=0.45\columnwidth]{DMPPmulti/HD67200_unpopular-lightcurve-kimasolution-87-93.png} \\
\caption{\hdsix{} TESS photometry showing the unbinned data (light grey), the $0.5$\,d binned points (black). Year 1 is plotted as odd and even sectors in rows 1 and 2. Year 3 is similarly plotted in rows 3 and 4. See Table \ref{tab:TESS_GP_params} for statistics and periodicities.}
    \label{fig:hd67200_lightcurve}
\end{figure*}

\begin{figure*}
%OLD version - before optimising Unpopular
%\includegraphics[trim=5mm 0mm 5mm 2mm, width=2.0\columnwidth]{DMPPmulti/HD118006_unpopular-lightcurve-kimasolution.png}
\includegraphics[trim=0mm 0mm 0mm 0mm, width=1.98\columnwidth]{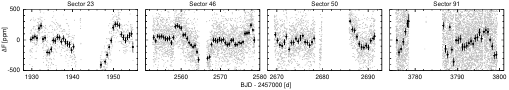}
\caption{\hdone{} TESS photometry.}
    \label{fig:hd118006_lightcurve}
\end{figure*}

\begin{figure*}
\includegraphics[trim=0mm 0mm 0mm 0mm, width=1.98\columnwidth]{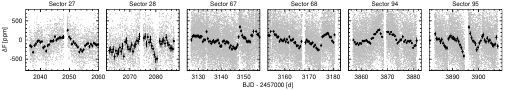}
\caption{\hdtwo{} TESS photometry.}
    \label{fig:hd2134_lightcurve}
\end{figure*}

%Each of the three targets were observed with the Transiting Exoplanet Survey Satellite (TESS) in multiple sectors and observing cycles. The TESS Simple Aperture Photometry (SAP) and Pre-Conditioned SAP (PDCSAP) lightcurves \citep{Twicken2020TESS} based on the shorter cadence observations are optimised for transit searches. As discussed in \citetalias{standing26dmpp}, the co-trending basis vectors used to obtain the PDCSAP data can lead to removal of stellar variability, while the potentially more suitable SAP data contain large systematic effects that makes them unsuitable for analysis of low-activity stars\footnote{see \url{https://archive.stsci.edu/missions-and-data/tess} and the TESS archive manual at \url{https://outerspace.stsci.edu/display/TESS/TESS+Archive+Manual}}. We instead performed photometric extraction of the TESS full frame images (FFIs) using the \texttt{unpopular} package \citep{hattori2022unpopular}. 
We plot the lightcurves obtained using \texttt{unpopular} \citep{hattori2022unpopular} in Figs \ref{fig:hd67200_lightcurve}, \ref{fig:hd118006_lightcurve} and \ref{fig:hd2134_lightcurve}. \texttt{Unpopular} uses distant pixels in the TESS FFIs to model and detrend systematic effects and employs regularisation to prevent over-fitting. We optimised tunable parameters and found similar settings to those adopted by \citet{boyle25} gave optimal results. Cutouts of $64 \times 64$ pixels were downloaded from the FFIs, and 200 predictive pixels with the ``Similar Brightness'' method were used. We found that L2 regularisation with $0.01$, or $0.1$ was suitable for our targets, which minimised the scatter while avoiding over-fitting \citep{hattori2022unpopular}. The periods discussed in \citetalias{standing26dmpp} and this paper are tabulated in Table \ref{tab:TESS_GP_params}. The $R_\textrm{var}$ column in Table \ref{tab:TESS_GP_params} lists the difference in ppm between the 95th and 5th percentile of the sorted flux after $3$\,{hr} binning. The relatively high ratio of $R_\textrm{var} / A$ may result from changing activity levels, but probably arises largely from TESS systematics. For \hdsix{}, the data span fairly constant ranges, with $297$\,ppm~$< R_\textrm{var} < 354$\,ppm). This variability is lower than the majority of K2 amplitudes with successful period determinations \citet{reinhold20} or TESS \citet{boyle25}. Only 0.24\% of the \citet{reinhold20} $32387$ K2 period determinations possessed $R_\textrm{var} < 359$\,ppm. Similarly, only 1.27\% of their period determinations were made for lightcruves with $R_\textrm{var} < R_\textrm{var}(\textrm{\hdone{}}) = 546$\,ppm. Although \cite{hattori25tess} were able to recover periods of $10-100$\,d, their sample also comprised stars with typically an order of magnitude greater variability than our targets.

\begin{table}
\renewcommand{\arraystretch}{1.3}
    \centering

\begin{tabular}{ccccc}
\hline
TESS            & Sectors & P   & A     & $R_\textrm{var}$  \\
year            &         & [d] & [ppm] & [ppm]             \\        
\hline\\[-12pt]

\multicolumn{5}{c}{{\hdsix{}}} \\[0pt]
%#1-13
$1$     &   $1$--$5$, $9$--$13$ (odd)        & 10.4  & 48       & 333 \\
%2-12
        &   $2$--$12$ (even)                &  6.23 & 30 (N/S)  & 354 \\
%#27-39
$3$     &   $27$--$39$ (odd)                & 13.7  & 47        & 297 \\
%#28-38
        &   $28$--$38$ (even)               & 10.7  & 46        & 297 \\
%#61-69
$5$     &   $61$--$64$,\,$66,\,68,\,69$    &  4.52 & 38        & 345 \\
%#87-93
$7$     &  $87$--$90,\,93,\,96$--$98$       & 3.65  & 49        & 359 \\

\multicolumn{5}{c}{{\hdone{}}} \\[0pt]

$2,4,7$ &   $23,46,50,91$                   &  21.5 & 111       & 546 \\

\multicolumn{5}{c}{{\hdtwo{}}} \\

$3,5,7$ &   $27,28,67,68,94,95$             &  26.4 & 105       & 535 \\
[2pt]
\hline   
\end{tabular}  
\caption{TESS Photometry results for each target. GLS periodogram analysis was used to recover the sinusoidal periodicities ($P$) and amplitudes ($A$). $R_\textrm{var}$ measures the $5$th--$95$th percentile span of the $3$\,hr-binned differential flux.}  
\label{tab:TESS_GP_params}
\end{table}

\FloatBarrier

%%%%%%%%%%%%%%%%%%%%%%%% HD67200 TESS photometry

%%%%%%%%%%%%%%%%%%%%%%%% HD2134 GP analysis of TESS lightcurves 

%If you want to present additional material which would interrupt the flow of the main paper, it can be placed in an Appendix which appears after the list of references.

%%%%%%%%%%%%%%%%%%%%%%%%%%%%%%%%%%%%%%%%%%%%%%%%%%

% Don't change these lines
\bsp	% typesetting comment
\label{lastpage}
\end{document}